\documentclass[12pt]{article}
 \pdfoutput=1

\usepackage{savesym,dsfont,ytableau}
\usepackage{cite}
\usepackage{picture}
\usepackage{wasysym}
\usepackage{tikz, tikz-cd}
\usepackage{amscd}
\usepackage{graphicx}
\usepackage{pifont}
\usepackage{float}
\usepackage{subfig}
\usepackage[all]{xy}
\usepackage{multicol}
\usepackage{amsfonts}
\usepackage{picture}
\usepackage{amssymb}
\savesymbol{iint}
\savesymbol{iiint}
\usepackage{amsmath}
\restoresymbol{TXF}{iint}
\restoresymbol{TXF}{iiint}
\usepackage{color}
\usepackage{clay}
\usepackage{hyperref}
\usepackage{slashed}
\hypersetup{colorlinks=true}
\hypersetup{linkcolor=black}
\hypersetup{citecolor=black}
\hypersetup{urlcolor=black}
\usepackage{setspace}
\usepackage{multirow}
\usepackage{wrapfig}
\usepackage{verbatim}
\usepackage{accents}
\numberwithin{equation}{section}

\newcommand\fro{{\overline{\underline{\Omega}}}}

\newcommand*{\dt}[1]{
  \accentset{\mbox{\large.}}{#1}}
\def\dot{\dt}

\begin{document}

\date{June, 2016}

\institution{IAS}{\centerline{${}^{1}$School of Natural Sciences, Institute for Advanced Study, Princeton, NJ, USA}}
\institution{PI}{\centerline{${}^{2}$Perimeter Institute for Theoretical Physics, Waterloo, Ontario, Canada N2L 2Y5}}
\institution{HarvardU}{\centerline{${}^{3}$Jefferson Physical Laboratory, Harvard University, Cambridge, MA, USA}}

\title{Infrared Computations of Defect Schur Indices}

\authors{Clay C\'{o}rdova\worksat{\IAS}\footnote{e-mail: {\tt clay.cordova@gmail.com}}, Davide Gaiotto\worksat{\PI}\footnote{e-mail: {\tt dgaiotto@perimeterinstitute.ca}},  and Shu-Heng Shao\worksat{\HarvardU}\footnote{e-mail: {\tt shuhengshao@gmail.com}} }

\abstract{We conjecture a formula for the Schur index of  four-dimensional  ${\mathcal N}=2$ theories in the presence of boundary conditions and/or line defects, in terms of the low-energy effective Seiberg-Witten description of the system together with massive BPS excitations. We test our proposal in a variety of examples for $SU(2)$ gauge theories, either conformal or asymptotically free. We use the conjecture to compute these defect-enriched Schur indices for theories which lack a Lagrangian description, such as Argyres-Douglas theories. We demonstrate in various examples that line defect indices can be expressed as sums of characters of the associated two-dimensional chiral algebra and that for Argyres-Douglas theories the line defect OPE reduces in the index to the Verlinde algebra.}

\maketitle

\setcounter{tocdepth}{3}
\tableofcontents

\section{Introduction}

The Schur index, introduced in \cite{Kinney:2005ej,Gadde:2011ik,Gadde:2011uv}, is a specialization of the superconformal index of four-dimensional ${\cal N}=2$ theories.  It depends on a single fugacity $q,$ and counts only superconformal multiplets which are quarter BPS. As a trace over the Hilbert space on $S^{3}$ it is defined as
\begin{equation}
\mathcal{I}(q)=\mathrm{Tr}\left[(-1)^{F}q^{\Delta-R}\right]~,
\end{equation}
where $\Delta$ is the scaling dimension and $R$ is the Cartan of the $SU(2)_{R}$ symmetry.\footnote{See Section  \ref{sec:IRReview} for a discussion of our specific choice of fermion number $(-1)^{F}.$}  Because of the enhanced supersymmetry, the Schur index is highly computable even for non-Lagrangian theories \cite{Gadde:2010te,Gaiotto:2012xa,Rastelli:2014jja,Buican:2015ina,Cordova:2015nma}.  For instance, in the context of class $\mathcal{S}$ theories, the Schur index equals the $q$-deformed topological Yang-Mills partition function \cite{Gadde:2009kb,Gadde:2011ik,Gadde:2011uv,Kawano:2012up,Fukuda:2012jr, Song:2015wta}. More recently, it was demonstrated that the local operators counted by this index secretly form a two-dimensional chiral algebra and correspondingly, $\mathcal{I}(q)$ is the vacuum character of that algebra \cite{Beem:2013sza} (see also \cite{Beem:2014rza,Lemos:2014lua ,Buican:2016arp,Xie:2016evu,Cecotti:2015lab,arakawa2015joseph,arakawa2016, BR} for further developments). 

One consequence of the extra supersymmetry of the Schur index is that it can be modified by adding to the system a variety of half-BPS defects, including in particular line defects and interfaces or boundary conditions.  Supersymmetric line defects have been previously studied in \cite{Kapustin:2005py, Gukov:2006jk, Kapustin:2006hi, Kapustin:2007wm, Drukker:2009tz, Gaiotto:2010be, Xie:2013lca, Aharony:2013hda, Xie:2013vfa,Coman:2015lna}, while BPS boundary conditions in $\mathcal{N}=2$ theories were investigated in \cite{DeWolfe:2001pq, Gaiotto:2008sa, Gaiotto:2008ak, Cecotti:2011iy, Dimofte:2011py, Dimofte:2012pd, Dimofte:2013lba}.   Indices enriched by these defects have been computed in \cite{Ito:2011ea, Dimofte:2011py, Gang:2012yr, Chang:2015ofn} and are the main objects of study in this work.

The structure that results strongly resembles that of the four-dimensional ellipsoid ($S^4_b$) partition function decorated by defects \cite{Pestun:2007rz,Hosomichi:2010vh,Drukker:2009id,Drukker:2010jp,Gomis:2011pf,Hama:2012bg}. Moreover, the relationship between the two indices is analogous to the relationship between the three-dimensional ellipsoid ($S^3_b$) partition function and the superconformal index of three-dimensional ${\cal N}=2$ theories \cite{Dimofte:2011py}. A striking feature of the ellipsoid partition function in both three and four dimensions is that although the original partition function can be defined for superconformal field theories by a conformal transformation from flat space, these partition functions can also be defined for non-conformal theories as well. 

The situation with the Schur index appears to be similar, and recent work by some of the authors \cite{Cordova:2015nma} strongly indicates that: 
\begin{itemize}
\item The Schur index is a meaningful quantity for non-conformal ${\cal N}=2$ field theories.
\item The Schur index can be computed in the IR using the Seiberg-Witten description \cite{Seiberg:1994rs,Seiberg:1994aj} as an Abelian gauge theory 
enriched by BPS particles. The details of the calculation depend on the choice of a chamber in the Coulomb branch but the result does not. 
\end{itemize}
Although the two statements are logically independent, they are closely related.  Once we accept that the index can be computed on the Coulomb branch, which spontaneously breaks both conformal symmetry and $U(1)_{r}$, it is natural to expect that these symmetries are not needed to define the index.\footnote{The problem of extending the Schur index to non-conformal systems is being investigated in \cite{DFZ}.}  

The general idea that wall-crossing invariant generating functions of BPS states in four-dimensional field theories are related to the local operators at the UV superconformal fixed point originated in \cite{Cecotti:2010fi,Iqbal:2012xm} following related ideas in \cite{Cecotti:1992rm}.   For instance, \cite{Cecotti:2010fi} found a relationship between the BPS spectrum and the $U(1)_{r}$ charges of chiral operators.  Our IR formulation of the Schur index and its generalization to line defects draws heavily from these works and their prescriptions for constructing such generating functions.

For non-conformal field theories there are two possible interpretations of the Schur index, which are not obviously equivalent.  It may be a counting function of certain supersymmetric local operators, or an index for an $S^{3}$ compactification of the theory.  We cannot use a conformal transformation to directly relate these two perspectives.  However, it is likely that although the physical theory depends on the conformal factor, the Witten index does not and thus both perspectives are valid. 

The operator counting perspective leads to an immediate UV definition for Lagrangian theories. It should be straightforward to reproduce the UV calculation by localization in a judicious supersymmetric compactification of the UV Lagrangian. On the other hand, as we discuss, the sphere compactification perspective gives an intuitive motivation for the IR formulation of the index using BPS states.  It would be interesting to justify the IR formula directly at the level of operator counting, perhaps by employing some map from local operators to 
fans of BPS particles as in \cite{Gaiotto:2015zna, Gaiotto:2015aoa}. In any case, in \cite{Cordova:2015nma} the agreement between the UV and IR calculations it was checked for various 
asymptotically free examples. 

There is a strong analogy between many computations in this paper and calculations in two-dimensional $(2,2)$ supersymmetric quantum field theories. Indeed, this analogy was a central motivation of \cite{Cecotti:2010fi} to define chamber independent combinations of BPS states. The dictionary proceeds as follows:
\begin{itemize}
\item The elliptic genus $\chi(y,p;\alpha) = \mathrm{Tr} (-1)^F y^{F_L} p^{L_0} \bar p^{\bar L_0} \alpha^{J_0}$ is the $2d$ version of the general superconformal index.  It does not depend on $\bar p$ and counts holomorphic operators.  
\item The specialization to $y=1$ given as $\chi(\alpha) = \mathrm{Tr} (-1)^F p^{L_0} \bar p^{\bar L_0} \alpha^{J_0}$ only receives contributions from chiral operators. It is analogous to the Schur index.
\item The specialization $\chi(\alpha)$ can be computed in mass-deformed or even asymptotically free theories. The Cecotti-Vafa formula \cite{Cecotti:1992rm} expresses $\chi(\alpha)$ in terms of the spectrum of IR BPS solitons. 
\item Although $\chi(\alpha)$ naturally counts chiral operators, in a non-conformal theory it can also be computed as the Witten index of a special twisted $S^1$ compactification, which gives an intuitive  understanding of the Cecotti-Vafa formula \cite{Cecotti:1992rm, Gaiotto:2015zna, Gaiotto:2015aoa}. 
\end{itemize}
In a separate work  we will discuss hybrid 2$d$/4$d$ systems at some length \cite{Cordova:2017ohl, Cordova:2017mhb} which involve coupling the $4d$ theory to surface defects.\footnote{The Schur index and chiral algebra in the presence of surface defects is also studied in \cite{BPR}.}

\subsection{The IR Formula for the Schur Index}
\label{sec:irSchurintro}
In the absence of defects, the formal IR expression for the Schur index is \cite{Cordova:2015nma}
\begin{equation}
\label{proposal}
\mathcal{I}(q) = (q)_\infty^{2r}~\text{Tr} \left[\mathcal{O}(q) \right]~,
\end{equation}
where $r$ is the rank of the Coulomb branch and the various symbols are defined as:
\begin{itemize}
\item $X$ stands for a quantum torus algebra 
\begin{equation}
X_{\gamma}X_{\gamma'}=q^{\frac{1}{2}\langle \gamma, \gamma' \rangle}X_{\gamma+\gamma'}~,\label{abelianope}
\end{equation}
of operators labelled by the IR charge lattice $\Gamma$ of the ${\cal N}=2$ gauge theory.
\item The charge lattice $\Gamma$ has a flavor sub-lattice $\Gamma_f$. The trace $\text{Tr}$ sets to $0$ all the $X_{\gamma}$ variables which carry non-zero gauge charge, i.e., such that $\gamma \notin \Gamma_f$. The surviving $X_{\gamma_f}$ variables commute and are identified with flavor fugacities.
\item $\mathcal{O}(q)$ is defined as a product of quantum Kontsevich-Soibelman (KS) factors over the whole BPS spectrum,
ordered along the phases of their central charges: 
\begin{equation}
\mathcal{O}(q) = \prod_{\gamma\in \Gamma}^{\curvearrowleft} K(q;X_\gamma;\Omega_{j}(\gamma))~. \label{IRschur}
\end{equation}
\end{itemize}
We will refer to the operator $\mathcal{O}(q) $ as the \textit{quantum monodromy operator}.\footnote{For notational simplicity, we suppress the $X$ dependence in the quantum KS operator $\mathcal{O}(q)$, and similarly for the quantum spectrum generator $\mathcal{S}_\vartheta(q),$ and the generating functions for the line defects $F(L,\vartheta)$ introduced below.} 
An important property of this formal IR expression for the Schur index is that it is invariant under wall-crossing: the quantum KS wall-crossing formula \cite{Kontsevich:2008fj, Dimofte:2009tm} guarantees that the ordered product of quantum KS factors remains unchanged across walls of marginal stability. This moduli invariance was also noted in \cite{Cecotti:2010fi} where traces of this type were first considered. Wall-crossing invariant is clearly necessary for \eqref{proposal} to make sense.  The generalization to traces of higher powers of $\mathcal{O}(q)$ and their $4d$ interpretation  have been considered in \cite{Cecotti:2015lab}, and an extension to five dimensions was explored in \cite{Papageorgakis:2016cej}.

The infrared formulation \eqref{proposal} of the index has a simple heuristic interpretation.  It is the Schur index as naively computed from the infrared QED description of the Coulomb branch physics.  Indeed, the factors $(q)_{\infty}^{2}$ are Schur indices of Abelian vector multiplets while KS factor contributions of BPS particles are the Schur indices for massive hypermultiplets with appropriate spin.  Finally, the trace selects the gauge invariant states which carry vanishing electric and magnetic charge.

Although this is a suggestive caricature of the physics encapsulated by the IR formula \eqref{proposal}, it ignores a crucial feature: the BPS particles carry both electric and magnetic charges so there is no local field description which includes these objects as fields.  This observation is intimately related to the appearance of non-commutative variables $X_{\gamma}$ and the need for a prescribed ordering of the KS factors to resolve ambiguities.   

To understand these issues it is helpful to unpack \eqref{IRschur} and interpret it on the sphere.  The quantum KS factors $K(q;X_\gamma;\Omega_{j}(\gamma))$ are graded Witten indices of Fock spaces of BPS particles of charge $\gamma$. We can decompose them by particle number 
\begin{equation}
K(q;X_\gamma;\Omega_{j}(\gamma)) = \sum_{n \geq 0} N(q;n, \Omega_{j}(\gamma))X_{n \gamma}~.
\end{equation}
Thus the Schur index can be written, at least formally, as a sum over Fock spaces of BPS particles 
\begin{equation}
\mathcal{I}(q) = \sum_{n_\gamma \geq 0} \left[\prod_{\gamma\in \Gamma} N(q;n, \Omega_{j}(\gamma))\right] (q)_\infty^{2r}~\text{Tr} \left[  \prod_\gamma^{\curvearrowleft} X_{n_\gamma \gamma} \right]~, \label{formalsum}
\end{equation}
The expression multiplying the Fock space degeneracies is just the Schur index of the free IR Abelian gauge theory, decorated by a collection of Abelian 't Hooft-Wilson line defects (the $X_{\gamma}$) which describe the coupling of the corresponding massive BPS particles to the low-energy Abelian gauge theory. The interpretation as line defects also helps to clarify their non-commutative nature.  A mutually non-local pair of line defects sources an electromagnetic field which contains angular momentum accounted for by  $q$ in \eqref{abelianope}.

\begin{figure}
\centering
\includegraphics[width=.6\textwidth]{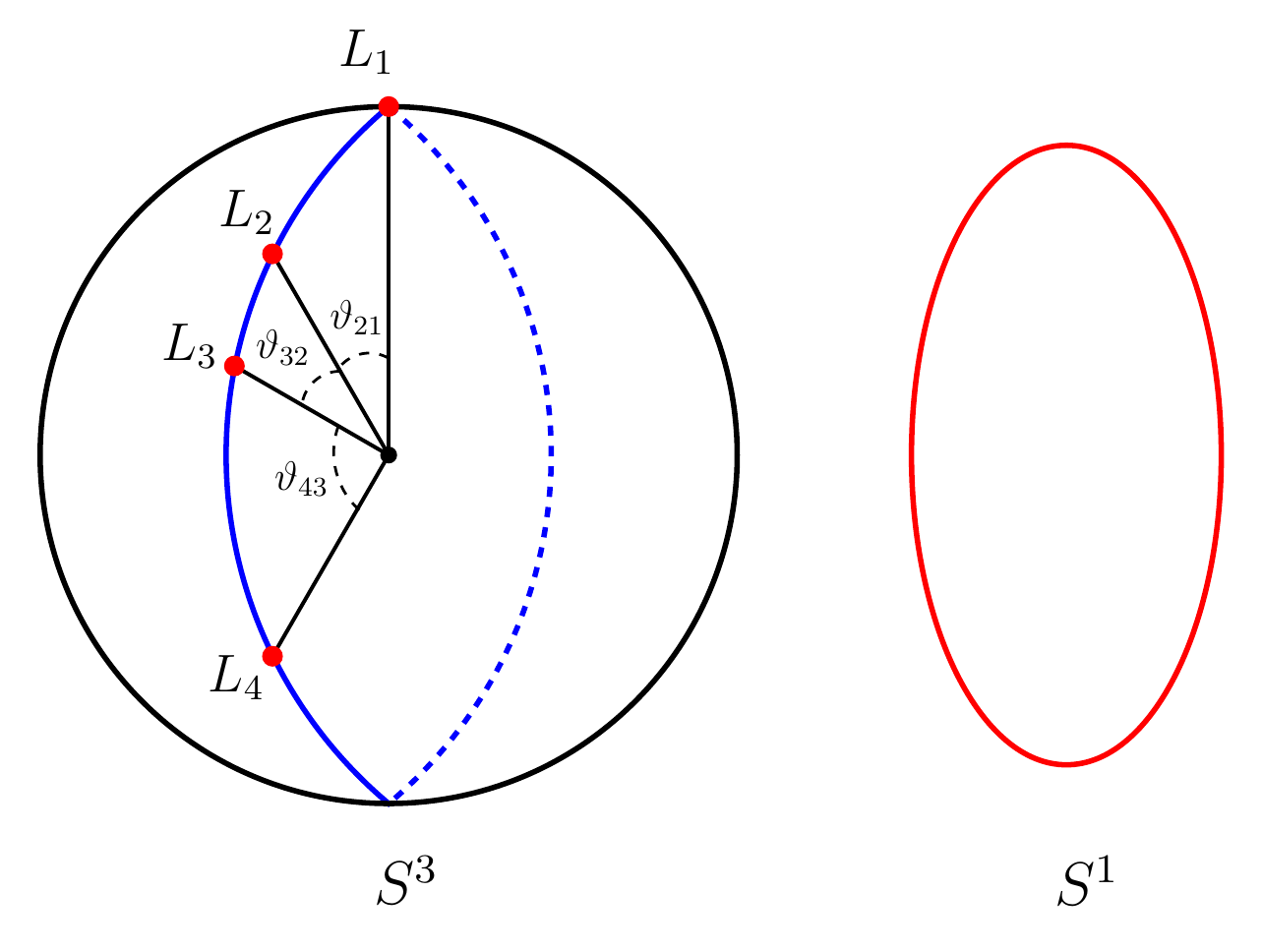}
\caption{The geometry of  line defects. When we conformally map to $S^3\times S^1$, each (half) line defect $L_i$ wraps around the $S^1$ and sits at a point on a great circle (blue line) on the $S^3$ according to their phases $\vartheta_i$. Here $\vartheta_{ij}=\vartheta_i-\vartheta_j$. The worldline of the line defect is shown in red.}\label{fig:S3S1}
\end{figure}

Another advantage of viewing the BPS particles as line defects on the sphere is that we can now understand the ordering prescription in \eqref{IRschur}.  Like BPS particles, half-BPS line defects $L$ preserve a combination of supercharges controlled by a phase $\vartheta$ (See Appendix \ref{sec:supercharge} for details).  When we conformally map to the sphere, this phase determines where they sit along a great circle. See Figure \ref{fig:S3S1}.  Therefore, the ordered product simply takes into account the position of the effective line defects along this circle.

We refer to these expressions as formal because the Schur index is expected to be a power series in $q$. On the other hand, the crucial factor $\text{Tr} \left[  \prod_{\gamma\in \Gamma}^{\curvearrowleft} X_{n_\gamma \gamma} \right]$ may produce powers of $q$ which can be arbitrarily negative as the gauge charges increase. These negative powers are supposed to cancel out in the final answer, but in general the summation in \eqref{formalsum} is conditionally convergent and hence the expression, and claimed cancellations, are ill-defined without a precise prescription for the order in which the $n_{\gamma}$ are summed.

This important technical point limited the checks of the IR calculation of the Schur index in \cite{Cordova:2015nma} to UV non-conformal gauge theories and non-Lagrangian examples, in special chambers only.  Here we resolve this issue and in Section \ref{sec:nf4schur} explicitly evaluate the IR formula for the Schur index of $SU(2)$ superconformal QCD.  We find a precise match with the index obtained by localization.

Specifically, we first by introduce the \textit{quantum spectrum generator} $\mathcal{S}(q)$ defined as the product in the $[0,\pi)$ sector of the quantum KS operator \cite{Kontsevich:2008fj}
\begin{equation}
\mathcal{S}(q)= \prod_{\arg(Z_{\gamma})\in [0,\pi)}^{\curvearrowleft} K(q;X_\gamma;\Omega_{j}(\gamma))~.
\end{equation}
Similarly the conjugate $\mathcal{\bar S}(q)$ is defined as the product in the $[\pi,2\pi)$ sector.

We then refine \eqref{IRschur}, and propose that for all well-defined ${\cal N}=2$ theories the coefficients of $X_\gamma$ in the quantum spectrum generator $\mathcal{S}(q)$ can be expanded as a power series in $q$, starting from a non-negative power which grows fast enough as $\gamma$ grows so that the trace 
\begin{equation}
\mathcal{I}(q) = (q)_\infty^{2r}~\text{Tr} \left[\mathcal{S}(q)\mathcal{\bar S}(q)  \right]~,
\end{equation}
is defined as a power series in $q$. We discuss further aspects of this proposal in Section \ref{sec:IRReview}. 

\subsection{The IR Formula for the Schur Index with Line Defects}

In Section \ref{sec:LineSchur} we introduce line defects and study aspects of their indices from both UV and IR points of view.  We concentrate on line defects inserted at a single point on $S^{3}$ which wrap the $S^{1}$ (see Figure \ref{fig:halffull}a).  After a decompactification of the $S^1$ and a conformal map to $\mathbb{R}^4$, these defects are rays which end at the origin.  Indices in the presence of such a defect may be thought of heuristically as counting gauge non-invariant local operators which can absorb the charge carried by the defect.

To extend our conjecture to include the insertions of line defects is straightforward.  As we have already discussed, Abelian line defects are represented in our setup by the quantum torus variables $X_{\gamma}$.  Thus, from the IR point of view it is trivial to generalize to include additional insertions of Abelian line defects.  Indeed, as described in \cite{Cecotti:2010fi} these are obtained by simply inserting a quantum torus variable $X_{\gamma}$ into the trace.  These insertions of IR line defects and their traces were computed \cite{Cecotti:2010fi} for a variety of examples.

From this point of view, the main physical input required to compute the Schur index in the presence of a line defect is the IR description of a UV line defect.  In general, a given UV line defect $L$ will map in the IR to a superposition of Abelian 't Hooft-Wilson line defects $X_{\gamma}$, each associated to some local ground state of the system in the presence of the UV defect. The Witten indices of these spaces of ground states are dubbed \textit{framed BPS degeneracies} $\fro(L,\gamma,q)$ defined in \cite{Gaiotto:2010be}, and further studied in \cite{Lee:2011ph, Chuang:2013wt, Cirafici:2013bha, Cordova:2013bza, Moore:2015qyu, Moore:2015szp, Gabella:2016zxu}.  These degeneracies are naturally collected into a generating function $F(L,\vartheta)$,
\begin{equation}
F(L,\vartheta)=\sum_{\gamma \in \Gamma}\fro(L,\vartheta,\gamma,q)X_{\gamma}~.
\end{equation}

We conjecture that the Schur index decorated by the defect $L$ with central charge phase $\vartheta$ can be computed by introducing $\mathcal{S}_\vartheta(q)$ defined as a the quantum spectrum generator for the phase range $[\vartheta,\vartheta+\pi)$ and computing the trace with the generating function $F(L,\vartheta)$ inserted
\begin{equation}
\mathcal{I}_{L}(q) = (q)_\infty^{2r}~\text{Tr} \left[ F(L,\vartheta)\mathcal{S}_\vartheta(q) \mathcal{S}_{\vartheta+\pi}(q)\right]~.\label{IRschurLine}
\end{equation}
This is well-defined if the coefficients of the spectrum generators have the expected $q$-expansion.  As a test of this conjecture, in Section \ref{sec:IRSchurExample} we check directly in $SU(2)$ superconformal QCD that this IR formula reproduces UV localization results for non-Abelian Wilson lines inserted at a point on $S^{3}$ and wrapping the $S^{1}$ circle.
 
A crucial feature of our conjecture \eqref{proposal} is that it is wall-crossing invariant.  Indeed, the framed BPS states which govern the decomposition of a UV defect $L$ jump as moduli are varied.  However the framed wall-crossing formula \cite{Gaiotto:2010be} ensures that these line defect indices are invariant.

In Section \ref{sec:LineAD} we use \eqref{IRschurLine} as a tool to compute line defect indices in the Argyres-Douglas theories \cite{Argyres:1995jj,Argyres:1995xn, Eguchi:1996vu}. These are non-Lagrangian theories arising from special loci on the moduli space of more familiar $\mathcal{N}=2$ gauge theories.  They can also be engineered from M5-brane compactifications \cite{Bonelli:2011aa, Xie:2012hs,Xie:2012jd,Xie:2013jc}. Their BPS spectra on the moduli space have been studied extensively in \cite{Shapere:1999xr,Cecotti:2010fi,Gaiotto:2010be,Cecotti:2011rv,Alim:2011ae,Alim:2011kw}.  The superconformal indices of these theories were considered in \cite{Buican:2015ina,Cordova:2015nma,Buican:2015tda,Song:2015wta, Maruyoshi:2016tqk}. The application to these examples illustrates the power of the conjecture: although these models are strongly interacting, in the IR it is possible to reconstruct the properties of their UV line operators using their known framed BPS spectra.  

We conclude our discussion of line defects in Section \ref{sec:verlinde} by discussing a variety of experimental connections between line defect indices and chiral algebras.  This relationship was first found in \cite{Cecotti:2010fi}, where it was pointed out that the insertion of IR line defects into the trace can yield characters of chiral algebras, and moreover that there is a connection between IR line defect OPEs and the $2d$ Verlinde algebra.  Indeed, this connection between chiral algebras and traces of the quantum monodromy was an important clue toward formulating the conjecture of \cite{Cordova:2015nma} relating the Schur index and the BPS spectrum.

We generalize these observations using our calculations of UV line defect indices.  In all models where we have obtained explicit expressions, the Schur index in the presence of a UV line defect produces a sum of characters of the chiral algebra associated to the $4d$ theory.  Moreover, we also find that for Argyres-Douglas theories the UV line defect operator product expansions, when inserted into the Schur index, reduce to the associated Verlinde algebra in the $q\to 1$ limit as anticipated in \cite{Cecotti:2010fi}.

As a specific example of these results, consider $SU(2)$ superconformal QCD and let $L$ be a half Wilson line in the doublet.  According to \cite{Beem:2013sza}, the chiral algebra is the affine Kac-Moody algebra $\widehat{so(8)}_{-2},$  and we find that line defect index $\mathcal{I}_{L}(q)$ can be written as the following linear combination of characters of $\widehat{so(8)}_{-2}$,
\begin{align}
\begin{split}
\mathcal{I}_L(q)& = \sum_{k=1}^\infty (-1)^{k} q^{ k^2+k-1\over2 } (1-q^k) \,\chi_{[-2k-1, 2k-1,0,0,0]}(q)~,
\end{split}
\end{align}
where $ \chi_{[a_0,a_1,a_2,a_3,a_4]}(q)$ is the affine character of $\widehat{so(8)}_{-2}$ with affine Dynkin labels $[a_0,a_1,a_2,a_3,a_4]$, and we have normalized the $\widehat{so(8)}_{-2}$ affine characters to start from order $q^0$.

We are able to explain aspects of these results in theories, like $SU(2)$ superconformal QCD, which are continuously connected to free theories, but leave a complete explanation of these phenomena as an open problem.

\subsection{The IR Formula for the Schur Half-Index}

Finally, in Section \ref{sec:boundary}  we propose an IR expression for the hemisphere index in the presence of some UV boundary condition. As before the key idea is to describe the boundary condition in the IR.  Typically the effective description involves a $3d$ ${\cal N}=2$ theory with a $U(1)^r$ global symmetry which is coupled at the boundary to the bulk IR Abelian gauge fields. Note that the choice of 3$d$ ${\cal N}=2$ subalgebra of the 4$d$ theory also selects a phase $\vartheta$. 

The 3$d$ index for the IR boundary degrees of freedom can be expanded in a charge basis including both electric charges and magnetically charged monopole operators.  We thus obtain a collection of formal $q$ power series $Z_\gamma(q)$ labelled by a bulk charge $\gamma$ and we collect them into a generating function 
\begin{equation}
Z^{IR}(q)[X] = \sum_{\gamma \in \Gamma} Z^{IR}_\gamma(q) X_\gamma~. \label{genhalfintro}
\end{equation}

Given this input, we conjecture that the Schur half-index in the presence of the given boundary condition is
\begin{equation}
\mathcal{II}(q) = (q)_\infty^{r}~\text{Tr} \left[Z^{IR}(q)[X] \mathcal{S}_{\vartheta+\pi}(q) \right]~. \label{IRschurboundary}
\end{equation}
Again, we demonstrate that this formula is wall-crossing invariant: as $\vartheta$ crosses a BPS ray, the IR boundary condition changes in a known way \cite{Dimofte:2013lba} and the 3$d$ index varies in the opposite way as $\mathcal{S}_\vartheta$.
Our formula can be decorated further by line defects in an obvious manner and extended to the case of interfaces.  We verify this formula for the examples of the Dirichlet  and the RG boundary conditions \cite{Dimofte:2013lba} for the pure $SU(2)$ gauge theory.  

The RG boundary conditions of \cite{Dimofte:2013lba} play a key role in our formula.  For any theory, the RG boundary condition has the property that it flows in the IR to simple Dirichlet boundary conditions.  In particular, this means that the 3$d$ index of the RG interface theory can interpreted as an invertible kernel which directly relates IR and UV Schur index calculations.  Moreover this kernel satisfies functional relations which imply the equality of IR and UV formulas for the bulk theory, either bare or decorated by any set of defects. 

As a consequence of this we obtain a novel interpretation of the quantum spectrum generator $\mathcal{S}(q)$.  The Schur half-index in the presence of the RG boundary condition can be expanded in a charge basis as in \eqref{genhalfintro}, and the resulting generating function is simply $(q)_{\infty}^{r}\mathcal{S}(q)$.  Thus the expansion of $\mathcal{S}(q)$  into quantum torus variables can be interpreted as the contribution of the bulk BPS hypermultiplets to the half-index with this boundary condition.  
In this way the IR formula for the Schur half-index gives the abstract quantum wall-crossing formalism a direct physical meaning.

\section{A Review of the Schur Index and its IR Formulation}
\label{sec:IRReview}

\subsection{The Schur Index}

Our discussion of the Schur index follows \cite{Gadde:2011ik,Gadde:2011uv}.  In general for a superconformal field theory with flavor symmetry of rank $n_{f}$ the Schur index is defined as a trace over the Hilbert space on $S^{3}$.  It depends on a single universal fugacity $q$ and may be refined to include flavor fugacities $z_{i}$
\begin{equation}
\mathcal{I}(q,z_{1},\cdots, z_{n_{f}})= \mathrm{Tr}\left[e^{2\pi i R} \,  q^{\Delta-R}\prod_{i=1}^{n_{f}}z_{i}^{f_{i}}\right]~. 
\end{equation}
The state operator correspondence implies that the same quantity may be computed by counting local operators.  These operators are quarter-BPS (annihilated by two $Q$'s and two $S$'s) and obey the following restrictions on their quantum numbers
\begin{equation}
\frac{1}{2} \left(  \Delta-  j_1-j_2 \right) -R=0~, \hspace{.5in}r+ j_1-j_2=0~, \label{schurshortening}
\end{equation}
where $j_{i}$ are  spins for the Lorentz group, $\Delta$ is the scaling dimension, and $R, r$ are Cartans of the $SU(2)_{R}\times U(1)_{r}$ symmetry.\footnote{We use conventions where the doublet of  $SU(2)$ has Cartan eigenvalues $\pm \frac{1}{2}$.}

Note that in our definition, we have chosen a slightly unconventional fermion number  
\begin{align}\label{fermionnumber}
(-1)^F=e^{2\pi i R}~,
\end{align}
 compared to \cite{Gadde:2011ik,Gadde:2011uv} in which $(-1)^F = e^{2\pi i (j_1+j_2)}$. The two conventions of the Schur index are related  by an $\mathbb{Z}_2$ flavor charge insertion $e^{2\pi i (j_1+j_2 +R)}$.  As a consequence of the shortening conditions \eqref{schurshortening} the two conventions are related by shifting $q^{\frac12} \to -q^{\frac12}$.

For theories with a Lagrangian description, the Schur index may be computed by simply counting gauge invariant local operators built out of the free fields.  The fact that it is an index then ensures that the result is correct even for an interacting theory.    This yields a simple matrix integral expression for the index.  The objects entering the expression are single letter partition functions for vector multiplets and hypermultiplets
\begin{equation}
f^{V}(q)=-\frac{2q}{1-q}~, \hspace{.5in} f^{\frac{1}{2}H}=-\frac{q^{1/2}}{1-q}~,
\end{equation}
as well as the plethyestic exponential 
\begin{equation}
P.E.[f(q,u,z)]=\exp\left[\sum_{n=1}^{\infty}\frac{1}{n}f(q^{n},u^n,z^{n})\right]~.
\end{equation}
Note that the sign in $f^{\frac12 H}$, compared to \cite{Gadde:2011ik,Gadde:2011uv}, comes from our choice of the fermion number $(-1)^F=e^{2\pi i R}$. 
We can also write 
\begin{equation}\label{fVH}
P.E.[f^V(q) u]=(q u;q)_\infty^2 ~, \hspace{.5in} P.E.[f^{\frac{1}{2}H}(q)u]=(-q^{1/2} u;q)_\infty^{-1} ~,
\end{equation}
where the Pochhammer symbol is defined as
\begin{equation}
(a;q)_n  
=\begin{cases}  1 & n=0~,\\ \prod_{j=0}^{n-1}(1-a q^{j}) & n>0~.\end{cases}
\end{equation}
We also define $(q)_n \equiv (q;q)_n$.

For the a Lagrangian theory with gauge group $G$ and matter in a representation $\mathbf{R}$ of $G$ and in a representation $\mathbf{F}$ of the flavor symmetry,   the Schur index is 
\begin{equation}
\mathcal{I}(q,z)=\int[du] \,P.E. \left[f^{V}(q)\chi_{G}(u)+f^{\frac{1}{2}H}(q)\chi_{\mathbf{R}}(u)\chi_{\mathbf{F}}(z)\right]~,\label{operatorSchur}
\end{equation}
where $\chi_{\alpha}$ are characters of the gauge and flavor group and $[du]$ is the Haar measure on the maximal torus of $G$.   Here $z$ collectively denotes the flavor fugacities.\footnote{In explicit calculations, it may be useful to convert some infinite products into infinite sums. For example, we can take half of the vector multiplet 
contribution to the gauge theory index and re-write it as a theta function by the Weyl-Macdonald identity. Similarly, the half-hypermultiplet contribution
can be manipulated as 
\begin{equation}
(q^{\frac12} z;q)^{-1}_\infty = \sum_{n=0}^{\infty}\frac{(q^{\frac{1}{2}}z)^{n}}{(q)_{n}}~.
\end{equation}
The integral over the gauge fugacities can then be done explicitly. }

Strictly speaking, our discussion so far involves superconformal field theories.  However, as elaborated on in the introduction the consistency of the IR formulation of the Schur index reviewed below strongly suggests that the Schur index may be defined for non-conformal $\mathcal{N}=2$ theories as well.  When we discuss such examples in the following, we take the operator counting formula \eqref{operatorSchur} as a working definition which applies to models with Lagrangians.

\subsection{An IR Formula for the Schur Index}
\label{sec:IRschur}

We now turn to the IR formula for the Schur index conjectured in \cite{Cordova:2015nma}.  This formulation can be made in any generic vacuum on the Coulomb branch where the theory is IR free.  At such a point the theory is described by a $U(1)^{r}$ gauge theory ($r$ is called the rank).  There is an integral charge lattice $\Gamma$ which is equipped with three structures:
\begin{itemize}
\item A Dirac pairing $\langle \cdot, \cdot \rangle$ which is bilinear, antisymmetric, integer-valued.  
\item A linear central charge function $\mathcal{Z}: \Gamma \rightarrow \mathbb{C}$.  The central charge function is the main output of the Seiberg-Witten solution \cite{Seiberg:1994rs,Seiberg:1994aj} of the low-energy dynamics.
\item A sublattice $\Gamma_f$ of ``flavor charges'' which has zero Dirac pairing with other charges. The Dirac pairing is non-degenerate on the quotient lattice 
$\Gamma_g = \Gamma/\Gamma_f$ of gauge charges. 
\end{itemize}
Associated to the lattice is a quantum torus algebra.   For each charge vector $\gamma \in \Gamma$ we introduce a variable $X_{\gamma}$ which obey
\begin{equation}
X_{\gamma}X_{\gamma'}=q^{\frac{1}{2}\langle\gamma, \gamma' \rangle}X_{\gamma+\gamma'}~. \label{torusalg}
\end{equation}
The torus algebra variables have a simple physical interpretation: they are line defects in the IR abelian gauge theory modeling infinitely massive source dyons with charge $\gamma$.  Note that this explains the algebra of these variables as well, since a pair of dyons (the left-hand side above) sources a electromagnetic fields carrying angular momentum, while a single dyon (the right-hand side) does not.  The variable $q$ is thus a fugacity for rotations and keeps track of this difference.

In order to compute the Schur index we require knowledge of the spectrum of supersymmetric massive excitations of the low-energy theory described by the BPS states.  Each massive BPS particle is a representation of the super little group which is $SU(2)_{J}\times SU(2)_{R}$.  After factoring out the center of mass degrees of freedom the one-particle Hilbert space for the charge sector $\gamma$ may be written as
\begin{equation}
H_{\gamma}=\left[(\mathbf{2},\mathbf{1})\oplus (\mathbf{1},\mathbf{2})\right] \otimes h_{\gamma}~.
\end{equation}
The degeneracies we require are integers $\Omega_{n}(\gamma)$ that are encoded in $h_{\gamma}$ as
\begin{equation}
\mathrm{Tr}_{h_{\gamma}}\left[y^{J}(-y)^{R}\right]=\sum_{n\in \mathbb{Z}}\Omega_{n}(\gamma)y^{n}~.
\end{equation}

From the above physical data we can now formulate the index.  We introduce the $q$-exponential, sometimes also called the quantum dilogarithm
\begin{equation}
E_{q}(z) = (-q^{\frac12} z;q)^{-1}_\infty = \prod_{i=0}^{\infty}(1+q^{i+\frac{1}{2}}z)^{-1}=\sum_{n=0}^{\infty}\frac{(-q^{\frac{1}{2}}z)^{n}}{(q)_{n}}~. \label{eqdef}
\end{equation}

For each charge vector $\gamma$ we then define a KS factor as
\begin{equation}
K(q;X_\gamma;\Omega_{j}(\gamma))=\prod_{n\in \mathbb{Z}}E_{q}((-1)^{n}q^{n/2}X_{\gamma})^{(-1)^{n}\Omega_{n}(\gamma)}~.
\end{equation}
Naively, we define the quantum KS operator as a product of these factors
\begin{equation}
\mathcal{O}(q) = \prod_{\gamma\in \Gamma}^{\curvearrowleft} K(q;X_\gamma;\Omega_{j}(\gamma))~. 
\end{equation}
Here the ordering in the product of non-commutative KS factors is defined using the central charge $\mathcal{Z}$.  If $\arg(\mathcal{Z}(\gamma_{1}))<\arg(\mathcal{Z}(\gamma_{2}))$ then $K(X_{\gamma_{1}})$ appears to the left of $K(X_{\gamma_{2}})$ in the product. This definition has severe convergence issues, 
which we will address momentarily. Up to these issues, according to the wall-crossing formula \cite{Kontsevich:2008fj}, as moduli are varied the individual factors $K(q;X_\gamma;\Omega_{j}(\gamma))$ may jump, but their formal product defining $\mathcal{O}(q)$ is invariant.

Finally, we can extract the Schur index from these ingredients.  Observe that flavor charges are those elements of $\Gamma$ with trivial Dirac pairings.  It then follows from the relations \eqref{torusalg} that the associated variables $X_{\gamma}$ are central elements of the torus algebra.  
We define a trace operation which projects the torus algebra onto these central flavor elements:   
\begin{equation}
\mathrm{Tr}[X_{\gamma}]=\begin{cases} X_{\gamma_f}  &\gamma = \gamma_f \in \Gamma_f,  \\
0 & \mathrm{else}~,\end{cases}
\end{equation}
The trace operation is then extended linearly to sums of the $X_{\gamma}$.

The flavor torus algebra generators in turn are identified with flavor fugacities appearing in the index according with the map between UV and IR global symmetries.
If we pick some basis of flavor fugacities $z_a$ and denote the corresponding components of the flavor charge as $\gamma_f^a$, we can write: 
\begin{equation}
\mathrm{Tr}(X_{\gamma_f})=  z_{\gamma_f} \equiv \prod_a  z_a^{\gamma^a_f}~.
\end{equation}

At last we can precisely state the conjecture of \cite{Cordova:2015nma}. It states that the Schur index may be calculated from the infrared as
\begin{equation}
\mathcal{I}(q,z)=(q)_{\infty}^{2r}~\mathrm{Tr}\left[\mathcal{O}(q)\right]~.\label{IRSchur2}
\end{equation}
By construction this formula is wall-crossing invariant.  In \cite{Cordova:2015nma} it was tested against non-conformal $SU(2)$ gauge theories (using the operator counting formula \eqref{operatorSchur} as a UV definition) as well as non-Lagrangian Argyres-Douglas models using comparisons with chiral algebra techniques.

%

For instance, the simplest example of \eqref{IRSchur2} is the case of a  free hypermultiplet.  The IR formula for the Schur index is (taking into account our choice of fermion number \eqref{fermionnumber})
\begin{equation}
\mathcal{I}(q,X_\gamma) = E_q(X_\gamma)E_q(X_{-\gamma})=(-q^{\frac12} X_\gamma;q)^{-1}_\infty(-q^{\frac12} X_{-\gamma};q)^{-1}_\infty~,
\end{equation}
which is equal to the UV answer  \eqref{fVH} upon the identification  $\mathrm{Tr}(X_{\gamma}) = z$.

\subsubsection{Convergence of the IR Formula}

Although the examples considered in \cite{Cordova:2015nma} provide significant evidence towards the validity of this conjecture, the expression \eqref{IRSchur2} suffers from an important problem: in general, as discussed in Section \ref{sec:irSchurintro}, the trace of the operator $\mathcal{O}(q)$ is not well-defined.  
 
This difficulty is closely related to the problem of giving a meaning to partial products of KS factors
\begin{equation}
\mathcal{S}_{V} = \prod_{\gamma\in \Gamma_V}^{\curvearrowleft} K(q;X_\gamma;\Omega_{j}(\gamma))
\end{equation}
restricted to charges such that the central charge $\mathcal{Z}$ lies in a radial sector $V$ of the complex plane.  If the sector $V$ has width less than $\pi$ (or has width $\pi$ but is open on one side), the product can be interpreted as an element in a group of formal power series in $X_\gamma$, as $\gamma$ and $-\gamma$ never belong simultaneously to $V$ and $V$ is closed under addition. The coefficients in the sum are rational functions of $q$. This does not work if the width of $V$ is greater than $\pi$. 

In particular, $\mathcal{O}(q)$ itself cannot be understood as a formal power series in $X_\gamma$, but the quantum spectrum generator $\mathcal{S}(q)$ defined as the product in the $[0,\pi)$ sector is well-defined, and so is the conjugate $\mathcal{\bar S}(q)$ defined as the product in the $[\pi,2\pi)$ sector.  To give meaning to the conjecture \eqref{IRSchur2} in general, we therefore write instead
\begin{equation}
\mathcal{I}(q) = (q)_\infty^{2r}~\text{Tr} \left[\mathcal{\bar S}(q) \mathcal{S}(q) \right]~. \label{refinedIRSchur}
\end{equation}

To be even more explicit about how this regulates the trace of $\mathcal{O}(q)$, let us fix a basis of charges $\gamma_{i}$ for $\Gamma$ such that $\mathcal{Z}(\gamma_{i})$ lies in the upper half-plane.  We can then define a truncated version of the quantum spectrum generator $\mathcal{S}_{N}(q)$ by setting to zero all torus algebra variables $X_{\gamma}$ such that their coefficients in this basis expansion are larger than $N$
\begin{equation}
X_{a_{i}\gamma_{i}} \mapsto 0 \hspace{.5in} \mathrm{if}~ ~~\sum_{i}a_{i}>N~.
\end{equation}
Plugging into \eqref{refinedIRSchur} we find a truncated version of the Schur index $\mathcal{I}_{N}(q).$  Then, we conjecture that the coefficient of $q^{k}$ in the Schur index $\mathcal{I}(q)$ can be obtained from that $\mathcal{I}_{N}(q)$ provided that $N$ is sufficiently large compared to $k$.\footnote{In the examples we have investigated it is sufficient to take $N$ to be larger than a certain (theory-dependent) linear function  of $k$. } In particular, the limit as $N$ tends to infinity of $\mathcal{I}_{N}(q)$ is the Schur index $\mathcal{I}(q)$.  We illustrate this procedure in Section \ref{sec:nf4schur}.

Notice that we could have split the spectrum in two halves in other ways, in terms of quantum spectrum generators  $\mathcal{S}_\vartheta(q)$ associated to other half planes $[\vartheta,\vartheta+\pi)$:
\begin{equation}
\mathcal{I}(q) = (q)_\infty^{2r}~\text{Tr} \left[\mathcal{S}_{\vartheta+\pi}(q) \mathcal{S}_\vartheta(q) \right]~,
\end{equation}
We expect that for a physical theory the coefficient of $X_\gamma$ in all possible spectrum generators will involve only non-negative, growing powers of $q$ and the above formula to be correct for all choices of $\vartheta$. This appears to be a somewhat non-trivial statement, especially because general $S_V$ for sectors of width smaller than $\pi$ definitely involve negative powers of $q$. 

Given this assumption, standard wall-crossing invariance will be automatic, as $\mathcal{S}_V(q)$ is invariant unless the central charge of BPS particles enters/exits $V$.

\subsection{$SU(2)$ Gauge Theory with $N_f=4$ Flavors}
\label{sec:nf4schur}

A simple example to which our formalism applies is  $SU(2)$ gauge theory with $N_f=4$ hypermultiplets in the fundamental representation.  This is a superconformal field theory where both the UV and IR formulas for the Schur index can be computed and compared.  As we shall see these two calculations yield perfect agreement.  This example is also significant because to properly evaluate the IR formula for the Schur index we require the regularization of the trace discussed in the previous Section.

The Schur index is readily computed from the formula \eqref{operatorSchur} and the UV Lagrangian.  This results in the following integral
\begin{align}
 \begin{split}\label{Nf4Schur}
\mathcal{I} (q,z_i)
&= {1\over \pi} \int^{2\pi}_0  d\theta\, \sin^2\theta\,
P.E.\left[
-2{q\over 1-q} (e^{2i\theta} +e^{-2i\theta} +1) 
-{q^{1\over2}\over 1-q} (e^{i\theta} +e^{-i\theta}) \sum_{i=1}^4 (z_i+ z_i^{-1} )
\right]~.
\end{split}
\end{align}
or 
\begin{align}
\begin{split}
&\mathcal{I}(q,z_i)\\
&=- \frac{1}{4 \pi i} \oint \frac{d u}{u} (u-u^{-1})^2 \frac{(q)_\infty^2(q u^2;q)_\infty^2(q u^{-2};q)_\infty^2}{\prod_{i=1}^4 ( - q^{\frac12}u z_i;q)_\infty( - q^{\frac12} u^{-1} z_i ;q)_\infty(  -  q^{\frac12} u z^{-1}_i;q)_\infty(  - q^{\frac12} u^{-1} z^{-1}_i;q )_\infty}\,.
\end{split}
\end{align}
Here, $z_i$ ($i=1,\cdots,4$) are the flavor fugacities for the $SO(2)^4$ Cartan subgroup of the $SO(8)$ flavor symmetry.\footnote{Readers who are interested in explicit expressions may use the relation 
\begin{equation}
(q)_\infty(q u^2;q)_\infty(q u^{-2};q)_\infty = \sum_{j=0}^\infty (-1)^n \chi_j(u) q^{j(j+1)/2}
\end{equation}
to decompose the vector multiplet contribution into $SU(2)$ characters and expand the hypermultiplet contribution explicitly into 
powers of gauge and flavor fugacities.}  Note again that the sign in front of each factor of $q^{1\over2}$ comes from our choice of the fermion number $(-1)^F=e^{2\pi i R}$ \eqref{fermionnumber}. 

To make the full $SO(8)$ flavor symmetry manifest, we perform the following change of variables,
\begin{align}
\begin{split}\label{SU(2)4}
&\eta_1 =  z_1\,,~~~~\eta_2  = z_1 z_2\,,~~~~\eta_3 =  \sqrt{z_1 z_2 z_3 z_4}\,,~~~~\eta_4  =  \sqrt{z_1z_2z_3\over z_4}\,,
\end{split}
\end{align}
where now $\eta_i$ ($i=1,\cdots,4$) are the flavor fugacities of $SO(8)$ in the convention that the power of $\eta_i$ is the the Dynkin label of the $i$-node.  We choose the second node to be the central one in the $SO(8)$ Dynkin diagram. For example, the character for the $\mathbf{8}_v$ is $\eta_1 + {\eta_2\over\eta_1} +{  \eta_3\eta_4\over\eta_2} + {\eta_4\over \eta_3} +{ \eta_3\over \eta_4 } +{ \eta_2 \over \eta_3\eta_4} +{\eta_1\over \eta_2} +{1\over \eta_1}$.  Incidentally, the Schur index of the $SU(2)$ with $N_f=4$ flavors theory equals to the vacuum character of the affine Lie algebra $\widehat{so(8)}_{-2}$ as established in \cite{Beem:2013sza}.

The BPS spectrum of this theory has been investigated in \cite{Gaiotto:2009hg,Alim:2011ae,Alim:2011kw}.
\begin{figure}[h!]
\begin{center}
\includegraphics[width=.35\textwidth]{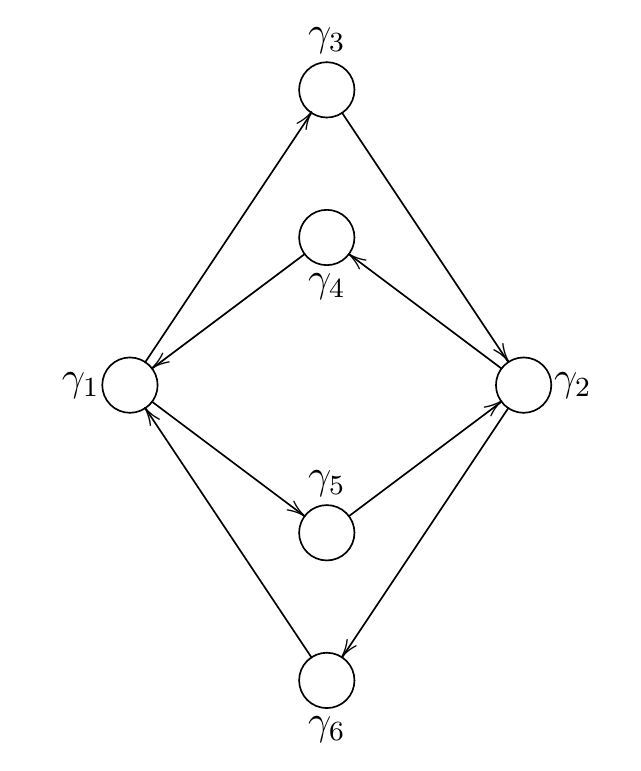}
\end{center}
\caption{ The BPS quiver for the $\mathcal{N}=2$ $SU(2)$ gauge theory with $N_f=4$ hypermultiplets in the fundamental representation.  The Dirac pairings $\langle \gamma_{i}, \gamma_{j}\rangle$ are given by the arrows between the nodes.}\label{fig:Nf4}
\end{figure}
The $SU(2)$ $N_f=4$  theory has a nice finite chamber where the BPS spectrum consists of $12$ hypermultiplets with various gauge and flavor charges, encoded in the 
BPS quiver in Figure \ref{fig:Nf4}. It is worth pointing out that such a convenient the chamber only exists upon mass deformation, which breaks the $SO(8)$ global symmetry 
to an $SU(2) \times SU(2) \times U(1) \times U(1)$ subgroup. 

The $12$  BPS hypermultiplets in increasing phase order are
\begin{align}
\gamma_1 , ~  \gamma_2, ~ \gamma_1+ \gamma_4 , ~ \gamma_1+ \gamma_6 , ~ \gamma_2 + \gamma_3 ,~\gamma_2+\gamma_5 ,~  \gamma_1+ \gamma_4+\gamma_6 ,~ \gamma_2+ \gamma_3 +  \gamma_5 ,~ \gamma_3,~\gamma_4,~\gamma_5,~\gamma_6\,.
\end{align}
In addition there are also the antiparticles to these BPS states. The BPS spectrum is organized into multiplets of that global symmetry, with $(\gamma_4, \gamma_6)$ and $(\gamma_3, \gamma_5)$ 
being doublets of the two unbroken $SU(2)$ global symmetries. 

We also have the following identification between the flavor quantum torus generators and $\eta_i$,
\begin{align}
\begin{split}\label{SO8fugacity}
&\eta_1  =   X_{{1\over2}  (\gamma_1 +\gamma_2 + \gamma_3+2\gamma_4  +  \gamma_6)}\,,\\
&\eta_2 =  X_{{1\over2} ( \gamma_1 + \gamma_2 + \gamma_3+\gamma_4 )}\,,\\
&\eta_3  =   X_{{1\over2} (\gamma_1 + \gamma_2  + 2\gamma_3 + \gamma_4 + \gamma_6)}\,,\\
&\eta_4  = X_{{1\over2}  (2\gamma_1 + 2\gamma_2  +\gamma_3  +\gamma_4  + \gamma_5 + \gamma_6)}\,.
\end{split}
\end{align}

The quantum spectrum generator is determined from the spectrum to be
\begin{align}
\begin{split}
\mathcal{S}(q)&  =E_q(X_{\gamma_1} ) E_q(X_{\gamma_2})
 E_q(X_{\gamma_1+\gamma_4}) E _q(X_{\gamma_1+\gamma_6}) E_q(X_{\gamma_2+\gamma_3})E_q(X_{\gamma_2+\gamma_5})\\
 &\times
 E_q(X_{\gamma_1+\gamma_4+\gamma_6}) E_q(X_{\gamma_2+\gamma_3+\gamma_5})
 E_q(X_{\gamma_3} ) E_q(X_{\gamma_4}) E_q(X_{\gamma_5}) E_q(X_{\gamma_6})\,.
\end{split}
\end{align}
After a somewhat lengthy rearrangement, we can write
\begin{align}
\begin{split}
\mathcal{S}(q)& =  \sum_{\substack{\ell_1,\cdots,\ell_6,\\ p_1,\cdots, p_6=0}}^\infty 
{(-1)^{\sum_{i=1}^6(\ell_i +p_i)}  q^{{1\over2} A }  \over (q)_{\ell_1} \cdots (q)_{\ell_6} (q)_{p_1} \cdots (q)_{p_6}} X_{  \sum_{i=1}^6 a_i \gamma_i}  \,,
\end{split}
\end{align}
where  
\begin{align}
\begin{split}
A&\equiv (p_3 - p_4 +p_5-p_6)(\ell_1-\ell_2) + (\ell_3 - \ell_4 +\ell_5 -\ell_6)(p_1-p_2) \\
&
+(\ell_3-\ell_4 +\ell_5 -\ell_6-2p_1+ 2p_2 )(\ell_1 -\ell_2 -p_3 +p_4-p_5+p_6)\\
&+(-p_3+p_4 -p_5 +p_6)(p_1-p_2)+ \sum_{i=1}^6 (\ell_i +p_i ) \,,
\end{split}
\end{align}
and
\begin{align}
\begin{split}
&a_1 = \ell_1+p_1+p_4+p_6\, ,~~~~  a_2 = \ell_2+p_2+p_3+p_5\, ,~~~~a_3= \ell_3+ p_2+p_3 \, ,\\
&a_4=   \ell_4 +p_1+p_4\,,~~~~~\,~~~~~ a_5= \ell_5+p_2+p_5\,,~~~~~~\,~~~~a_6= \ell_6+ p_1+p_6\,.
\end{split}
\end{align}

To determine the Schur index from these expressions we now use the regularization procedure discussed in Section \ref{sec:IRschur}.  We find that compute terms of order $q^{k}$ in the Schur index we must compute $\mathcal{S}_{N}(q)$ where $N \geq 6k$. For instance $\mathcal{S}_{6}(q)$ expanded to order $q$ is
\begin{eqnarray}
\label{S(q)}
\mathcal{S}_{6}(q)&=& 1 -  q^{1\over2}  \sum_{i=1}^6 X_{\gamma_i} + q \Big(\, X_{\gamma_1 +\gamma_2}  + \sum_{\substack{i,j=3\\ i<j}}^6 X_{\gamma_i +\gamma_j } +   \sum_{i=1}^6 X_{2\gamma_i} + X_{\gamma_1+\gamma_2+\gamma_3+\gamma_4}  +X_{\gamma_1+\gamma_2 + \gamma_4 +\gamma_5 }  \nonumber \\
&+& X_{\gamma_1+\gamma_2  +\gamma_3+\gamma_6 }  + X_{\gamma_1+\gamma_2 + \gamma_5 +\gamma_6} + X_{\gamma_1+\gamma_2+ \gamma_3 +\gamma_4 +\gamma_5+\gamma_6} \,\Big) +\mathcal{O}(q^{3\over2})~,
\end{eqnarray}
and $\overline{ \mathcal{S}}_{6}(q)$ is simply given by replacing every $\gamma_i$ by $-\gamma_i$ in $\mathcal{S}_{6}(q)$.  The IR formula for the Schur index is then, to order $q$
\begin{align}
\begin{split}
&\mathcal{I}(q,\eta_i)=(q)_\infty^2 \text{Tr} [\overline{\mathcal{ S}}_6(q)\mathcal{ S}_6(q)]\\
& = 1 + q\,\Big( 4 + X_{\gamma_1+\gamma_2}  + X_{-\gamma_1-\gamma_2}   +X_{\gamma_3+\gamma_4}  + X_{-\gamma_3-\gamma_4} +X_{\gamma_3-\gamma_5}  + X_{-\gamma_3+\gamma_5}  \\
&+X_{\gamma_4+\gamma_5}  + X_{-\gamma_4-\gamma_5} +X_{\gamma_3+\gamma_6}  + X_{-\gamma_3-\gamma_6} +X_{\gamma_4-\gamma_6}  + X_{-\gamma_4+\gamma_6} +X_{\gamma_5+\gamma_6}  + X_{-\gamma_5-\gamma_6}  \\
&
+X_{\gamma_1+\gamma_2+\gamma_3+\gamma_4}     +X_{-\gamma_1-\gamma_2-\gamma_3-\gamma_4}       
+X_{\gamma_1+\gamma_2+\gamma_4+\gamma_5}      +X_{-\gamma_1-\gamma_2-\gamma_4-\gamma_5}   \\
&
+X_{\gamma_1+\gamma_2+\gamma_3+\gamma_6}    +X_{-\gamma_1-\gamma_2-\gamma_3-\gamma_6}                  
+X_{\gamma_1+\gamma_2+\gamma_5+\gamma_6}    +X_{-\gamma_1-\gamma_2-\gamma_5-\gamma_6}
\\
&+X_{\gamma_1+\gamma_2+\gamma_3+\gamma_4+\gamma_5+\gamma_6}
+X_{-\gamma_1-\gamma_2-\gamma_3-\gamma_4-\gamma_5-\gamma_6}
\Big) +\mathcal{O}(q^2)~,\\
& = 1  + \chi_{\mathbf{28}}(\eta_i) q +\mathcal{O}(q^2)~,
\end{split}
\end{align}
where we have used the relations between the flavor $X_\gamma$ with the $SO(8)$ fugacities \eqref{SO8fugacity}.  Here $\chi_{\mathbf{28}}(\eta_i)$ is the character of the $ \mathbf{28}$ of $SO(8)$.  Including the higher order terms in $\mathcal{S}(q)$, we have computed the trace of the quantum monodromy operator to $q^4$ order,
\begin{align}
\begin{split}
&(q)_\infty^2 \text{Tr} [\overline{\mathcal{ S}}_{24}(q) \mathcal{S}_{24}(q)]  = 1 + \chi_\mathbf{28} \, q + (\chi_\mathbf{1} + \chi_\mathbf{28} + \chi_\mathbf{300})\, q^2
+ (\chi_\mathbf{1}  + 2\chi_{\mathbf{28}}  +  \chi_\mathbf{300}  + \chi_\mathbf{350}  + \chi_\mathbf{1925})\,q^3 \\
&~~~+ ( 2\chi_\mathbf{1}  +  3\chi_\mathbf{28}
   + \chi_{\mathbf{35}_v}+ \chi_{\mathbf{35}_s}+ \chi_{\mathbf{35}_c}  + 3\chi_\mathbf{300}   + \chi_\mathbf{350}  
+\chi_{\mathbf{1925}} +\chi_{\mathbf{4096}} +\chi_{\mathbf{8918}})\, q^4+\mathcal{O}(q^5)~.
\end{split}
\end{align}
  This agrees perfectly  with the UV integral expression for the Schur index \eqref{Nf4Schur}.

\section{Line Defects and Their Schur Indices}\label{sec:LineSchur}

In this section we study supersymmetric line defects and their indices in $\mathcal{N}=2$ field theories.  These include 't Hooft-Wilson lines in gauge theories as well as their generalizations to non-Lagrangian field theories.  See \cite{Kapustin:2005py, Gukov:2006jk, Kapustin:2006hi, Kapustin:2007wm, Drukker:2009tz, Gaiotto:2010be, Xie:2013lca, Aharony:2013hda, Xie:2013vfa} for further background.

\subsection{Defect Junctions and the Schur Index}\label{sec:defectintro}

The class of defects of interest can be characterized by the symmetries that they preserve.  The most symmetric situation occurs when the defect is point-like in space and extended along time.  It is then stabilized by the following odd and even generators:
\begin{itemize}
\item Four supercharges (thus the defect is half-BPS).
\item The group $SU(2)_{J}$ of spatial rotations about the defect, the R-symmetry $SU(2)_{R}$, and time translations (but no other translations).
\end{itemize}
We refer to any object preserving these symmetries as a \textit{full line defect} $L$.  

Implicit in this definition is a parameter $\zeta \in \mathbb{C}^{*}$ which characterizes which four supercharges are preserved by the defect.  The $U(1)_{r}$ symmetry (which is broken by the defect) rotates $\zeta$ by a phase.  In the special case where $|\zeta|=1$ we express it in terms of a phase as $\zeta =\exp(-i\vartheta)$.  In this case the symmetry algebra of the line defect has a simple physical interpretation: it is the symmetries of a massive BPS particle at rest, where $\vartheta$ is the phase of the central charge.  This interpretation is implicit in the following.

In the special case of a conformal field theory, we can strengthen the requirements on line defects to promote the translations along the defect to a full $SL(2,\mathbb{R})$ symmetry.  Alternatively we may characterize the same objects as supersymmetric boundary condition on $AdS_{2}\times S^{2}$ \cite{Kapustin:2005py,Kapustin:2006pk}.

In addition to these line defects extended along time, there are other configurations of defects which will be significant to us.  Specifically, it is useful to also consider line defects which extend along a ray in $\mathbb{R}^4$ and terminate at the origin.  We sometimes refer to these configurations as \textit{half line defects} to distinguish them from the full line defects defined above.  These half defects are also supersymmetric and the preserved supercharges are given in Appendix \ref{sec:supercharge}.   The origin where the half line defect terminates can support a variety of endpoint operators and we seek to count these in an index. 

\begin{figure}[h!]
\centering
\subfloat[]{
\includegraphics[width=.7\textwidth]{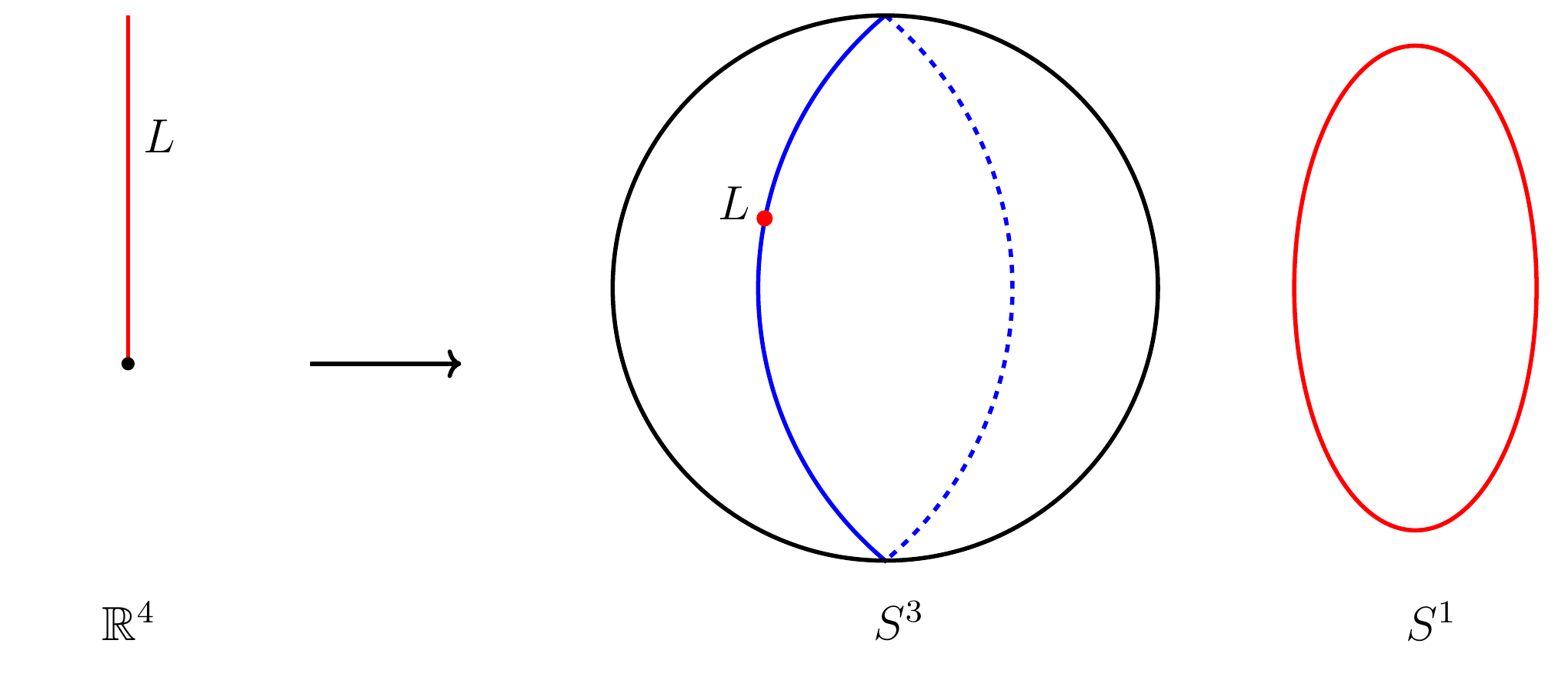}}\\
\subfloat[]{
\includegraphics[width=.7\textwidth]{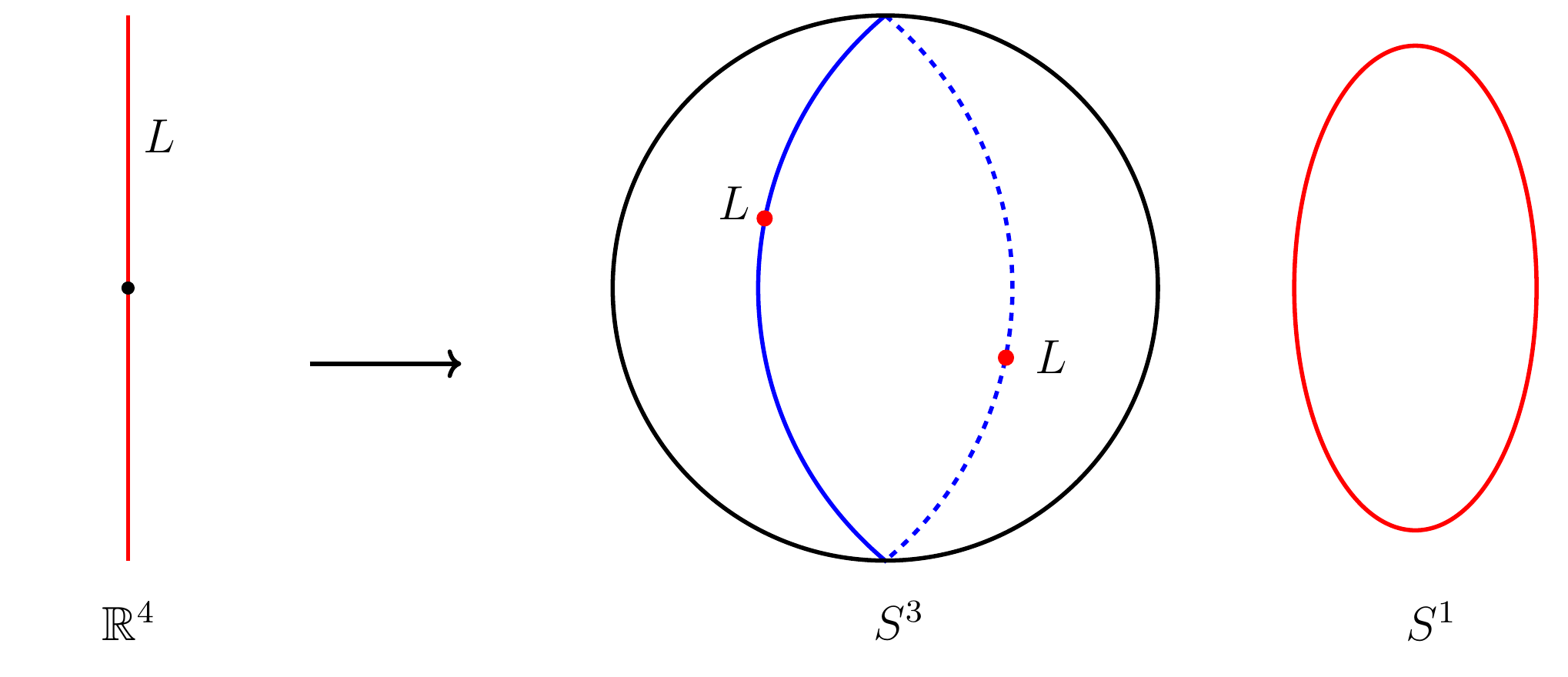}}
\caption{The conformal map (together with the compactification of $\mathbb{R}$ to $S^1$) of (a) a half line defect and (b) a full line defect from $\mathbb{R}^4$ to $S^3\times S^1$.  The worldline of the line defect is shown in red.}\label{fig:halffull}
\end{figure}

It is instructive to consider both the full and half defects on $S^{3}\times \mathbb{R}$ via conformal mapping.  The latter is simplest, it marks the sphere at a single point associated to the defect $L$, and thereby modifies the radially quantized Hilbert space.  Similarly, in the case of a full defect one modifies the $S^{3}$ at two antipodal points by insertion of $L$ and its CPT conjugate defect $\bar{L}$.  See Figure \ref{fig:halffull} for the distinction between a full and a half line defect.

Because of this geometry, a full line defect can be thought simply as a junction of two half line defects.  More generally, we can consider a junction of an arbitrary number of radial half line defects. There is however an important constraint on such junctions.  As we demonstrate in Appendix \ref{sec:supercharge}, in order to preserve supersymmetry all of the half lines must lie in a common two-plane in Euclidean space.  Moreover, the angle of a given half line defect in the plane is exactly the same as the central charge phase $\vartheta$ of the defect.  Conformally mapping to the sphere the defect insertions then lie along a fixed great circle.  See Figure \ref{fig:S3S1}.

The symmetry preserved by these junctions consists of $SU(2)_{R},$ as well as a $U(1)$ rotation in the plane transverse to the rays defining the junction.  The supercharges preserved by this configuration are compatible with those use to define the Schur index (see Appendix \ref{sec:supercharge}) and allow us to extend the definition Schur index  to include these insertions \cite{Dimofte:2011py}:
\begin{equation}\label{Lines}
\mathcal{I}_{L_{1}(\vartheta_{1}) L_{2}(\vartheta_{2})\cdots L_{n}(\vartheta_{n})}(q)=\mathrm{Tr}\left[e^{2\pi i R}\,q^{\Delta-R}\right]~.
\end{equation}
Here the trace is over the Hilbert space on $S^{3}$ with defects $L_{i}$ inserted at angle $\vartheta_{i}$ along a great circle.  Note that this index does not depend continuously on the parameters $\vartheta_{i}$, but does depend on the relative ordering of the points along the circle.  In practice we will mostly focus on the case of a single half line defect insertion in the index in which case the $\vartheta$ dependence can be suppressed.

In theories with a UV Lagrangian formulation, the localization formula \eqref{operatorSchur} can be simply extended to include line defects  \cite{Gang:2012yr}.  Consider first the case of a Wilson line in a representation $\mathbf{R}$ of the gauge group, and let $\chi_{\mathbf{R}}(u)$ denote the character of this representation.  The index in the presence of the half Wilson line $L_{\mathbf{R}}$ is then
\begin{equation}
\mathcal{I}_{L_{\mathbf{R}}} (q, z) = \int [du]~ \chi_{\mathbf{R}}(u)Z( q,u,z) ~, \label{wilsonlineschur}
\end{equation}
where $Z(q,u,z)$ is the integrand in the absence of the line defect.

The above may be readily generalized to include multiple half Wilson lines in general representations.  In this case the various half lines are all mutually local and thus their relative ordering on the great circle in $S^{3}$ does not effect the index.  To add these to the index we simply include a separate character factor $\chi_\mathbf{R}$ for each of the half Wilson lines.  In particular, for the specific case of two half lines, which is equivalent to the insertion of a full unbroken line defect in a representation $\mathbf{R},$ we add the character of the representation $\mathbf{R}$, associated to a defect at the north pole of $S^{3},$ and the character of $\overline{\mathbf{R}}$, associated to the defect insertion at the south pole.

One interesting aspect of the localization formula \eqref{wilsonlineschur} is that it gives a more intuitive description of what exactly is being counted in the line defect index.  In the absence of the character, the integral \eqref{wilsonlineschur} counts gauge invariant local operators satisfying the Schur shortening conditions \eqref{schurshortening}.  With the character $\chi_{\mathbf{R}}(u)$ it counts ``gauge non-invariant local operators" (i.e. words in the free field variables) which satisfy the same shortening conditions and transform in the representation $\mathbf{R}.$  Indeed, these are exactly the objects that may end on the defect and absorb its charge.  

We can generalize from junctions of Wilson lines to a localization formula for the Schur index in the presence of a 't Hooft-Wilson line half-defect $L$ \cite{Dimofte:2011py}. 
This requires introducing some new notations. 

It is useful to interpret the integral expression for the Schur index in terms of an inner product in a space of functions of gauge fugacities and magnetic charges, denoted as 
\begin{equation}
({\mathcal A}, \mathcal{B}) \equiv \sum_{\vec m} \int [du]_{\vec{m}} {\mathcal A}_{-\vec m}(u) {\mathcal B}_{\vec m}(u) \,,
\end{equation}
where $[du]_{\vec{m}}$ is a certain shifted Haar measure with magnetic charge $\vec{m}$.
The usual Schur index is written as an inner product
\begin{align}
\mathcal{I}(q,z)  =  (\Pi^N, \Pi^S) 
\end{align}
of two \textit{half-indices} $\Pi^{N,S}_{\vec{m}}(q,u,z)$ associated with the two hemispheres.

Concretely, the gauge theory integrand is evenly distributed between the two half-indices
\begin{equation}
\Pi^{N,S}_{\vec{m}}(q,u,z)=\delta_{\vec m,0} P.E. \left[\frac12 f^{V}(q)\chi_{G}(u)+f^{\frac{1}{2}H}(q)\chi_{\mathbf{R}\times \mathbf{F}}^{N,S}(u,z)\right]~,\label{operatorSchur}
\end{equation}
where we pick any Lagrangian splitting of the hypermultiplets into two sets of half-hypermultiplets with characters $\chi_{\mathbf{R}\times \mathbf{F}}^{N,S}(u,z)$.

The Schur index decorated by a line defect $L$ can be written as follows:
\begin{align}
\mathcal{I}_L(q,z)  =  (\Pi^N, \hat{{O}}_L \, \Pi^S)~, \label{Thooftschur}
\end{align}
where $\hat{O}_L$ is a certain difference operator acting on functions of $q, u,\vec{m}$. The specific form of $\hat{O}_L$ follows from localization computations as in 
\cite{Gomis:2011pf}. It can also be obtained with the help of the AGT correspondence
\cite{Drukker:2009id,Alday:2009fs} (see also the relation to quantization of the Coulomb branch of the circle-compactified theory \cite{Braverman:2016pwk,Bullimore:2015lsa}).
To include more half line defects, we include more difference operators $\hat{{O}}_{L_{i}}$,
\begin{align}
\mathcal{I}_{(L_i)}(q,z)  =  (\Pi^N, \hat{{O}}_{L_1} \cdots \hat{{O}}_{L_n} \, \Pi^S)~, \label{Thooftschur2}
\end{align}
Now the order of insertion on the circle matters and is captured by the order in which the operators act.

In this formula, the line defects are inserted along a southern quarter of the great circle which goes from the equator to the south pole of the three-sphere. 
There is a second set of difference operators, $\hat{O}'_L$, which represents an insertion along the other southern quarter of the great circle and commute with 
the first set. It is convenient to represent the action of the second set of operators from the right, writing 
\begin{align}
\mathcal{I}_{(L_i),(L'_i)}(q,z)  =  (\Pi^N, \hat{{O}}_{L_1} \cdots \hat{{O}}_{L_n} \, \Pi^S \, \hat{{O}}'_{L_1} \cdots \hat{{O}}'_{L_{n'}})~, \label{Thooftschur3}
\end{align}
The inner product is defined in such a way that 
\begin{equation}
({\mathcal A}, \hat{{O}}_L \mathcal{B}) = ({\mathcal A} \hat{{O}}'_L, \mathcal{B})\,, \qquad \qquad ({\mathcal A}, \mathcal{B}\hat{{O}}'_L) = ( \hat{{O}}_L{\mathcal A}, \mathcal{B})\,,
\end{equation}
while the half-indices satisfy 
\begin{equation}
\hat{{O}}_L \Pi^S = \Pi^S \hat{{O}}'_L\,, \qquad \qquad \hat{{O}}_L \Pi^N = \Pi^N \hat{{O}}'_L\,.
\end{equation}
These relations encode the fact that the location of a line defect can be moved freely along the great circle. 

In an Abelian theory, writing the gauge fugacity as $u$, the 't Hooft-Wilson lines are represented by monomials in the operators\footnote{Here we wrote the operators in a manner suitable for the Schur index with the fermion number choice $(-1)^F=e^{2 \pi i R}$, in order to facilitate the comparison with the quantum torus algebra. In the standard Schur index convention where $(-1)^F =  e^{2\pi i (j_1+j_2)}$,  we should replace $q^{\frac12} \to - q^{\frac12}$. The two conventions can also be related by the fugacity re-definition 
$u \to (-1)^m u$.}
\begin{equation}
x = q^{\frac m2} u, \qquad p = (m \to m+1, u \to q^{\frac12} u) ,\qquad x' = q^{-\frac m2} u, \qquad p' = (m \to m+1, u \to q^{-\frac12} u)\,.
\end{equation}

We can map a function $\mathcal{A}_m(u)$ to a generating function 
\begin{equation}
\mathcal{A}[X] = \sum_{m\in \mathbb{Z}}:\mathcal{A}_m(X_\gamma) X_{-m \gamma'}:
\end{equation}
where $:X_a X_b: \equiv X_{a+b}$ and $\langle \gamma', \gamma\rangle=1$. Then 
\begin{align}
&(x \mathcal{A})[X] = X_\gamma \mathcal{A}[X] \,, ~~~~~~\, (p \mathcal{A})[X] = X_{\gamma'} \mathcal{A}[X]  \,,\notag\\ 
&(x' \mathcal{A})[X] = \mathcal{A}[X] X_\gamma \,, ~~~~~~(p' \mathcal{A})[X] = \mathcal{A}[X]X_{\gamma'} \,.\end{align}
Furthermore, $\Pi^{N,S} = \delta_{m,0}$ for a pure Abelian gauge theory. This explains why Abelian Schur index calculations can be expressed as traces over the quantum torus algebra.

As with the Schur index, the localization formulas, \eqref{wilsonlineschur}-\eqref{Thooftschur} and their interpretation as traces strictly speaking apply only to conformal field theories.  Consistency of the conjectures to follow strongly suggests that these concepts have a more universal definition with the localization formulas describing Lagrangian non-conformal theories as well.  We thus continue to apply these formulas to non-conformal systems.

\subsection{Examples of the Line Defect Schur Index}\label{sec:UVSchur}

In this subsection we compute various line defect indices using the UV localization formula in the pure $SU(2)$ gauge theory and in the $SU(2)$ superconformal QCD.  Our methods follow directly from \cite{Dimofte:2011py,Gang:2012yr}.  We will later reproduce these line defect indices from an IR calculation  in Section \ref{sec:IRSchur} and Section \ref{sec:IRSchurExample}.

\subsubsection{Wilson Lines in $SU(2)$ Gauge Theory}

Let us consider various explicit calculations of the index in the presence of half line defects in $SU(2)$ gauge theory.  The Schur index of the pure $SU(2)$ gauge theory with a half Wilson line defect $L_{0,n}$ (in the representation of dimension $n+1$) is
\begin{equation}
\mathcal{I}_{L_{0,n}} (q)= {1\over \pi} \int^{2\pi}_0  d\theta\, \sin^2\theta\,\left( {e^{i (n+1) \theta }-e^{-i(n+1) \theta}\over   e^{i \theta }-e^{-i \theta}}\right)P.E.\left[
-{2q\over 1-q} (e^{2i\theta} +e^{-2i\theta} +1) 
\right]~.
\end{equation}
If $n$ is odd, $\mathcal{I}_{L_{0,n}} (q)$ vanishes, while if $n$ is even we obtain non-trivial results. For example,
\begin{eqnarray}
\mathcal{I}_{L_{0,2}} & = & -2q +q^{2}-2q^{4}+q^{6}-2q^{9}+q^{12}+\mathcal{O}(q^{13})~, \nonumber\\
\mathcal{I}_{L_{0,4}} & = & q^{2}+2q^{3}-2q^{4}+q^{6}+2q^{7}-2q^{9}+q^{12}+\mathcal{O}(q^{13})~, \label{su(2)ans}\\
\mathcal{I}_{L_{0,6}} & = & -2q^{4}-q^{6}+2q^{7}-2q^{9}-2q^{11}+q^{12}+\mathcal{O}(q^{13})~.\nonumber
\end{eqnarray}

\subsubsection{'t Hooft Lines in $SU(2)$ Gauge Theory}

Let us consider the Schur index with the presence of a full 't Hooft line with minimal magnetic charge in the pure $SU(2)$ gauge theory. The magnetic charge $m$ rangess over non-negative half integers, 0, ${1\over2}$,\,1,.... The shifted Haar measure $[du]_m$ is 
\begin{align}
[du]_m ={1\over2\pi} \left( 1-{1\over2}\delta_{m,0}\right) q^{-m} (1-q^{m} e^{2i\theta} ) (1- q^m e^{-2i\theta})\,,
\end{align}
where we have written $u=e^{i\theta}$. 
The half-index $\Pi_m$ is 
\begin{align}
\Pi^{N,S}_m(q,\theta) = \delta_{m,0} \, P.E. \left[ -{q\over 1-q} (e^{2i\theta} +e^{-2i\theta}+1)\right]\,.
\end{align}

The difference operator $\hat{O}_{1,0}$ for a 't Hooft line with minimal magnetic charge can be read off from that of the $\mathcal{N}=4$ $SU(2)$ gauge theory by decoupling the adjoint hypermultiplet \cite{Dimofte:2011py}. In terms of the Abelian difference operators $x,p$ defined above it is \begin{align}
\hat{O}_{1,0} ={i\over x -x^{-1}} ( p^{1\over2}-p^{-{1\over2}} )\,.
\end{align}

The  Schur index with a full 't Hooft line is
\begin{align}
\begin{split}
\mathcal{I}_{L_{1,0}} (q)& = \sum_m \int[du]_m \, \Pi^N_{-m}(q,\theta) \left( \hat{O}_{1,0} \right)^2 \Pi^S_m(q,\theta)\,.
\end{split}
\end{align}
Since $\Pi^N_{-m}$ is nonzero only when $m=0$, it suffices to compute the term in  $\left( \hat{O}_{1,0} \right)^2 \Pi_m(q,\theta)$ that comes with $\delta_{m,0}$,{
\begin{align}
\left( \hat{O}_{1,0} \right)^2 \Pi_m(q,\theta)
 =  -   \delta_{m,0}\,
{
q^{1\over2} (1+q) \over
(1-q e^{2i\theta} ) (1-q e^{-2i\theta})
}
\Pi_0(q,\theta)+ \cdots\,.
\end{align}
It follows that 
\begin{align}
\begin{split}\label{thooft}
\mathcal{I}_{L_{1,0}} (q)& = -  {1\over \pi}\int_0^{2\pi} d\theta \, \sin^2\theta\, {
q^{1\over2} (1+q) \over
(1-q e^{2i\theta} ) (1-q e^{-2i\theta})
} 
P.E.\left[
-{2q\over 1-q} (e^{2i\theta}+e^{-2i\theta} +1)
\right]\\
&=  -  q^{1\over2} + q^{5\over2} - q^{7\over2} - q^{9\over2} +q^{13\over2} + q^{15\over2} - q^{17\over2} -q^{19\over2}+\cdots\,.
\end{split}
\end{align}

\subsubsection{Wilson Lines in $SU(2)$ Gauge Theory with $N_f=4$ Flavors}\label{sec:Nf4Line}

In the conformal case $N_f=4$, the half Wilson line defect $L_{0,n}$ in the $(n+1)$-dimensional irreducible representation of $SU(2)$ is \cite{Gang:2012yr}
\begin{align}
 \begin{split}
\mathcal{I}_{L_{0,n}} (q,z_i)
&= {1\over \pi} \int^{2\pi}_0  d\theta\, \sin^2\theta\,
\left( {e^{i (n+1) \theta }-e^{-i(n+1) \theta}\over   e^{i \theta }-e^{-i \theta}}\right)
\,\\
&\times
P.E.\left[
-2{q\over 1-q} (e^{2i\theta} +e^{-2i\theta} +1) 
-{q^{1\over2}\over 1-q} (e^{i\theta} +e^{-i\theta}) \sum_{i=1}^4 (z_i+ z_i^{-1} )
\right]\,,
\end{split}
\end{align}
where $z_i$ ($i=1,\cdots,4$) are the flavor fugacities for $SO(2)^4$ Cartan subgroup of the $SO(8)$ flavor symmetry. They are related to  the  $SO(8)$ fugacities $\eta_i$ ($i=1,\cdots,4$) by \eqref{SU(2)4}.  The first few terms of the doublet half Wilson line defect index are
\begin{align}\label{Nf4Doublet}
\mathcal{I}_{L_{0,1}} (q,\eta_i )  = - \chi_{[1,0,0,0]} q^{1\over2}  -  \chi_{[1,1,0,0]} q^{3\over2}-
(\chi_{[1,0,0,0]} + \chi_{[0,0,1,1]}  + \chi_{[1,1,0,0]}  + \chi_{[1,2,0,0]})q^{5\over2}\cdots \,,
\end{align}
where $\chi_{[a_1,a_2,a_3,a_4]}(\eta_i)$ is the (finite) $SO(8)$ character for the representation  $[a_1,a_2,a_3,a_4]$. 

It is instructive to enumerate the operators that are counted by the doublet half Wilson line defect index \eqref{Nf4Doublet} in the free limit of zero gauge coupling.  The defect operators living at the end of a half Wilson line transform as doublets under the gauge $SU(2)$ and satisfy the Schur operator conditions.  For the $SU(2)$ $N_f=4$ theory at the free coupling point, the single-letter operators that contribute to the Schur index are the complex scalars $H^{ia}$ of the 8 half-hypermultiplets, 2 components of the gauginos $\rho^{\pm A}$ (the $\pm$ denotes their $U(1)_r$ charges) in the vector multiplet, and 1 derivative $\partial\equiv \partial_{-\dot{+}}$ \cite{Gadde:2011uv}.  Here $i=1,\cdots,8$ is the index for the $\mathbf{8}_v$ of the flavor $SO(8)$, while $a=1,2$ and $A=1,2,3$ are the indices for the $\mathbf{2}$ and $\mathbf{3}$ of the gauge $SU(2)$, respectively.  The representations of the single-letter Schur operators  under the flavor $SO(8)$ and gauge $SU(2)$ and their contributions to the Schur index are summarized below
\begin{align}
\left.\begin{array}{|c|c|c|c|}\hline  & SO(8) & SU(2) & \text{Schur index} \\\hline H^{ia} & \textbf{8}_v & \textbf{2} & q^{1/2} \\\hline \rho^{\pm A}& \textbf{1} & \textbf{3} & -q \\\hline \partial & \textbf{1} & \textbf{1} & q \\\hline \end{array}\right.
\end{align} 
At the zero coupling point, the line defect index \eqref{Nf4Doublet} is counting all the composite operators of the  single-letter Schur operators that are in the doublet of the gauge $SU(2)$.

At $q^{1\over2}$ order in the line defect index \eqref{Nf4Doublet}, the only contributing operator is $H^{ia}$, which transforms in the $\mathbf{8}_v$ of the flavor $SO(8)$. Hence the coefficient of $q^{1\over2}$ in the defect index is minus the (finite) $SO(8)$ character of  of $\mathbf{8}_v$.  The sign comes from our our choice of the fermion number $(-1)^F=e^{2\pi i R}$ and the fact that $H^{ia}$ has $R=1/2$.   At $q^{3\over2}$ order, the doublet Schur operators are
\begin{align}\label{level32}
\left.\begin{array}{|c|c|c|}\hline  & SO(8) & SU(2) \\\hline \partial H^{ia} & \mathbf{8}_v & \mathbf{2} \\\hline (\rho^{\pm A} H^{ib})(T^A)^a_{~b}  & \mathbf{8}_v & \mathbf{2} \\\hline H^{ia}H^{jb}H^{kc} & \mathbf{8}_v\oplus\mathbf{160}_v & \mathbf{2} \\\hline \end{array}\right.
\end{align}
where $(T^A)^a_{~b}$ are the generators of $SU(2)$ in the doublet representation, i.e., the Pauli matrices. Adding their contributions together, we indeed find the $q^{3\over2}$ coefficient to be minus the character of $\mathbf{160}_v$ of $SO(8)$.  

We also record the  defect indices $\mathcal{I}_{L_{0,n}}$ of  half Wilson lines in the $(n+1)$-dimensional representation with  flavor fugacities set to be 1,
\begin{align}\label{Nf4Lines}
\mathcal{I}_{L_{0,1}} (q , \eta_i=1)& =   -\Big( 8q^{1\over2} +160q ^{3\over2}+ 1624q^{5\over2}+11768q^{7\over2}+68376q^{9\over2}+339408q^{11\over2}+1493064q^{13\over2}\notag\\
&+5965192q^{15\over2}+
22015936q^{17\over2}+76007904q^{19\over2}\Big)+\mathcal{O}(q^{21\over2})\,,\notag\\
\mathcal{I}_{L_{0,2}} (q , \eta_i=1)&=34 q+567 q^2 +5236 q^3+35476 q^4 +196072 q^5+935334 q^6+3982598 q^7\notag \\
&+15480618 q^8
+55804008 q^9+188738978  q^{10}+604269758 q^{11}+ 1844182063 q^{12}\notag\\
&+5395212272 q^{13}+15199582939 q^{14}
+\mathcal{O}(q^{15})\,,\notag\\
\mathcal{I}_{L_{0,3}}(q,\eta_i=1)&= -\Big(104 q^{3\over2}+1560 q^{5\over2}+13512 q^{7\over2}+87312 q^{9\over2}+465072 q^{11\over2}+2152584 q^{13\over2}\notag\\
&+8936536 q^{15\over2}+33990704 q^{17\over2} + 120232216 q^{19\over2} + 399908120 q^{21\over2} \notag\\
& + 1261401360 q^{23\over2}\Big)
+\mathcal{O}(q^{25\over 2})\,,\\
\mathcal{I}_{L_{0,4}}(q,\eta_i=1)&= 259 q^2+3634 q^3+30112 q^4+187994 q^5+974050 q^6+4405014 q^7+17928498 q^8\notag\\
&+67023502 q^9 + 233484050 q^{10}+ 766078486 q^{11} + 2386860955 q^{12}+\mathcal{O}(q^{13})\,\notag.
\end{align}

\subsection{An IR Formula for the Line Defect Schur Index}\label{sec:IRSchur}

In this section, we generalize the IR formula for the Schur index to include line defects.  The basic intuition is easy to explain.  The IR formula for the Schur index \eqref{refinedIRSchur} can be interpreted as the index of an Abelian gauge theory with independent fields for each BPS hypermultiplet.  It is straightforward to generalize such a formula to include infrared line defects: we simply include the expression $X_{\gamma}$ in the trace.  Thus to extract a correct UV line defect index all that we require is the data of how a UV line defect $L$ is decomposed in the IR into a sum of Abelian defects $X_{\gamma}$.  As we now review, this is precisely the data captured by framed BPS states.

Consider a full line defect $L$ at a point in space and extended along time.  On the Coulomb branch of the theory, $L$ modifies the Hilbert space and there is a new class of BPS states, so-called framed BPS states, which may be viewed as ordinary particles bound to the defect.  The framed BPS Hilbert space $\mathcal{H}_{L}$  is graded by electromagnetic charges
\begin{equation}
\mathcal{H}_{L}=\bigoplus_{\gamma \in \Gamma}\mathcal{H}_{L,\gamma}~.
\end{equation}
We encode the framed BPS states in a framed protected spin character
\begin{equation}
\fro(L,\gamma,q)=\mathrm{Tr}_{\mathcal{H}_{L,\gamma}}q^{J}(-q)^{R}~,
\end{equation}
The framed BPS Hilbert spaces, as well as the framed protected spin characters, jump at walls of marginal stability.  

The framed BPS states also characterize the defect renormalization group flow.  In the infrared, the defect $L$ is described by a collection of defects in the IR abelian gauge theory.  These are simply dyons characterized by their electromagnetic charges and are represented exactly by the quantum torus variables $X_{\gamma}$.  The $X_{\gamma}$ and the OPE \eqref{abelianope} provide a convenient way of describing the ultraviolet line defects in terms of the infrared data.  For each $L$ we introduce the generating function \cite{Gaiotto:2010be,Cordova:2013bza}
\begin{equation}\label{generating}
F(L,\vartheta)=\sum_{\gamma \in \Gamma}\fro(L,\gamma,q)X_{\gamma}~.
\end{equation}
The above expression specifies how the ultraviolet defect $L$ is decomposed into infrared pieces.  Again as a consequence of wall crossing, this decomposition will jump.  

With these ingredients, we now formulate our conjecture for the Schur index in the presence of line defect $L$ with central charge phase $\vartheta$.  It is simply a modified trace with the defect generating function inserted at the appropriate phase,\footnote{We will often choose the central charge phase $\vartheta$ of the line defect to the right of all ordinary BPS states and suppress the $\vartheta$ dependence in $F(L,\vartheta)$ and in $\mathcal{S}_\vartheta(q)$.}
\begin{equation}\label{lineproposal}
\mathcal{I}_{L}(q) = (q)_\infty^{2r}~\text{Tr} \left[ F(L,\vartheta) \mathcal{S}_\vartheta(q) \mathcal{S}_{\vartheta+\pi}(q)\right]~,
\end{equation}
where $\mathcal{S}_\vartheta(q)$ is the quantum spectrum generator associated to the half plane $[\vartheta , \vartheta +\pi)$.  

As advocated before, since the line defect index is originally defined in the UV, its IR formula \eqref{lineproposal} should be framed wall-crossing invariant.  The framed wall-crossing phenomenon occurs when the central charge $Z_\gamma = |Z_\gamma|e^{i\vartheta_\gamma}$ of an ordinary BPS state crosses the central charge of the line defect $\zeta$ as we vary the moduli.  The framed wall-crossing formula for the generating function $F(L,\vartheta)$ is  \cite{Gaiotto:2010be}
\begin{align}
F(L , \vartheta_\gamma- \epsilon ) = E_q(X_\gamma)  F(L , \vartheta_\gamma + \epsilon) E_q(X_{\gamma})^{-1}\,,
\end{align}
where $\epsilon>0$.  As the central charge $Z_\gamma$ of a BPS state crosses the line defect central charge $\zeta$ from above, the quantum spectrum generator $\mathcal{S}_\vartheta(q)$ also jumps
\begin{align}
\mathcal{S}_{ \vartheta_\gamma-\epsilon} (q) = E_q(X_\gamma) S_{\vartheta_\gamma+\epsilon} (q) E_q(X_{-\gamma})^{-1}\,.
\end{align}
Similarly, the quantum spectrum generator for the other half plane $\mathcal{S}_{\vartheta +\pi}(q)$ also jumps discontinuously, $\mathcal{S}_{ \vartheta_\gamma+\pi -\epsilon} (q) = E_q(X_{-\gamma}) S_{\vartheta_\gamma+\pi+\epsilon} (q)E_q(X_{\gamma})^{-1}$.  Combining the above transformations, we conclude that our IR formula \eqref{lineproposal} for the Schur index is framed wall-crossing invariant,
\begin{align}
 (q)_\infty^{2r}~\text{Tr} \left[ F(L,\vartheta) \mathcal{S}_\vartheta(q) \mathcal{S}_{\vartheta+\pi}(q)\right]\Big|_{\vartheta= \vartheta_\gamma-\epsilon}
 = (q)_\infty^{2r}~\text{Tr} \left[ F(L,\vartheta) \mathcal{S}_\vartheta(q) \mathcal{S}_{\vartheta+\pi}(q)\right]\Big|_{\vartheta= \vartheta_\gamma+\epsilon}\,.
 \end{align}

More generally, for an arbitrary junction of half line defects $L_i$ with central charge phases $\vartheta_1<\vartheta_2 <\cdots <\vartheta_n$ all chosen to be on the right of the ordinary BPS particles and to the left of the anti-particles, we propose the following IR formula for the line defect index \eqref{Lines},
\begin{align}
\mathcal{I}_{L_{1}(\vartheta_{1})L_{2}(\vartheta_{2})\cdots L_{n}(\vartheta_{n})}(q)
 =  (q)_\infty^{2r}~\text{Tr} \left[ F(L_1, \vartheta_1)F(L_2, \vartheta_2)\cdots F(L_n, \vartheta_n) \mathcal{S}_{\vartheta_n}(q) \mathcal{S}_{\vartheta_n+\pi}(q)\right].
\end{align}
The IR formula for the more general central charge phases $\vartheta_i$ assignment can be obtained by applying the framed wall-crossing formula.

\subsection{Examples of the IR Formula}\label{sec:IRSchurExample}

In this subsection we will apply the IR formula introduced in Section \ref{sec:IRSchur} to various line defects in the pure $SU(2)$ gauge theory and in the $SU(2)$ gauge theory with $N_f=4$ flavors.  In particular, we will reproduce the line defect indices obtained in Section \ref{sec:UVSchur} from the trace of the quantum KS operator and the generating function $F(L,\vartheta)$ of the line defect.

\subsubsection{Wilson Lines in $SU(2)$ Gauge Theory}

\begin{figure}[h!]
\begin{center}
\includegraphics[width=.2\textwidth]{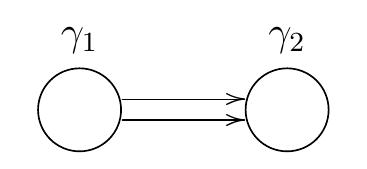}
\end{center}
\caption{ The BPS quiver for the $\mathcal{N}=2$ pure $SU(2)$ gauge theory.}\label{fig:Nf0}
\end{figure}

We begin by testing our IR formula  in the case of Wilson lines in the pure $SU(2)$ gauge theory.  Let $L_{0,n}$ denote the Wilson line in the irreducible $n+1$-dimensional representation of $SU(2).$  To describe the resulting generating functions $F(L_{0,n},\vartheta),$ it is helpful to introduce centrally symmetric $q$-binomial coefficients as \cite{Cordova:2013bza}
\begin{equation}
\binom{m}{r}_{q} \equiv q^{-\frac{1}{2}r(m-r)}\frac{(1-q^{m})(1-q^{m-1})\cdots (1-q^{m-r+1})}{(1-q^{r})(1-q^{r-1})\cdots (1-q)}~.
\end{equation}
Then (assuming the defect phase $\vartheta$ is chosen to lie to the right of all ordinary BPS states) the generating functions are
\begin{equation}
F(L_{0,n},\vartheta)= \sum_{s=0}^{n}\sum_{r=0}^{s}\binom{n-r}{n-s}_{q}\binom{s}{r}_{q}X_{\left(r-\frac{n}{2}\right)\gamma_{1}+\left(s-\frac{n}{2}\right)\gamma_{2}}~. \label{wilsonlines}
\end{equation}
Here, charges $\gamma_{1}$ and $\gamma_{2}$ are chosen such that $\langle \gamma_{1}, \gamma_{2}\rangle=2$, and the standard electric charge in the lattice is $\frac{1}{2}(\gamma_{1}+\gamma_{2})$.  Thus in the sum above, the terms with $r=s$ describe the expected decomposition of the representation of $SU(2)$ into electrically charged states on the Coulomb branch.  The terms with $r\neq s$ carry non-vanishing magnetic charge. 

The general formula \eqref{proposal} for the defect index reduces to
\begin{align}  \mathcal{I}_{L_{0,n}} (q) = (q)_\infty^2\text{Tr} \left[  \,F(L_{0,n}, \vartheta) E_q(X_{\gamma_1})E_q(X_{\gamma_2})   E_q(X_{\gamma_1}^{-1})E_q(X_{\gamma_2}^{-1})\,\right]\,.
\end{align}
The trace involves a linear combination of the following quantity
\begin{align}
\begin{split}
&\text{Tr}\left[
 X_\gamma E_q(X_{\gamma_1})E_q(X_{\gamma_2})E_q(X_{\gamma_1}^{-1})E_q(X_{\gamma_2}^{-1})
\right]\\
&= \sum_{k_1,k_2,\ell_1,\ell_2=0}^\infty
{(-1)^{k_1+k_2+\ell_1+\ell_2 } q^{{1\over2} (k_1+k_2+\ell_1+\ell_2)  }  \over (q)_{k_1} (q)_{k_2} (q)_{\ell_1} (q)_{\ell_2}     }
\text{Tr}
\left[
\,
   X_\gamma 
X_{\gamma_1}^{\ell_1}  X_{\gamma_2} ^{\ell_2}X_{\gamma_1}^{-k_1} X_{\gamma_2}^{-k_2}
\,
\right]\\
&
= \sum_{k_1,k_2,\ell_1,\ell_2=0}^\infty
{(-1)^{k_1+k_2+\ell_1+\ell_2 } q^{{1\over2} (k_1+k_2+\ell_1+\ell_2)  }  \over (q)_{k_1} (q)_{k_2} (q)_{\ell_1} (q)_{\ell_2}     }
q^{ a_1a_2  + 2(\ell_1+a_1) (k_2-a_2)}  \, \delta_{\ell_1+a_1 , k_1}\delta_{\ell_2, k_2-a_2}\\
&= (-1)^{a_1+a_2 }q^{{1\over2} (a_1+a_2) +a_1a_2 }    \sum_{\substack{\ell_1 =\text{max}( \lceil -a_1\rceil ,0 )\\\ell_2=\text{max}( \lceil -a_2\rceil ,0 )}}^\infty
{q^{\ell_1+\ell_2+2\ell_1\ell_2+2a_1\ell_2 }\,  \over  (q)_{\ell_1} (q)_{\ell_2} (q)_{\ell_1+a_1} (q)_{\ell_2+a_2}    }\,,
\end{split}
\end{align}
where $\gamma= a_1\gamma_1 +a_2\gamma_2$.
Thus, all that remains is to sum these quantities as dictated by generating functions $F(L_{0,n})$ which may be found in \eqref{wilsonlines}.  Carrying out this straightforward calculation we find for instance
\begin{align}
\begin{split}
\mathcal{I}_{L_{0,2}} & =  -2q +q^{2}-2q^{4}+q^{6}-2q^{9}+q^{12}+\mathcal{O}(q^{13})~, \\
\mathcal{I}_{L_{0,4}} & =  q^{2}+2q^{3}-2q^{4}+q^{6}+2q^{7}-2q^{9}+q^{12}+\mathcal{O}(q^{13})~,
\end{split}
\end{align}
in exact agreement with the  UV defect indices \eqref{su(2)ans} computed from localization.

\subsubsection{'t Hooft Lines in $SU(2)$ Gauge Theory}

For the 't Hooft line $L_{1,0}$ with minimal magnetic charge in the pure $SU(2)$ gauge theory, the generating function of the frame BPS states is very simple
\begin{align}
F(L_{1,0}) = X_{-{1\over 2}\gamma_2}\,.
\end{align}
The general formula \eqref{proposal} for a full 't Hooft line $L_{1,0}$ can be computed in a similar way as in the Wilson line defect cases,
\begin{align}
\begin{split}
\mathcal{I}_{L_{1,0}}(q)  &=  (q)^2_\infty \, \text{Tr} \left[\,
 F(L_{1,0})^2 \, 
E_q(X_{\gamma_1} )  E_q(X_{\gamma_2}) \, E_q(X_{\gamma_1}^{-1})  E_q(X_{\gamma_2}^{-1}) \,
\right]\,\\
&=-q^{1\over2} +q^{5\over2} -q^{7\over2} -q^{9\over2} +q^{13\over2} +q^{15\over2} -q^{17\over2} -q^{19\over2} +\cdots\,,
\end{split}
\end{align}
which equals the  't Hooft line defect index \eqref{thooft} computed from UV localization. 


\subsubsection{Wilson Lines in $SU(2)$ Gauge Theory with $N_f=4$ Flavors}

Let us move on to the half Wilson line index in the $SU(2)$ superconformal QCD. The framed BPS degeneracies for the line defect can be read off, say, from 
the class $\mathcal{S}$ description, given i.e. in \cite{Gaiotto:2010be} in a slightly different chamber. For illustrative purposes, we will reproduce the answer for the Wilson line in the doublet using the representation theory of the framed BPS quiver, i.e., an extended BPS quiver with one extra node representing the defect \cite{Cordova:2013bza}.\footnote{As explained in \cite{Cordova:2013bza}, this method, which is referred to as the Higgs branch calculation, generally over counts the framed BPS state degeneracies. For example, in the pure $SU(2)$ gauge theory, while the Higgs branch calculation gives the correct framed BPS state degeneracies for the doublet Wilson line,  it yields incorrect answers for Wilson lines in the $n$-dimensional representation if $n>2$.  In the $SU(2)$ superconformal QCD, the situation is similar and the Higgs branch calculation gives the correct generating function $F(L_{0,1})$ for the doublet Wilson line.}
 
In the chamber shown in Figure \ref{fig:Nf4}, the core charge for a Wilson line in the $\mathbf{2}$ of $SU(2)$ is
\begin{align}
\gamma_c =  -{1\over2}  ( \gamma_1 + \gamma_2  + \gamma_3  +\gamma_5)\,.
\end{align}
The framed BPS quiver for this Wilson line is shown in Figure \ref{fig:Nf4Framed}, with the square node representing the doublet Wilson line.  
\begin{figure}[h]
\centering
\includegraphics[width=.35\textwidth]{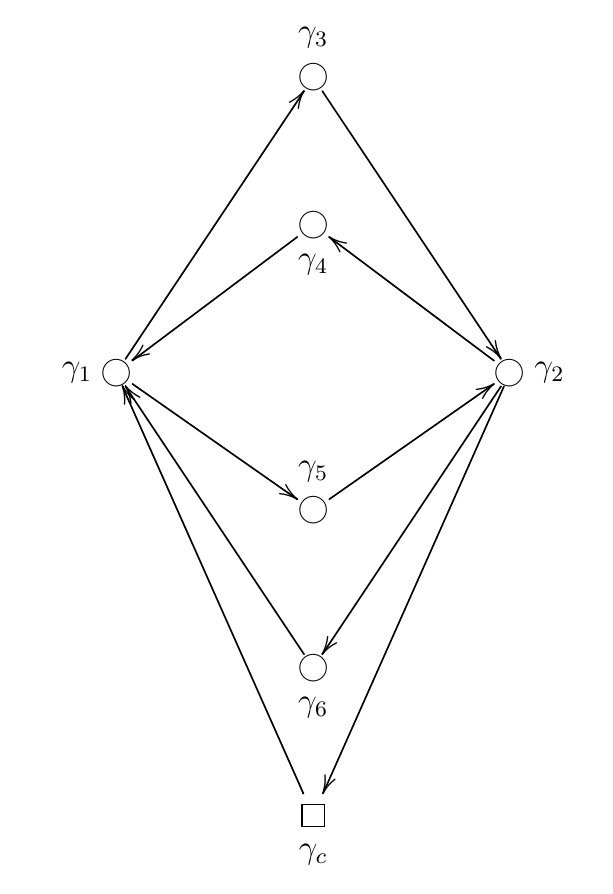}
\caption{The framed quiver for a Wilson line in the $\mathbf{2}$ in the $SU(2)$ gauge theory with $N_f=4$ flavors. The core charge of the doublet Wilson line is given by $\gamma_c=  -{1\over2}  ( \gamma_1 + \gamma_2  + \gamma_3  +\gamma_5)$. }\label{fig:Nf4Framed}
\end{figure}

 The generating function $F(L_{0,1})$ for the Wilson line in the $\mathbf{2}$ of $SU(2)$ is determined from the framed BPS states. The framed BPS states of the framed quiver can in turn be obtained by mutations as in Figure \ref{fig:Nf4Mutation},
\begin{equation}\label{Nf4generating}
F(L_{0,1} ) = X_{\gamma_c}  +  X_{\gamma_c+\gamma_2  } +X_{\gamma_c+\gamma_2+\gamma_3} + X_{\gamma_c+\gamma_2+\gamma_5} +X_{\gamma_c+\gamma_2+\gamma_3+\gamma_5}+X_{\gamma_c+\gamma_1+\gamma_2+\gamma_3+\gamma_5}\,.
\end{equation}
Where the central charge of the framed node is to the right of the vanilla BPS states.

The line defect index \eqref{Nf4Doublet} can be reproduced from the trace of $F(L_{0,1})$.  For example, the leading term $-\chi_{[1,0,0,0]}(\eta_i) q^{1\over2}$ in \eqref{Nf4Doublet} can be  reproduced from the $q^{1\over2}$ term in $\mathcal{S}(q)$  \eqref{S(q)} with the help of \eqref{SO8fugacity},
\begin{align}
&(q)_\infty^2  \text{Tr} [F(L_{0,1})  \, \mathcal{S}(q) \, \overline{\mathcal{ S}}(q)]   = 
(q)_\infty^2  \text{Tr} \left[ \,F(L_{0,1})  \,  (1- q^{1\over2}\sum_{i=1}^6 X_{\gamma_i } )(1- q^{1\over2}\sum_{j=1}^6 X_{-\gamma_j }) \, \right] + \mathcal{O}(q)\notag\\
&= - q^{1\over2}  \Big(\, 
X_{{1\over2}(\gamma_1 +\gamma_2 +\gamma_3-\gamma_5)}+X_{{1\over2}(-\gamma_1 -\gamma_2 +\gamma_3-\gamma_5)}  
+ X_{{1\over2}(\gamma_1 +\gamma_2 -\gamma_3+\gamma_5)}+X_{{1\over2}(-\gamma_1 -\gamma_2 -\gamma_3+ \gamma_5)}   \notag\\
&
+ X_{ {1\over2}(\gamma_1 +\gamma_2 +\gamma_3   +2\gamma_4    +\gamma_5)}
+X_{{1\over2}(-\gamma_1 -\gamma_2 -\gamma_3-2\gamma_4-\gamma_5)}  
+ X_{{1\over2}(\gamma_1 +\gamma_2 +\gamma_3+\gamma_5+2\gamma_6 }+X_{{1\over2}(-\gamma_1 -\gamma_2 -\gamma_3-\gamma_5-2\gamma_6)}  
  \,\Big)\notag\\
  &+\mathcal{O}(q^{3\over2})\,.
\end{align}

\newpage

\begin{figure}[h!]
\centering
\includegraphics[width=.92\textwidth]{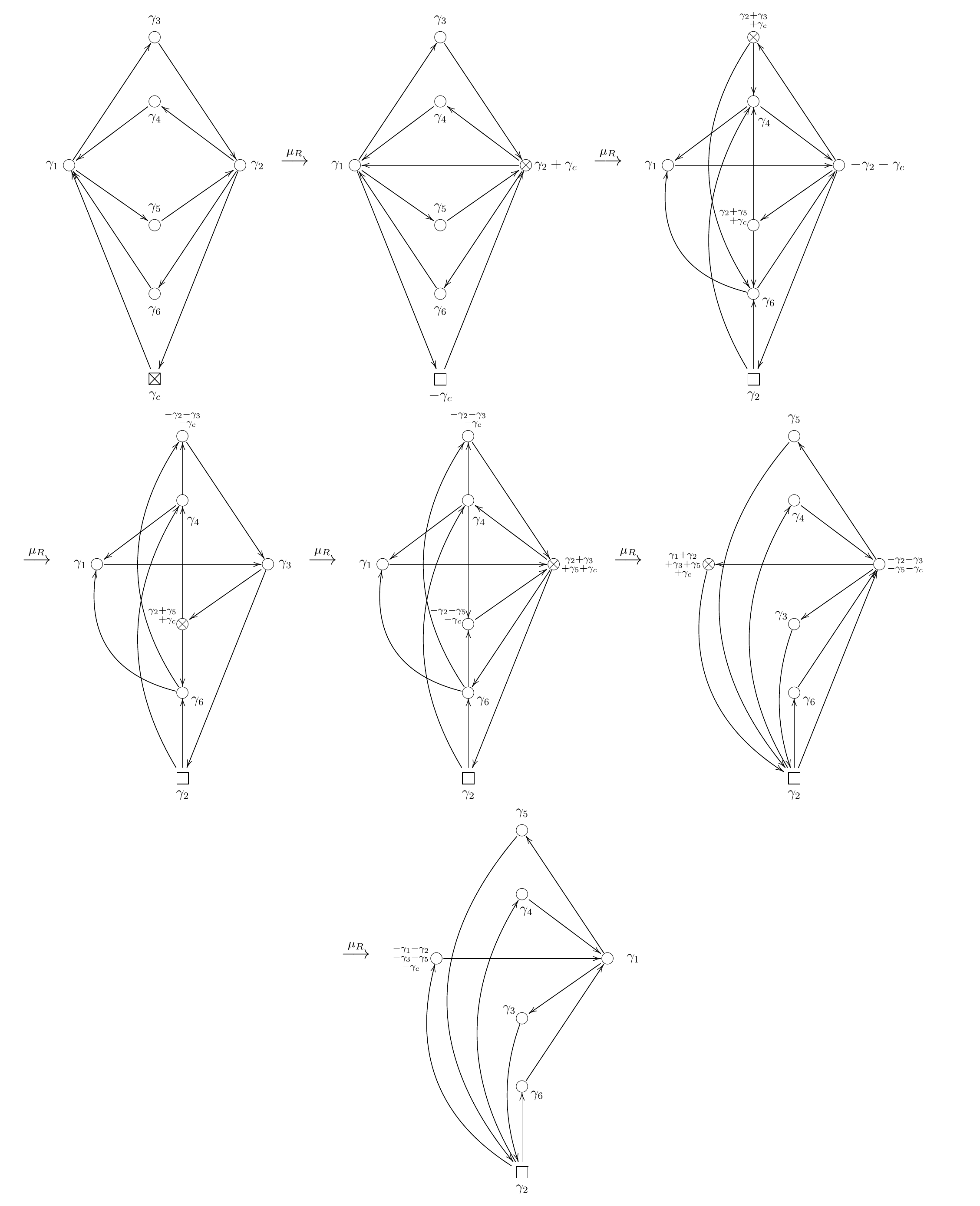}
\caption{Mutations of the framed quiver for the doublet Wilson line defect in the $SU(2)$ gauge theory with $N_f=4$ flavors. The core charge for the doublet Wilson line is $\gamma_c = -{1\over 2} (\gamma_1+\gamma_2 + \gamma_3 +\gamma_5)$. The crossed  denotes the node that is about to be right-mutated.}\label{fig:Nf4Mutation}
\end{figure}

\newpage
We have further computed the trace of $L_{0,1}$ to  $q^{3\over2}$ order,\begin{align}
&(q)_\infty^2  \text{Tr} [F(L_{0,1})  \, \mathcal{S}(q) \, \overline{\mathcal{ S}}(q)]  =  - \chi_{[1,0,0,0]} (\eta_i )\, q^{1\over2}  - \chi_{[1,1,0,0]} (\eta_i) \, q^{3\over2}+\mathcal{O}(q^{5\over2})\,,
\end{align}
where we have used \eqref{SO8fugacity} to express the flavor $X_\gamma$ in terms of the $SO(8)$ fugacities $\eta_i$.  Indeed, we see that the trace of $L_{0,1}$ nicely agrees with the line defect index \eqref{Nf4Doublet}.  In the case when the flavor fugacities are off, $\eta_i=1$, we have computed the trace of $L_{0,1}$ up to order $q^{7\over2}$,
\begin{align}
&(q)_\infty^2  \text{Tr} [F(L_{0,1})  \, \mathcal{S}(q) \, \overline{\mathcal{ S}}(q)] \Big|_{\eta_i=1} =   -8q^{1\over2} - 160q^{3\over2}  - 1624q^{5\over2} - 11768 q^{7\over2}+\mathcal{O}(q^{9\over2})\,,
\end{align}
which is equal to the  doublet Wilson line defect index $\mathcal{I}_{L_{0,1}}(q,\eta_i=1)$ in \eqref{Nf4Lines} obtained from localization.

\section{Half-Indices and Boundary Conditions}\label{sec:boundary}

The Schur index can be generalized further by the insertion of half-BPS boundaries or interfaces along the equator of the sphere. 
These boundary conditions will preserve $3d$ ${\cal N}=2$ supersymmetry.\footnote{The interfaces can likely be moved to a generic parallel on $S^3$. This is useful in order to discuss collisions of interfaces, 
but the space-time interpretation as conical defects is somewhat more cumbersome.} 

We have already encountered the simplest example in the form of the half-index 
\begin{equation}
\Pi^{S}_{\vec{m}}(q,u,z),
\end{equation}
which corresponds to a choice of Dirichlet boundary conditions for the UV gauge fields. 
Remember that we made a choice of Lagrangian splitting of the hypermultiplet scalar fields, 
which determines which half of the scalars has Dirichlet boundary condition and which half has Neumann boundary condition \cite{Dimofte:2012pd}.
The bulk gauge symmetry becomes a 3$d$ global symmetry at a Dirichlet boundary and thus the fugacity $u$ 
and magnetic charge $\vec{m}$ should be interpreted as associated to that 3$d$ global symmetry.

Remember that the general index is written as
\begin{align}
\mathcal{I}(q,z)  = (\Pi^N,\Pi^S) =  \sum_{\vec{m}} \int [du]_{\vec{m}}\,  \Pi^N_{-\vec{m}}(q,u,z) \, \Pi^S_{\vec{m}}(q,u,z)\,.
\end{align}
We can interpret the a sum over magnetic fluxes and the integral over gauge fugacities as following from the fact that the index is glued from two hemisphere indices with Dirichlet boundary condition
by restoring the gauge fields at the equator. Of course, only the term $\vec{m}=0$ contributes unless we add line defects in the two hemispheres. 

The half index for a more general  boundary condition, with Neumann boundary condition for the gauge group and general boundary matter fields is written analogously as 
\begin{align}
\mathcal{II}_{\vec{n}}(q,z,\xi)  = (\Pi^N,Z) = \sum_{\vec{m}} \int [du]_{\vec{m}}\,  \Pi^N_{-\vec{m}}(q,u,z) \, Z_{\vec{m}, \vec{n}}(q,u,z,\xi) \,,
\end{align}
where $Z_{\vec{m}, \vec{n}}(q,u,z,\xi)$ is the 3$d$ index of the boundary matter fields. 
We have included fugacities $\xi$ and magnetic flux $\vec{n}$ for 
possible other global symmetries of the 3$d$ matter fields. It is also possible to gauge at the boundary only a subgroup of the original 
gauge group, by restricting appropriately the $u$ integral and $\vec{m}$ sum. 

Again, only the term $\vec{m}=0$ contributes unless we add bulk line defects in the hemisphere, as in\footnote{Notice that the specific form of $\hat{\mathcal{O}}_L$ depends on the choice of hypermultiplet splitting.}
\begin{align}
\mathcal{II}^L_{\vec{n}}(q,z,\xi)  = (\Pi^N,\hat{O}_L  Z)  \,.
\end{align}

The line defect can be brought to any position along the great circle of the sphere, which intersects the equator at two points $N$ and $S$, 
the poles of the boundary $S^2$. In a purely 3$d$ context, the index $Z_{\vec{m}, \vec{n}}(q,u,z,\xi)$ often satisfies 
difference equations which arise from the insertion of line defects at $N$ or $S$. More precisely, $Z$ satisfies two sets of 
difference equations built from the two commuting sets of difference operators $x,p$ and $x',p'$. 

We propose the following IR description of the Schur half-index,
\begin{equation}\label{IRhalfindex}
\mathcal{II}_{\vec{n}}(q,z,\xi) = (q)^r_\infty \mathrm{Tr} \, \left[  Z^{IR}_{\vec{n}}(q,\xi)[X] \,\overline{\mathcal{S}}(q)  \right]\,,
\end{equation}
and for the half-index with line defect insertion,
\begin{equation}
\mathcal{II}^L_{\vec{n}}(q,z,\xi) = (q)^r_\infty \mathrm{Tr}\, \left[ F(L) Z^{IR}_{\vec{n}}(q,\xi)[X] \,\overline{\mathcal{S}}(q) \,\right]\,,
\end{equation}
where 
\begin{equation}
 Z^{IR}_{\vec{n}}(q,\xi)[X] \equiv \sum_\gamma  Z^{IR}_{\gamma,\vec{n}}(q,\xi) X_\gamma
\end{equation}
is a formal generating function for the 3$d$ indices of the IR degrees of freedom living on the domain wall, 
expressed in a charge basis for the Abelian symmetries which are coupled to the bulk Abelian gauge fields.  And as usual, $r$ is the rank of the Coulomb branch.

We can give an intuitive interpretation of formula \eqref{IRhalfindex} by interpreting it in the IR effective QED description of the Coulomb branch.  Indeed, If we pretend the BPS particles are free and mutually local, this would be a Lagrangian splitting for the bulk hypermultiplets, which is expected to be such that the bulk fields which survive at the boundary are those whose charges appear in the $\mathcal{S}(q)$ product. 

The full story is likely more complex and requires some way to define some kind of effective action, both in the bulk and boundary, analogous to what is done in 2$d$ in \cite{Gaiotto:2015zna, Gaiotto:2015aoa}.  At the level of the index, though, this approximate perspective is expected to be sufficient. 

With this caveat, the wall-crossing behavior of the IR boundary conditions is well-understood \cite{Dimofte:2013lba}.
Across a wall of marginal stability for a BPS particle, where some BPS ray exits the half plane associated to $\mathcal{S}(q)$ and 
the opposite ray enters it, the canonical choice of boundary condition also flips. 
The boundary degrees of freedom change in such a way  as to compensate the change in boundary condition.

For BPS hypermultiplets, the flip of boundary condition adds an extra chiral field to the boundary degrees of freedom. 
This multiplies $Z^{IR}_\gamma$ by the Fourier modes of 
\begin{equation}
Z_m^{\mathrm{chiral}}(z) = 
 \frac{\prod_{n=0}^\infty (1+z^{-1} q^{-m/2+n+\frac12})}{\prod_{n=0}^\infty (1+z q^{-m/2+n+\frac12})} \,,
\end{equation}
i.e. changes $Z^{IR}[X]$ to 
\begin{equation}
 \tilde Z^{IR}[X] = \prod_{n=0}^\infty (1+X_{-\gamma} q^{n+\frac12}) Z^{IR}[X] \frac{1}{\prod_{n=0}^\infty (1+ X_\gamma q^{n+\frac12})}  = E^{-1}_q(X_{-\gamma})Z^{IR}[X]E_q(X_{\gamma})\,.
\end{equation}

Then the candidate Schur index is invariant 
\begin{equation}
\mathcal{II}_{\vec{n}}(q,z,\xi) = (q)_\infty^r \mathrm{Tr} \left[\,\mathcal{S}(q) \, Z^{IR}_{\vec{n}}(q,\xi)[X]\,\right]= (q)^r_\infty \mathrm{Tr} \left[ \, \tilde{ \mathcal{S}}(q)\, \tilde Z^{IR}_{\vec{n}}(q,\xi)[X]\,\right]\,,
\end{equation}
as $\mathcal{S}(q) = E_q(X_{\gamma}) \tilde {\mathcal{S}}(q) E^{-1}_q(X_{-\gamma})$.  Vice versa, adding a chiral of the opposite charge to cross the wall backwards gives 
\begin{equation}
Z^{IR}[X] = E_q(X_{-\gamma})  \tilde Z^{IR}[X] E^{-1}_q(X_{\gamma})\,.
\end{equation}

We expect these relations to hold for BPS particles of every spin. It would be interesting to understand which boundary degrees of freedom 
are added or removed in that case, but as higher spin BPS particles usually come together with infinite cohorts of hypermultiplet particles (see e.g. \cite{Galakhov:2013oja}), 
individual wall-crossing events are perhaps less physically meaningful. 

\subsection{Indices and RG Interfaces}
There is a special class of boundary conditions/interfaces which is very useful in relating BPS quantities in the UV and IR description 
of ${\cal N}=2$ gauge theories: \textit{RG interfaces}. These are special interfaces between the UV theory and its IR effective description, 
obtained by applying the IR effective description to one side only of the identity interface in the UV \cite{Gaiotto:2012np, Dimofte:2013lba}. 

A very useful properties of RG interfaces is that they intertwine between the IR and UV description of several BPS objects, included line defects and boundary conditions. 
Thus the IR description of a UV boundary condition is obtained by acting on it with the RG interface, and vice versa. This implies a precise relation between the 
indices $Z^{UV}_{\vec m}(q,u)$ and $Z^{IR}[X]$ of the UV and IR boundary degrees of freedom. Similar considerations apply to line defects. We will write down 
these relations momentarily. 

We will call the interface degrees of freedom defining an RG interface the \textit{RG theory}. Formally, the interface degrees of freedom
 can be obtained by starting from UV Dirichlet boundary conditions and 
flowing to the IR: the result should be the RG theory coupled to the IR Abelian gauge theory. 
Vice versa, the UV boundary condition defined by the RG theory will flow to Dirichlet boundary conditions for the IR theory.
Of course, the RG theory depends on a choice of hypermultiplet splitting in the UV and a chamber as well as an electromagnetic duality frame for the IR theory.  The RG theory transforms appropriately as these choices are modified.  Explicit examples of conjectural RG theories were described in \cite{Dimofte:2013lba}. 

At the level of the indices, the UV Dirichlet boundary condition gives us the half-index and thus we expect
\begin{equation}
\Pi^S_{\vec{m}}(q,u,z) = (q)_\infty^r \mathrm{Tr}  \left[ \, K_{\vec{m}}(q,u,z)[X] \, {\mathcal{ S}}_{\vartheta+\pi}(q) \,\right]\,,
\end{equation}
where $K_{\gamma, \vec{m}}(q,u,z)$ is the 3$d$ index of the RG theory.  Building the UV boundary condition which flows to a 
simple Dirichlet IR boundary condition gives us an inverse relation:
\begin{equation}
(q)_\infty^r \mathcal{S}_\vartheta(q) = (\Pi^N,K[X]) = \sum_{\vec{m}} \int [du]_{\vec{m}}\, \Pi^N_{-\vec{m}}(q,u,z)  \,K_{\vec{m}}(q,u,z)[X]\,.
\end{equation}
In particular, this gives a direct physical meaning of the quantum spectrum generator $\mathcal{S}(q)$ as a generating function for Schur indices in the presence of the RG interface boundary condition.

Notice also that these relations are compatible and imply the identity between the UV and IR bulk Schur indices: 
\begin{equation}
(\Pi^N,\Pi^S)= (q)_\infty^r \mathrm{Tr} \left[\, (\Pi^N,K[X])\, {\mathcal{ S}}_{\vartheta+\pi}(q)\,\right] = (q)_\infty^{2r} \mathrm{Tr}\left[ \,  \mathcal{S}_\vartheta(q) {\mathcal{ S}}_{\vartheta+\pi}(q)\,\right]\,.
\end{equation}
Furthermore, the existence of the Kernel $K_{\vec{m}}(q,u,z)[X]$ also implies our formulae for line defects: 
the index of the RG interface satisfies difference equations of the form 
\begin{equation} \label{eq:rec1}
 \hat{O}_L K[X] = F(L,\vartheta) K[X] \,,
\end{equation}
and
\begin{equation}\label{eq:rec2}
 K[X]  \hat{O}'_L=K[X]  F(L,\vartheta+\pi) \,,
\end{equation}
The above two relations  imply our proposal for the line defect indices \eqref{lineproposal},
\begin{align}
(\Pi^N,\hat{O}_L \Pi^S)&= (q)_\infty^r \mathrm{Tr}\left[\, (\Pi^N,\hat{O}_L K[X])\,{\mathcal{ S}}_{\vartheta+\pi}(q)\,\right] = (q)_\infty^r \mathrm{Tr}  \left[\, F(L,\vartheta)\, (\Pi^N, K[X])\,{\mathcal{ S}}_{\vartheta+\pi}(q)\,\right]\notag\\
& = (q)_\infty^{2r} \mathrm{Tr}\left[\, F(L,\vartheta) \mathcal{S}_\vartheta(q){\mathcal{S}}_{\vartheta+\pi}(q)\,\right]\,.
\end{align}

Indeed, one can argue that equations \eqref{eq:rec1} and \eqref{eq:rec2} are the truly crucial relationships. For example,  they imply a recursion relation 
\begin{equation}
F(L,\vartheta)  \, (\Pi^N,K[X]) = (\Pi^N,K[X]) \,F(L,\vartheta+\pi) \,,
\end{equation}
which in turn implies its identification with $(q)_\infty^r \mathcal{S}_\vartheta(q)$ up to an overall function of $q$
and similarly for $(q)_\infty^r \mathrm{Tr}\left[\, K_{\vec{m}}(q,u,z)[X] {\mathcal{ S}}_{\vartheta+\pi}(q)\,\right]$.

Similar considerations apply in the presence of boundary conditions.  If $Z^{UV}$ is the partition function in the presence of a UV boundary condition and $Z^{IR}$ the partition function of the IR boundary condition to which it flows, we expect the relations 
\begin{equation} \label{eq:rec3}
(Z^{UV},K[X]) = Z^{IR}[X]~, \qquad \qquad  Z^{UV} = \mathrm{Tr} \left[\, Z^{IR}[X] \bar K[X]\,\right]~.
\end{equation}
These then imply
\begin{equation}
(\Pi^N,Z^{UV})= \mathrm{Tr} \left[\, Z^{IR}[X] (\Pi^N,\bar K[X]) \,\right] = (q)_\infty^r \mathrm{Tr} \left[\,Z^{IR}[X]  \mathcal{S}_{\vartheta+\pi}(q)\,\right]\,.
\end{equation}

\subsection{A Free Hypermultiplet}
The free hypermultiplet index in IR conventions is 
\begin{equation}
\mathcal{I}_{\mathrm{hyper}} = \frac{1}{\prod_{n=0}^\infty (1+z q^{n+\frac12})(1+z^{-1} q^{n+\frac12})} = E_q(z) E_q(z^{-1})~.
\end{equation}
This is an obvious example of the BPS formula.

The half-indices for the two possible half-BPS boundary conditions are 
\begin{equation}
\mathcal{II}_{\mathrm{hyper},\pm} = \frac{1}{\prod_{n=0}^\infty (1+z^\pm q^{n+\frac12})} = E_q(z^\pm) ~.
\end{equation}
This is a neat example of the BPS formula for half-indices, which is already somewhat non-trivial. 

If we pick the phase of the hypermultiplet mass in such a way that $\mathcal{S}(q) = E_q(z)$, we have that 
$\mathcal{II}_{\mathrm{hyper},+}$ simply equals $\mathcal{S}(q)$. This makes sense: we are already using the ``canonical''
choice of IR boundary condition for the bulk hypermultiplet. 

On the other hand, we have 
\begin{equation}
\mathcal{II}_{\mathrm{hyper},-} =  E_q(z^{-1}) = \mathcal{S}(q) E^{-1}_q(z) E_q(z^{-1})\,.
\end{equation}
We  recognize the expression $E^{-1}_q(z) E_q(z^{-1})$ as the 3$d$ index of a 
3$d$ chiral field with no magnetic flux on the two-sphere. Indeed, the two boundary conditions can be 
related by adding a boundary chiral field with appropriate boundary superpotential coupling to the 
hypermultiplet  \cite{Dimofte:2011py,Dimofte:2012pd,Dimofte:2013lba}. 

\subsection{Pure $SU(2)$ Gauge Theory}
The RG  theory for pure $SU(2)$ gauge theory consists  of 
a $SU(2)$ doublet of chiral fields, transforming with charge $-1$ under a $U(1)$ global symmetry \cite{Dimofte:2013lba}. 

We can immediately compute  
\begin{align}
\mathcal{II}^{RG}_{n}(q,\xi) &= -\frac{1}{4 \pi i} \int \frac{d u}{u}\,(u - u^{-1})^2 \left[(q;q)_\infty (q u^2;q)_\infty (q u^{-2} ;q)_\infty  \right]\notag\\
&\times\left[ \frac{(-q^{\frac12+\frac n2} \xi u^{-1};q)_\infty}{(-q^{\frac12+\frac n2} \xi^{-1} u;q)_\infty}\frac{(-q^{\frac12+\frac n2} \xi u;q)_\infty}{(-q^{\frac12+\frac n2} \xi^{-1} u^{-1};q)_\infty}\right]\,\,,
\end{align}
where $\xi$ and $n$ are the fugacity and the magnetic flux, respectively, for the 3$d$ $U(1)$ global symmetry of the RG theory. We have separated in the integrand the contributions from the hemisphere and from the boundary doublet. The superscript $RG$ indicates that we are computing the half index in the RG boundary condition. 

Explicit calculation at finite order in $q$ suggests that this complicated contour integral has a dramatically simple answer: 
\begin{equation}
\mathcal{II}^{RG}_{n}(q,\xi) = (q)_\infty \sum_{e=\mathrm{max}(0,-n)}^{\infty} \xi^{2e} \frac{q^{e(e+n)}}{(q)_e (q)_{e+n}}   \,,
\end{equation}

The corresponding generating function is 
\begin{align}
\mathcal{II}^{RG}(q;X)& \equiv \, :\sum_n\, (-1)^nq^{\frac n2} \, \mathcal{II}^{RG}_{n}(q,\xi = q^{\frac12}X_\gamma)X_{- n \gamma'}:\notag\\
&= (q)_\infty \sum_n \sum_{e=\mathrm{max}(0,n)}^{\infty}  (-1)^n q^{e-\frac n2} \frac{q^{e(e-n)}}{(q)_e (q)_{e-n}} X_{n \gamma' + 2 e \gamma}  \,,
\end{align}
where we denote the electric charge and the magnetic charge by $\gamma$ and $\gamma'$, respectively. The insertion of $(-1)^n$ is interpreted as a convenient shift of the fermion number of monopole operators, while the insertion of $q^{n\over2}$ and the factor of $q^{1\over2}$ in $\xi$ are useful re-definitions of the boundary R-charge. The latter essentially assigns trivial R-charge and fermion number to the 
bosonic components of the boundary chiral multiplet.\footnote{Intuitively, the doublet encodes the direction of the breaking of $SU(2)$ to $U(1)$ and thus should have 
scaling dimension $0$ in the IR.}  Importantly, the R-charge shift $\xi \to q^{1/2} \xi$ has to be performed \textit{after} the contour integral in $u$, not before.

The generating function for the half index can be written as
\begin{equation}
\mathcal{II}^{RG}(q;X) = (q)_\infty \sum_{e\geq 0} \sum_{e' \geq 0} (-q^{\frac12})^{e+e'}\frac{q^{e e'}}{(q)_e (q)_{e'}} X_{- e' \gamma' +e (\gamma' +  2 \gamma)}  = (q)_\infty E_q( X_{2 \gamma+\gamma'})E_q(X_{-\gamma' }) \,,
\end{equation}
where we set the Dirac pairing $\langle \gamma',\gamma\rangle=1$.  
This is the same as $(q)_\infty \mathcal{S}(q)$ if we choose the following electromagnetic duality frame $\gamma_1 =2\gamma+ \gamma'$ and $\gamma_2 = -\gamma'$ for the two nodes in the BPS quiver of the pure $SU(2)$ theory!  We can thus identify 
\begin{equation}
K[X] \equiv\, : \sum_n  (-1)^n q^{n\over2} \left[ \frac{(-q^{1+\frac n2} X_\gamma u^{-1};q)_\infty}{(-q^{\frac n2} X_{-\gamma} u;q)_\infty}\frac{(-q^{1+\frac n2}X_\gamma u;q)_\infty}{(-q^{\frac n2} X_{-\gamma} u^{-1};q)_\infty}\right]X_{- n \gamma'}: \,.
\end{equation}

Similarly, if we insert a Wilson loop operators
\begin{align}
\mathcal{II}^{RG;W}_{n}(q,\xi) &\equiv -\frac{1}{4 \pi i} \int \frac{d u}{u}\,(u - u^{-1})^2 (u + u^{-1})\left[(q)_\infty (q u^2;q)_\infty (q u^{-2} ;q)_\infty  \right]\notag\\
&\times\left[ \frac{(-q^{\frac12+\frac n2} \xi u^{-1};q)_\infty}{(-q^{\frac12+\frac n2} \xi^{-1} u;q)_\infty}\frac{(-q^{\frac12+\frac n2} \xi u;q)_\infty}{(-q^{\frac12+\frac n2} \xi^{-1} u^{-1};q)_\infty}\right]\,\,,
\end{align}
we get a neat expression
\begin{equation}
\mathcal{II}^{RG;W}_{-n}(q,\xi)=  (q)_\infty \sum_{e=\mathrm{max}(0,n)}^{\infty} \xi^{1-2e} \frac{q^{e^2-e n - e + \frac{n+1}{2}}-q^{e^2-e n + \frac{n+1}{2}}-q^{e^2-e n - \frac{n-1}{2}}}{(q)_e (q)_{e-n}} \,,
\end{equation}
The corresponding generating function is 
\begin{equation}
\mathcal{II}^{RG;W}(q;X) = (q)_\infty \sum_{e\geq 0} \sum_{e' \geq 0} (-q^{\frac12})^{e+e'-1} \frac{q^{(e-\frac12)(e'-\frac12) + \frac{1}{4}}(1-q^e-q^{e'})}{(q)_e (q)_{e'}} X_{e' \gamma' -e (\gamma' + 2 \gamma)+ \gamma}  \,,
\end{equation}
which can be manipulated to 
\begin{equation}
\mathcal{II}^{RG;W}(q;X) = (q)_\infty E_q(X_{\gamma' })E_q( X_{-2 \gamma-\gamma'})  X_{ \gamma} - (q)_\infty E_q(X_{\gamma' }) X_{-\gamma} E_q( X_{-2 \gamma-\gamma'})  \,,
\end{equation}
which yields a reasonable framed BPS degeneracy:
\begin{equation}
\mathcal{II}^{RG;W}(q;X) =(q)_\infty \mathcal{S}(q) \left[X_{\gamma} -   X_{- \gamma} -   X_{-3 \gamma-\gamma'}\right] \,.
\end{equation}

We can deal in a similar manner with 't Hooft lines. Consider for example $\hat O_{1,0}$\begin{equation}
\frac{i}{q^{\frac m2}u - q^{-\frac m2}u^{-1}}(p^{\frac12} - p^{-\frac12}) 
\end{equation}
so that ($n$ has to be half-integral here)
\begin{align}
&\mathcal{II}^{RG;\hat O_{1,0}}_{n}(q,\xi) \equiv -\frac{1}{4 \pi } \int \frac{d u}{u}\,(u - u^{-1}) \left[(q)_\infty (q u^2;q)_\infty (q u^{-2} ;q)_\infty  \right] \cr
&\left[ \frac{(- q^{\frac n2}\xi u^{-1};q)_\infty}{(-q^{\frac12+\frac n2} \xi^{-1} u;q)_\infty}\frac{(-q^{1+\frac n2} \xi u;q)_\infty}{(-q^{\frac12+\frac n2} \xi^{-1} u^{-1};q)_\infty} -  \frac{(-q^{1+\frac n2} \xi u^{-1};q)_\infty}{(-q^{\frac12+\frac n2} \xi^{-1} u;q)_\infty}\frac{(-q^{\frac n2} \xi u;q)_\infty}{(-q^{\frac12+\frac n2} \xi^{-1} u^{-1};q)_\infty}\right]\,\,,
\end{align}
which becomes 
\begin{equation}
\mathcal{II}^{RG;\hat O_{1,0}}_{n}(q,\xi) \equiv \xi q^{\frac n2} \mathcal{II}^{RG}_{n+\frac12}(q,q^{\frac14}\xi)\,\,,
\end{equation}
i.e. 
\begin{equation}
\mathcal{II}^{RG;\hat O_{1,0}}(q;X) \equiv  (q)_\infty \mathcal{S}(q) X_{\frac{\gamma'}{2}+\gamma}\,\,.
\end{equation}

Similarly, for the 't Hooft-Wilson line, the difference operator  $\hat O_{1,1}$  is
\begin{equation}
\frac{i}{q^{\frac m2}u - q^{-\frac m2}u^{-1}}(q^{\frac m2}u p^{\frac12} - q^{-\frac m2}u^{-1} p^{-\frac12}) 
\end{equation}
so that ($n$ has to be half-integral here)
\begin{align}
&\mathcal{II}^{RG;\hat O_{1,1}}_{n}(q,\xi) \equiv -\frac{1}{4 \pi i} \int \frac{d u}{u}\,(u - u^{-1}) \left[(q)_\infty (q u^2;q)_\infty (q u^{-2} ;q)_\infty  \right] \notag\\
&\left\{ u\left[ \frac{(- q^{\frac n2}\xi u^{-1};q)_\infty}{(-q^{\frac12+\frac n2} \xi^{-1} u;q)_\infty}\frac{(-q^{1+\frac n2} \xi u;q)_\infty}{(-q^{\frac12+\frac n2} \xi^{-1} u^{-1};q)_\infty}\right]-u^{-1}\left[ \frac{(-q^{1+\frac n2} \xi u^{-1};q)_\infty}{(-q^{\frac12+\frac n2} \xi^{-1} u;q)_\infty}\frac{(-q^{\frac n2} \xi u;q)_\infty}{(-q^{\frac12+\frac n2} \xi^{-1} u^{-1};q)_\infty}\right] \right\}\,\,,
\end{align}
which becomes 
\begin{equation}
\mathcal{II}^{RG;\hat O_{1,1}}_{n}(q,\xi) \equiv \mathcal{II}^{RG}_{n+\frac12}(q,q^{\frac14}\xi)\,\,,
\end{equation}
i.e. 
\begin{equation}
\mathcal{II}^{RG;\hat O_{1,1}}(q,X) \equiv q^{-\frac14} (q)_\infty \mathcal{S}(q) X_{\frac{\gamma'}{2}}\,\,.
\end{equation}

Next, we should compute the IR index for Dirichlet boundary condition
\begin{align}
(q)_\infty \mathrm{Tr} \left[\,\overline{\mathcal{ S}}(q) K_{m}(q,u)[X] \,\right]&= (q)_\infty \sum_{n=-\infty}^\infty \sum_{e=\text{max}(0,n)}^\infty q^{2e-n} \frac{ q^{e(e-n)}}{(q)_e(q)_{e-n}}\notag\\
&\times \frac{1}{2 \pi i} \oint \frac{d\xi}{\xi} \xi^{2e}\left[ \frac{(-q^{\frac12-\frac n2+\frac m2} \xi^{-1} u^{-1};q)_\infty}{(-q^{\frac12-\frac n2+\frac m2} \xi u;q)_\infty}\frac{(-q^{\frac12-\frac n2-\frac m2} \xi^{-1} u;q)_\infty}{(-q^{\frac12-\frac n2-\frac m2} \xi u^{-1};q)_\infty}\right]
\end{align}
The $q^{2e-n}$ factor is due again to the choice of quantum numbers for the boundary theory. 

Amazingly, the sum vanishes unless $m=0$ and at $m=0$ it reproduces the expected answer:
\begin{equation}
(q)_\infty \mathrm{Tr} \left[\,\overline{\mathcal{S}}(q) K_{m}(q,u)[X]\,\right] = \delta_{m,0} (q)_\infty (q u^2 ;q)_\infty(q u^{-2};q)_\infty = \Pi^S_{m}(q,u) \end{equation}

Of course, all these miraculous-looking relations are somewhat demystified by the recursion relations satisfied by  $K[X]$, which  
can be easily seen to intertwine between the difference operators associated to UV line defects and the corresponding generating functions of IR framed BPS degeneracies. 

%
%
%

\section{Defect Indices in Argyres-Douglas Theories}\label{sec:LineAD}

One important application of our IR formula  in Section \ref{sec:IRSchur} is a prediction for the line defect Schur indices in the strongly-coupled Argyres-Douglas theories, where a UV localization calculation is not available.  In this section, we will compute the Schur indices of the $A_2,A_3,A_4$ Argyres-Douglas theories with the presence of line defects using the conjectural formula \eqref{lineproposal}.  We will also discuss how the  OPEs between the  defects are respected by the line defect indices $\mathcal{I}_L(q)$.  We will defer the discussion on the relation between these defect OPEs with the Verlinde algebra of the associated chiral algebra to Section \ref{sec:verlinde}.

\subsection{Line Defect OPEs and Schur Indices}

The UV line defects satisfy a non-commutative defect OPE that takes the form \cite{Gaiotto:2010be,Cordova:2013bza}
\begin{align}
L_\alpha L_\beta \equiv \lim_{\epsilon\to 0}L_\alpha (\vartheta) L_\beta (\vartheta+\epsilon)= \sum_\gamma c_{\alpha \beta}^\gamma(q) L_\gamma(\vartheta)\,,
\end{align}
where the coefficients $c_{\alpha \beta}^\gamma(q) $ are valued in $\mathbb{Z}_{\ge0}[q^{1\over2} ,q^{-\frac12}]$.  We have restored the central charge phase $\vartheta$ dependence to indicate their positions on $S^3\times S^1$. 
The defect OPE can be intuitively understood as bringing two  line defects, which are points on a great circle in $S^3\times S^1$,  close to each other to form a composite line defect, which then admits the above expansion in terms of simple defects. This configuration  preserves the $U(1)$ rotation transverse to the great circle and the $SU(2)_R$ symmetry, and we can turn on the variable $q$ to keep track of these quantum numbers. The resulting OPE is non-commutative because the first line defect can approach the second one either from above or from below on the great circle.   See Figure \ref{fig:OPE}.

\begin{figure}[h!]
\centering
\includegraphics[width=.9\textwidth]{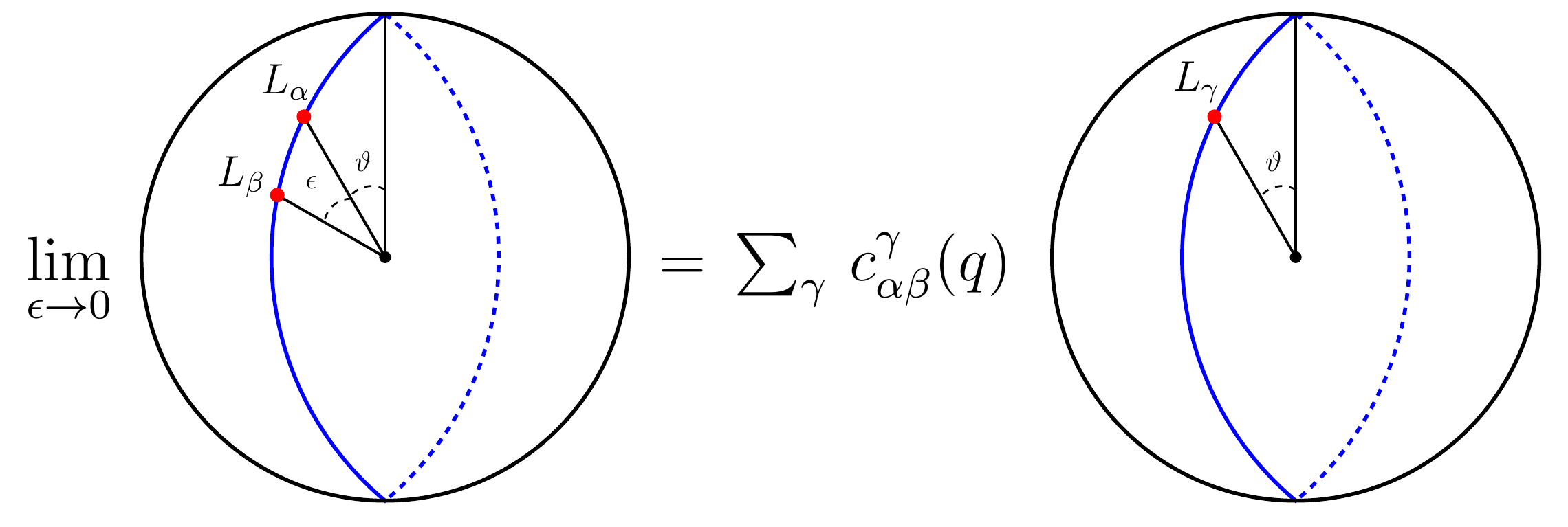}
\caption{The  OPE for line defects. As we bring two line defects near each other on the $S^3$ while they both wrap around the $S^1$, the composite defect can be expanded into a sum of simple defects. }\label{fig:OPE}
\end{figure}

 The simplest example of the line defect OPE is that between the IR Abelian defects $X_\gamma$ given in \eqref{abelianope},
\begin{align*}
X_{\gamma}X_{\gamma'}=q^{\frac{1}{2}\langle \gamma, \gamma' \rangle}X_{\gamma+\gamma'}~,
\end{align*}
where the Dirac pairing $\langle \gamma,\gamma'\rangle$  captures the angular momentum of the composite defect.

Because of supersymmetry, the defect OPE is independent of the distance separated between the two defects, and thus the OPE computed in the UV should agree with that computed in the IR.  This implies that the IR description of the UV line defect, i.e., the generating function \eqref{generating} 
$$F(L)=\sum_{\gamma \in \Gamma} \fro(L,\gamma,q)X_{\gamma},$$
has to obey the same OPE
\begin{align}\label{nonabelianope}
\lim_{\epsilon\to 0}F(L_\alpha,\vartheta) F(L_\beta,\vartheta+\epsilon)  =  \sum_\gamma c_{\alpha \beta}^\gamma(q) F(L_\gamma,\vartheta)\,,
\end{align}
where the product on the lefthand side is given by the non-commutative product of  IR Abelian  line defects in \eqref{abelianope}. This provides a strong consistency check on the framed BPS state degeneracies $\fro(L,\gamma,q)$.  Since  the  defect OPE can be computed in the UV, the coefficients $c_{\alpha \beta}^\gamma(q)$ must be wall-crossing invariant, while the individual generating functions $F(L_\alpha, \vartheta)$ are not.

One consequence of our IR formula  for the line defect Schur index \eqref{lineproposal},
\begin{equation}\label{lineproposal2}
\mathcal{I}_{L}(q) = (q)_\infty^{2r}~\text{Tr} \left[ F(L,\vartheta) \mathcal{S}_\vartheta(q)\mathcal{S}_{\vartheta+\pi}(q)\right]~,
\end{equation}
is that it manifestly respect the defect OPE because $F(L,\vartheta)$ does.  That is,
\begin{align}
\mathcal{I}_{L_\alpha L_\beta} (q) =  \sum_\gamma c_{\alpha \beta}^\gamma(q)\,  \mathcal{I}_{L_\gamma}(q)\,.
\end{align}
In addition, we will obtain more relations between the defect indices than those descended from the defect OPEs in the case of the  Argyres-Douglas theories.

\subsection{$A_2$ Argyres-Douglas Theory}\label{sec:A2}

\begin{figure}[h!]
\begin{center}
\includegraphics[width=.25\textwidth]{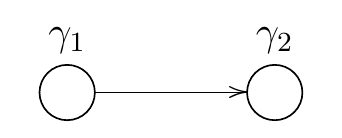}
\end{center}
\caption{ The BPS quiver for the $A_2$ Argyres-Douglas theory.}\label{fig:A2}
\end{figure}

The $A_2$ Argyres-Douglas theory arises from special points on the moduli space of the pure $SU(3)$ gauge theory or from the $SU(2)$ SQCD with $N_f=1$ flavor \cite{Argyres:1995jj,Argyres:1995xn}.  
The UV line defects in the $A_2$ Argyres-Douglas theory are generated by five defects $L_i$'s together with the unit operator. Assuming the defect phase $\vartheta$ is chosen to lie to the right of all ordinary BPS states, the generating functions for $L_i$'s are \cite{Gaiotto:2010be,Cordova:2013bza},
\begin{align}
\begin{split}
&F(L_1) = X_{\gamma_1}\,,\\
&F(L_2) = X_{\gamma_2} +X_{\gamma_1+\gamma_2}\,,\\
&F(L_3)= X_{-\gamma_1}  +X_{-\gamma_1+\gamma_2 } +X_{\gamma_2}\,,\\
&F(L_4) =  X_{-\gamma_1-\gamma_2}  + X_{-\gamma_1} \,,\\
&F(L_5)=  X_{-\gamma_2}\,.
\end{split}
\end{align}
It follows that the five $L_i$'s satisfy the OPE algebra,
\begin{align}\label{A2OPE}
L_i  L_{i+2}  = 1+q^{1\over2} L_{i+1}\, ,
\end{align}
which also implies $L_i  L_{i-2}  =1+q^{-{1\over2}} L_{i-1}$. We have taken the index $i$ to be periodic mod 5. 

Following our proposal \eqref{IRschurLine}, the Schur index  with the insertion of $L_i$ is computed by the trace of $F(L_i)$, i.e., $\mathcal{I}_{L_i} = (q)_\infty^2 \text{Tr}[F(L_i) \mathcal{S}(q) \overline{\mathcal{ S}}(q)]$. After a similar calculation as in the pure $SU(2)$ gauge theory, $\mathcal{I}_L$ can be computed to be
\begin{align}
&\mathcal{I}_{L_i}(q)= (q)^2_\infty \,\text{Tr}\left[\, F(L_i) \, E_q(X_{\gamma_1})E_q(X_{\gamma_2})\, E_q(X_{\gamma_1}^{-1})E_q(X_{\gamma_2}^{-1})\,\right]\\
&~~~~= - q^{{1\over2}} (1+q^3 +q^4 +q^5 +q^6+q^7+2q^8+2q^9+3q^{10}+\cdots)\,,~~~~
\text{for all} ~~i=1,\cdots,5\,.\notag
\end{align}
Notice that the dependence on the index $i$  is washed out inside the trace, reflecting the $\mathbb{Z}_5$ symmetry of the $A_2$ Argyres-Douglas theory. We can therefore define $\mathcal{I}_L$ unambiguously as, 
\begin{align}
\mathcal{I}_L  \equiv \mathcal{I}_{L_1} = \mathcal{I}_{L_2}=\cdots = \mathcal{I}_{L_5}\,,
\end{align}
More explicitly, for example when $i=1$, the trace of $L_i$ can written as
\begin{align}
\mathcal{I}_L&= 
(q)_\infty^2 \, \sum_{\ell_1,\ell_2=0}^\infty 
{ (-1) q^{\ell_1+2\ell_2+\ell_1\ell_2 + {1\over2} } \over (q)_{\ell_1} [(q)_{\ell_2}]^2 (q)_{\ell+1} }\\
&
= - q^{-{1\over2}} (1+q^3 +q^4 +q^5 +q^6+q^7+2q^8+2q^9+3q^{10}+\cdots)\,,~~~~
\text{for all} ~~i=1,\cdots,5\,.\notag
\end{align}

We further observe the following  relations among the line defect Schur indices (no sum in the indices), 
\begin{align}\label{LL}
\begin{split}
&\mathcal{I}_{ L_i L_{i-2}} =\mathcal{I} + q^{-{1\over2} } \mathcal{I}_{L}\,,\\
&\mathcal{I}_{ L_i L_{i-1} } =q^{-1}\mathcal{I} + q^{-{3\over2} } \mathcal{I}_{L}\,,\\
&\mathcal{I}_{L_i L_{i}} =q^{-1}\mathcal{I} + q^{-{3\over2} } \mathcal{I}_{L}\,,\\
&\mathcal{I}_{ L_i L_{i+1}} =\mathcal{I} + q^{-{1\over2} } \mathcal{I}_{L}\,,\\
&\mathcal{I}_{L_i L_{i+2}} =\mathcal{I} + q^{{1\over2} } \mathcal{I}_{L}\,,
\end{split}
\end{align}
which hold true for any $i=1,\cdots,5$. Here $\mathcal{I}$ is the Schur index without the insertion of line defects.  Note that the first and the last relations simply follow from the UV line defect OPE \eqref{A2OPE}. The third relation was already noticed in \cite{Cecotti:2010fi} in the case of the inverse of the quantum KS operator, and a connection with the Verlinde algebra was observed. We will give a similar proposal in Section \ref{sec:verlinde}.

\subsection{$A_3$ Argyres-Douglas Theory}\label{sec:A3}

\begin{figure}[h!]
\begin{center}
\includegraphics[width=.4\textwidth]{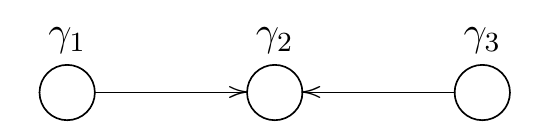}
\end{center}
\caption{ The BPS quiver for the $A_3$ Argyres-Douglas theory.}\label{fig:A3}
\end{figure}

The $A_3$ Argyres-Douglas theory arises from special points on the moduli space of the  $SU(2)$ SQCD with $N_f=2$ flavors \cite{Argyres:1995xn}.   The $A_3$ Argyres-Douglas theory has an $SU(2)$ flavor symmetry which corresponds to the direction $\gamma_1-\gamma_3$ in the charge lattice.

The core charges of the line defects in the $A_3$ Argyres-Douglas theory can  be derived from the seeds of the BPS quiver and their dual cones \cite{Gaiotto:2010be,Cordova:2013bza}.   We present the details of the derivation of the framed BPS quivers in Appendix \ref{sec:A3Cones} following the logic of \cite{Cordova:2013bza}.  The result is that the  line defects in the $A_3$ Argyres-Douglas theory  are generated by  six defects, $A_i$, $B_i$, $i=1,2,3$, one flavor defect $C$, and the unit operator.

The generating functions for these line defects can then  be obtained straightforwardly from the associated framed BPS quivers, say, by the mutation method \cite{Alim:2011kw}.   Assuming the defect phase $\vartheta$ is to the right of all the ordinary BPS state phases, the generating functions for the above six defects are 
\begin{align}\label{generatingA3}
\begin{split}
&F(A_1) =X_{\gamma'}\,,\\
&F(A_2 )= X_{-\gamma'-\gamma_2}+X_{-\gamma'} \,,\\
&F(A_3) = X_{-\gamma'} +(z+z^{-1})X_{\gamma_2} +X_{\gamma'+\gamma_2} +X_{-\gamma'+\gamma_2}\,,\\
&F(B_1 )=  X_{-\gamma_2}\,,\\
&F(B_2)=  X_{\gamma_2} +(z+z^{-1})X_{-\gamma'} +(z+z^{-1})X_{-\gamma'+\gamma_2}  + (q^{1\over2} +q^{-{1\over2}} ) X_{-2\gamma'}  +X_{-2\gamma'+\gamma_2} +X_{-2\gamma'-\gamma_2 }\, ,\\
&F(B_3) = X_{\gamma_2} +(z+z^{-1})X_{\gamma'+\gamma_2}  +X_{2\gamma'+\gamma_2}\,,\\
&F(C) =z+z^{-1}\,.
\end{split}
\end{align}
Here $z$ is the $SU(2)$ flavor fugacity that is related to the flavor generator as
\begin{align}
\text{Tr} [ X_{\gamma_1-\gamma_3\over2} ] = z\,.
\end{align}
$\gamma'$ is defined as
\begin{align}
\gamma' \equiv {\gamma_1+\gamma_3\over2}\,.
\end{align}
Note that $\gamma'$ has the same Dirac pairings with every charge vector as those of $\gamma_1$ and $\gamma_3$.  Note that the UV flavor defect $C$ is a Wilson line in the $\mathbf{2}$ of the flavor $SU(2)$ symmetry, whose insertion into the path integral is just an overall multiplication of $z+z^{-1}$.

The  defect OPE can be readily derived from the above generating functions,
\begin{align}\label{A3OPE}
\begin{split}
&A_i A_{i+1} = 1+q^{-{1\over2}} B_i\,,\\
&B_i B_{i+1} = 1+q^{-{1\over2}}C A_{i+1} + q^{-1} A_{i+1}^2\,,\\
&A_i B_{i+1} =  C+q^{-{1\over2} } A_{i+1} + q^{1\over2} A_{i+2}\,,
\end{split}
\end{align}
with the flavor defect $C$ commuting with everything. Here we view the index $i$ as periodic mod 3.

To compute the line defect indices of $A_i,B_i$, we will repeatedly encounter the insertion of an IR line defect $X_{a\gamma'+b\gamma_2}$,
\begin{align}
\begin{split}\label{X}
& (q)_\infty^2 \, \text{Tr}
[ \, X_{a\gamma'+b\gamma_2}\, \mathcal{S}(q) \overline{\mathcal{S}}(q)]\\
& =(q)_\infty^2 \, \text{Tr}
[ \, X_{a\gamma'+b\gamma_2} \,  E_q(X_{\gamma'} )E_q(X_{\gamma_3} )E_q(X_{\gamma_2})E_q(X_{\gamma'}^{-1})E_q(X_{\gamma_3}^{-1}) E_q(X_{\gamma_2} ^{-1}) \,]\\
&=(q)_\infty^2 \, \sum_{\substack{\ell_1,\ell_2,\ell_3,\\ k_1,k_2,k_3=0}}^\infty
{ (-1)^{a+b} q^{\ell_1+\ell_2+\ell_3 +\ell_2(\ell_1+\ell_3) +{1\over2} (a+b+ab) +a\ell_2  }\over (q)_{\ell_1} (q)_{\ell_2} (q)_{\ell_3} (q)_{k_1} (q)_{k_2} (q)_{k_3} }\,z^{2(l_1-k_1) +a} \delta_{k_2,\ell_2+b}\, \delta_{k_1+k_3 ,\ell_1+\ell_3+a}\,.
\end{split}
\end{align}
Following the conjecture \eqref{lineproposal2}, together with \eqref{generatingA3} and \eqref{X}, we obtain the line defect Schur indices for $A_i$ and $B_i$,
\begin{align}
\mathcal{I}_{A_i}(q,z)&=(q)_\infty^2\text{Tr}[F(A_i)\mathcal{S}(q)\overline{\mathcal{S}}(q)]\notag\\
&=  -q^{{1\over2}} \left[\, \chi_{\bf 2}+ \chi_{\bf 4}\,q + \left(  \chi_{\bf 2}+ \chi_{\bf 4}+ \chi_{\bf 6}\right)q^2 +\left( 2 \chi_{\bf 2}+2 \chi_{\bf 4}+\chi_{\bf 6}+ \chi_{\bf 8} \right)q^3 
\right.\notag  \\
&\left. +\left(3\chi_{\bf2}+ 3\chi_{\bf4}+ 3\chi_{\bf6} +\chi_{\bf8}+\chi_{\bf10}\right)q^4 + \cdots\right]\,,
~~~~~~~~~~~~~~~~~\text{for all}~~~i=1,2,3\,,\\
\mathcal{I}_{B_i}(q,z)&=(q)_\infty^2\text{Tr}[F(B_i)\mathcal{S}(q)\overline{\mathcal{S}}(q)]  \notag\\
&  =  -q^{{1\over2}} \left[\,\chi_{\bf 1}+ \chi_{\bf 3}\,q^2 + \left(  \chi_{\bf 1}+ \chi_{\bf 3}\right)q^3 +\left(  \chi_{\bf 1}+ \chi_{\bf 3}+\chi_{\bf 5} \right)q^4 
+\left(\chi_{\bf1}+ 2\chi_{\bf3}+\chi_{\bf5}\right)q^5 + \cdots\right]\,,\notag\\
&~~~~~~~~~~~~~~~~~~~~~~~~~~~~~~~~~~~~~~~~~~~~~~~~~~~~~~~~~~~~~~~~~~~~~~~~~~\text{for all}~~~i=1,2,3\,,
\end{align}
where $\chi_{\mathbf{n}}$ is the  character  for the $n$-dimensional irreducible representation of  $SU(2)$, normalized such that $\chi_\mathbf{2} = z+ z^{-1}$. 
As in the $A_2$ Argyres-Douglas theory, the dependence on the index $i$ is washed out inside the trace. We can therefore define $\mathcal{I}_A$ and $\mathcal{I}_B$ unambiguously as
\begin{align*}
&\mathcal{I}_A \equiv\mathcal{I}_{A_1}= \mathcal{I}_{A_2} =\mathcal{I}_{A_3}\,,\\
&\mathcal{I}_B \equiv\mathcal{I}_{B_1}= \mathcal{I}_{B_2} =\mathcal{I}_{B_3}\,.
\end{align*}

We observe the following  relations between the line defect indices (no sum in the indices),
\begin{align}\label{LLA3}
\begin{split}
&\mathcal{I}_{A_i A_i} =  \mathcal{I}_{A_j A_{j+1} } = \mathcal{I} +q^{-{1\over2}}\mathcal{I}_B\,,\\
&\mathcal{I}_{ B_iB_i } =  \mathcal{I}_{ B_j B_{j+1}} = \mathcal{I}+q^{-{1\over2}} (z+z^{-1})\, \mathcal{I}_A+q^{-1} \mathcal{I}_{A_kA_k }\,,\\
&q\mathcal{I}_{ A_i B_i } = \mathcal{I}_{ A_j B_{j+2}} = \mathcal{I}_{ A_k B_{k+1} } =  (z+z^{-1})\,\mathcal{I}+(q^{1\over2}+q^{-{1\over2}} ) \,\mathcal{I}_A\,,
\end{split}
\end{align}
which hold true for all values of $i,j,k=1,2,3$. Note that the rightmost equalities in the above relations are implied by the UV line defect OPE \eqref{A3OPE}. 

\subsection{$A_4$ Argyres-Douglas Theory}\label{sec:A2}

\begin{figure}[h]
\begin{center}
\includegraphics[width=.5\textwidth]{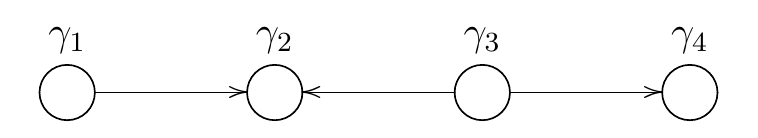}
\end{center}
\caption{ The BPS quiver for the $A_4$ Argyres-Douglas theory.}\label{fig:A4}
\end{figure}

For the $A_4$ Argyres-Douglas theory, we present a similar (but much more tedious) derivation of the core charges of the line defects in Appendix \ref{sec:A4Cones} again following the method of \cite{Cordova:2013bza}.  The result is that there are fourteen generators $A_i$, $B_i$ $(i=1,\cdots ,7)$ together with the unit operator for the defects.    Assuming the defect phase $\vartheta$ is chosen to lie to the right of all ordinary BPS states, the generating functions for the  line defects are,
\begin{align}\label{A4generating}
&F(A_1) = X_{-\gamma_2+\gamma_4}\,,\notag\\
&F(A_2) = X_{\gamma_2-\gamma_4} +X_{\gamma_1+\gamma_2-\gamma_4}\,,\notag\\
&F(A_3)=X_{-\gamma _1-\gamma _2+\gamma _4}+ X_{-\gamma _1+\gamma _4}+X_{-\gamma _1+\gamma _3+\gamma _4}\,,\notag\\
&F(A_4)=X_{\gamma _1-\gamma _3-\gamma _4}+ X_{\gamma _1-\gamma _3} \,,\notag\\
&F(A_5)=  X_{-\gamma_1+\gamma_3}\,,\notag\\
&F(A_6)=X_{\gamma _1-\gamma _3}+X_{\gamma _1-\gamma _3+\gamma _4}  +X_{\gamma _1+\gamma _4}\,,\notag\\
&F(A_7)=X_{-\gamma _1-\gamma _4}+X_{-\gamma _1+\gamma _2-\gamma _4}+X_{\gamma _2-\gamma _4}\,,\notag\\
&F(B_1)=X_{-\gamma _1}+X_{\gamma _2}+X_{\gamma _2-\gamma _1}+X_{\gamma _2+\gamma _3}+X_{-\gamma _1+\gamma _2+\gamma _3}\,,\\
&F(B_2)=X_{-\gamma _3-\gamma _4}+X_{\gamma _2}+X_{\gamma _1+\gamma _2}+X_{\gamma _2-\gamma _3}+X_{\gamma _1+\gamma _2-\gamma _3}+X_{-\gamma _3}+X_{\gamma _2-\gamma _3-\gamma _4}+X_{\gamma _1+\gamma _2-\gamma _3-\gamma _4}\,,\notag\\
&F(B_3)=X_{-\gamma _2-\gamma _3}+X_{-\gamma _3}+X_{\gamma _4}+X_{\gamma _4-\gamma _3}+X_{-\gamma _2-\gamma _3+\gamma _4}\,,\notag\\
&F(B_4)=X_{-\gamma _1-\gamma _2}X_{-\gamma _1}\,,\notag\\
&F(B_5)= X_{-\gamma_4}\,,\notag\\
&F(B_6) = X_{\gamma_1} \,,\notag\\
&F(B_7) = X_{\gamma_4 } +X_{\gamma_3+\gamma_4}\,\notag.
\end{align}
The fourteen generators for the line defects  satisfy the following defect OPE,
\begin{align}\label{A4OPE}
\begin{split}
&A_i A_{i+1}=1+ q^{1\over2} B_{2 i+4}\,,  \\
&B_i B_{i+2}=1+q^{1\over2} A_{4 i-1} B_{i+1}\,,\\
&B_i B_{i+3}=A_{4 i-4} A_{4 i-1}+A_{4 i+1}\,,\\
&A_i B_{2 i}=q^{1\over2} A_{i-2}+B_{2 i+1}\,,\\
&A_i B_{2 i-1}=q^{-{1\over2}}A_{i+2}+B_{2 i-2}\,.
\end{split}
\end{align}
 We have taken the index $i$ to be periodic mod 7. 

The Schur indices  with  insertions of $A_i$ and $B_i$  computed from our IR formula \eqref{lineproposal2}  are,
\begin{align}
\mathcal{I}_{A_i}(q)&=
 (q)^4_\infty \,\text{Tr}\left[ \,  F(A_i) \mathcal{S}(q)\overline{\mathcal{S}}(q)\,   \right]\notag\\
&= q +q^4 +q^5 +q^6 + 2q^7 +2q^8  +3q^9 +4q^{11} +5q^{12} +7q^{13}+8q^{14}+\mathcal{O}(q^{15})\,,\notag\\
&~~~~~~~~~~~~~~~~~~~~~~~~~~~~~~~~~~~~~~~~~~~~~~~~~~~~~~~~~~~~~~~~~~~~~~~
\text{for all} ~~i=1,\cdots,7\,,\\
\mathcal{I}_{B_i}(q)&=
 (q)^4_\infty \,\text{Tr}\left[ \,  F(B_i)  \mathcal{S}(q)\overline{\mathcal{S}}(q)\,   \right]\notag\\
&= -q^{1\over2}(1 +q^2 +q^3 +q^4 + 2q^5 +3q^6  +3q^7 +4q^8 +5q^{9} +7q^{10}+8q^{11}+11q^{12}+\mathcal{O}(q^{13}))\,,\notag\\
&~~~~~~~~~~~~~~~~~~~~~~~~~~~~~~~~~~~~~~~~~~~~~~~~~~~~~~~~~~~~~~~~~~~~~~~
\text{for all} ~~i=1,\cdots,7\,,
\end{align}
where for the $A_4$ Argyres-Douglas theory 
\begin{align}
 \mathcal{S}(q)\overline{\mathcal{S}}(q)=E_q(X_{\gamma_1}) E_q(X_{\gamma_2})  E_q(X_{\gamma_3})E_q(X_{\gamma_4 })  E_q(X_{\gamma_1}^{-1})  E_q(X_{\gamma_2}^{-1})E_q(X_{\gamma_3}^{-1})E_q(X_{\gamma_4}^{-1})\,.
\end{align} 
Notice that the dependence on the index $i=1,\cdots,7$  is washed out inside the trace, reflecting the $\mathbb{Z}_7$ symmetry of the $A_4$ Argyres-Douglas theory. We can therefore define $\mathcal{I}_ A(q)$ and $\mathcal{I}_ B (q)$ unambiguously as $(q)_\infty^6\text{Tr}\left[ \,  F(A_i)  \mathcal{S}(q)\overline{\mathcal{S}}(q)\,   \right]$ and $(q)_\infty^6\text{Tr}\left[ \,  F(B_i)  \mathcal{S}(q)\overline{\mathcal{S}}(q)\,   \right]$ for any choice of $i$, respectively.

\section{Chiral Algebra and Line Defects}\label{sec:verlinde}

To every 4$d$ $\mathcal{N}=2$ superconformal field theory, we can associate a 2$d$  chiral algebra a la the work of  \cite{Beem:2013sza}. The states in the vacuum module of the 2$d$ chiral algebra are in one-to-one correspondence with the protected operators in  the 4$d$ theory that contribute to the Schur index.  It is then natural to ask whether one has access to the states in the other modules of the chiral algebra from the 4$d$ physics.

In Appendix \ref{sec:supercharge}, we show that when the line defects are extended on a plane, say the 12-plane, transverse to the  chiral algebra plane, say, the 34-plane, the combined system preserves two supercharges \eqref{fancyQ}.  The incidence geometry of the line defects and the chiral algebra plane is shown in Figure \ref{fig:IncidenceGeometry2}.  Given that the line defects share some common supercharges with the chiral algebra plane, it is tempting to speculate that the defect operators are related to the states in the other modules of the chiral algebra.  

\begin{figure}[h!]
\centering
\includegraphics[width=.5\textwidth]{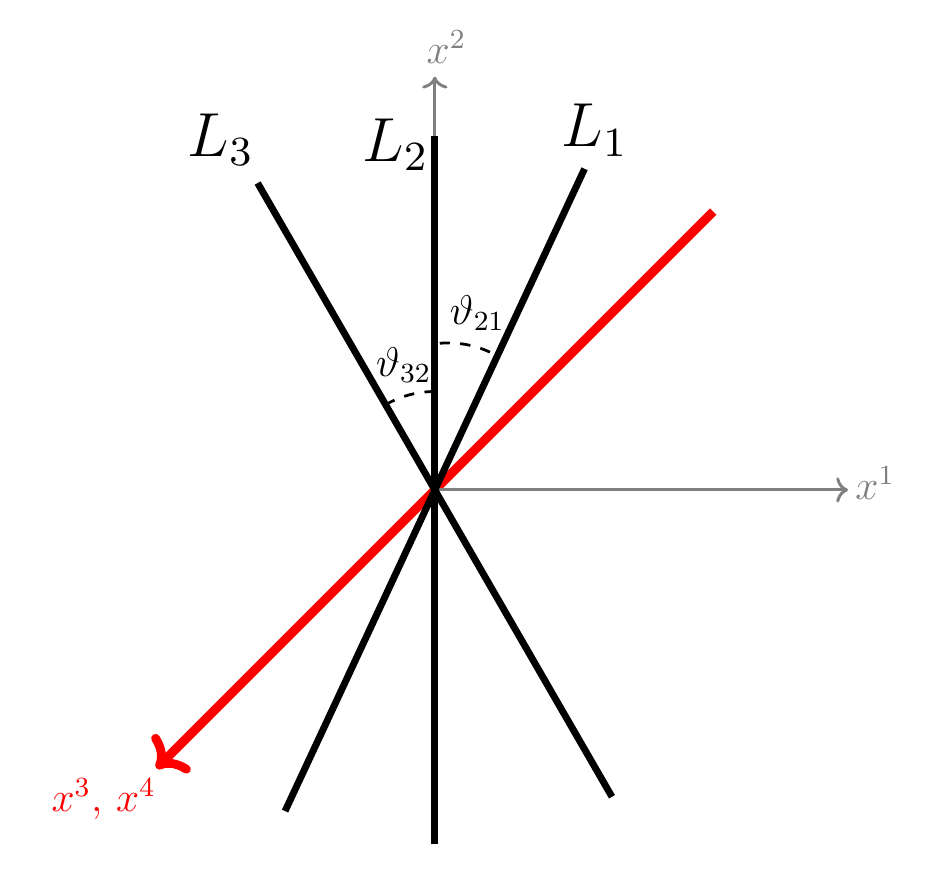}
\caption{The incidence geometry of  line defects and the chiral algebra plane.  We suppress one dimension of the chiral algebra plane (the 34-plane) and represent it as the red line above. The black lines are the line defects lying on the 12-plane. The line defects are oriented on the 12-plane by their central charge phases $\vartheta_i$.  Here $\vartheta_{ij} = \vartheta_i -\vartheta_j$.}\label{fig:IncidenceGeometry2}
\end{figure}

In this section we demonstrate in several examples, including the Argyres-Douglas theories and $SU(2)$ gauge theory with $N_f=4$ flavors, that the line defect indices can indeed be written as linear combinations of the characters for the other modules in the chiral algebra. This should not come as a surprise in the case of Lagrangian theories. Indeed, in Section \ref{sec:Nf4chiralalgebra} we  show explicitly how the defect Schur operators at the zero coupling point of $SU(2)$ SQCD organize themselves into modules of the associated chiral algebra.  However a general abstract derivation which holds also for non-Lagrangian theories is still lacking and is an open problem for future research.  It would also be interesting to explore the modular properties of the resulting sums of characters such as \eqref{doubletwilson} and \eqref{adsumeq} as in \cite{Razamat:2012uv, arakawa2015joseph, arakawa2016, BR}.

In the case of Argyres-Douglas theories, we go further and describe how the fusion rule, or the Verlinde algebra, of the 2$d$ chiral algebra can be realized from line defect indices in 4$d$ in the $q\rightarrow 1$ limit. We will start with a general proposal in Section \ref{sec:generalverlinde} and demonstrate it with the examples of the $A_2$,  $A_3$, and $A_4$ Argyres-Douglas theories.  This connection between the 2$d$ Verlinde algebra and the 4$d$ defect indices was first observed in \cite{Cecotti:2010fi}.

\subsection{$SU(2)$ Gauge Theory with $N_f=4$ Flavors}\label{sec:Nf4chiralalgebra}

The chiral algebra associated to the 4$d$ $SU(2)$ gauge theory with $N_f=4$ flavors is $\widehat{so(8)}_{-2}$ \cite{Beem:2013sza}. 
The Schur index without any insertion of defect is reproduced by the vacuum character of  $\widehat{so(8)}_{-2}$.  It is natural to speculate that the defect indices discussed in Section \ref{sec:Nf4Line} can be related to the other characters of $\widehat{so(8)}_{-2}$. Indeed, we observe that the  line defect index for a half Wilson in the $\mathbf{2}$ can be written as the following linear combinations of the characters for $\widehat{so(8)}_{-2}$,
\begin{align}\label{doubletwilson}
\begin{split}
\mathcal{I}_{L_{0,1}}(q,\eta_i=1)& = \sum_{k=1}^\infty (-1)^{k} q^{ k^2+k-1\over2 } (1-q^k) \,\chi_{[-2k-1, 2k-1,0,0,0]}(q,\eta_i=1)\,
\end{split}
\end{align}
where $ \chi_{[a_0,a_1,a_2,a_3,a_4]}(q,\eta_i)$ is the affine character\footnote{This is not to be confused with the character $\chi_{[a_1,a_2,a_3,a_4]}(\eta_i)$ of the finite $SO(8)$, which is labeled by four (instead of five) Dynkin labels.} of $\widehat{so(8)}_{-2}$ with affine Dynkin labels $[a_0,a_1,a_2,a_3,a_4]$.  We have normalized the $\widehat{so(8)}_{-2}$ affine characters to start from 1.  We present the details of the calculation for the affine characters of $\widehat{so(8)}_{-2}$ in Appendix \ref{sec:AffineCharacter}. The $SO(8)$ flavor fugacities have been set  to  1 for simplicity and we have checked the above relation to $\mathcal{O}(q^{19\over2})$.  

The vacuum representation of $\widehat{so(8)}_{-2}$ has certain null states that can be translated into null relations among the Kac-Moody currents $J^{[ij]}(z)$. It is perhaps more natural to consider representations that respect these null relations.  Among all the highest weight irreducible representations of $\widehat{so(8)}_{-2}$, there are four of them, in addition to the vacuum representation, that respect these null relations \cite{arakawa2015joseph}. Their highest weights are $[0,-2,0,0,0]$, $[0,0,-1,0,0]$, $[0,0,0,-2,0]$, and $[0,0,0,0,-2]$. We do not know whether the line defect indices can be decomposed into sum of these four characters together with the vacuum character. It would be very interesting to pursue this direction further.

The appearance of  non-vacuum characters in the Wilson line indices of Lagrangian theories is not a surprise.  Indeed, by going to the free theory point on the moduli space, it is easy to see why the line defect Schur operators organize themselves into modules of $\widehat{so(8)}_{-2}$.  Let us denote the holomorphic operators in the chiral algebra by the same symbols as their 4$d$ Schur operators.  In the case of $SU(2)$ gauge theory with $N_f=4$ flavors, we have a fermionic current  $\rho^A_\pm(z)$ with dimension 1 from the vector multiplet, and a bosonic current $H^{ia}(z)$ with dimension $\frac12$ from the hypermultiplets.  Here $a$, $A$, and $i$ are the indices for the $SU(2)$ doublet, $SU(2)$ triplet, and the $\mathbf{8}_v$ of $SO(8)$, respectively.  The dimensions of $H^{ia}$ and $\rho^{\alpha A}$ are determined by their 4$d$ quantum numbers $\Delta-R$, the exponent of $q$ in the Schur index. Their  2$d$ OPEs can be obtained from the 4$d$ OPEs following the construction of the chiral algebra in \cite{Beem:2013sza},
\begin{align}
\begin{split}\label{Nf4OPE}
&\rho^{\alpha A} (z) \rho^{\beta B} (0)\sim {\delta^{AB} \epsilon^{\alpha\beta} \over z^2}\,,\\
&H^{ia} (z) H^{jb}(0) \sim {\delta^{ij} \epsilon^{ab}\over z}\,,\\
&H^{ia} (z) \rho^{\beta B}(0)\sim 0\,,
\end{split}
\end{align}
where $\epsilon^{\alpha \beta}$ is the two-by-two antisymmetric tensor normalized such that $\epsilon^{+-}=+1$.  We will also use $\epsilon_{\alpha\beta}$ which is the inverse of $\epsilon^{\alpha \beta}$, i.e. $\epsilon_{+-}=-1$.

We can then construct the dimension 1 $\widehat{so(8)}_{-2}$ currents $J^{[ij]} (z)$,
\begin{align}
J^{[ij]} (z)  =  \mathcal{N}\epsilon_{ab} \,:H^{ia} H^{jb}:(z)\,,
\end{align}
where $\mathcal{N}$ is a normalization constant.  Using \eqref{Nf4OPE}, one can show that $J^{[ij]}(z)$ satisfies the $\widehat{so(8)}$ current algebra at level $-2$.  

The half Wilson line index $\mathcal{I}_{L_{0,1}}$ counts ``gauge non-invariant" Schur operators in the doublets of $SU(2)$.  As discussed in Section \ref{sec:Nf4Line}, we can enumerate these operators at the free theory point, which are normal-ordered operators  in the doublets made out of $H^{ia}$ and $\rho^{\alpha A}$.  Now since the $\widehat{so(8)}_{-2}$ current $J^{[ij]}(z)$  is a singlet under $SU(2)$, any descendant $J_{-n_1}^{[i_1j_1]}\cdots J_{-n_k}^{[i_kj_k]}\cdot \mathcal{O}^a(z)$ of a doublet $\mathcal{O}^a(z)$ is still a doublet made out of $H^{ia}$ and $\rho^{\alpha A}$. Hence if  $\mathcal{O}^a(z)$ is counted by the  line defect index, so are all its current algebra descendants.  It follows that the line defect Schur operators fall into modules of the chiral algebra $\widehat{so(8)}_{-2}$.

We can see explicitly how this works for the low dimension operators.  At dimension $1/2$, the only line defect Schur operator is $H^{ia}(z)$. It is a current algebra primary since it is annihilated by all the $J^{[ij]}_{+n}$ with $n>0$.

At dimension $3/2$,  the Schur operators are $\partial H^{ia}$, $(\rho^{\pm A} H^{ib})(T^A)^a_{~b}$, $(H^3)_{\mathbf{8}_v}$, and $(H^3)_{\mathbf{160}_v}$ listed in \eqref{level32}. Here $(H^3)_\mathbf{R}$ denotes the part of the normal-ordered operator\footnote{The normal-ordered product of more than two operators are defined  recursively taking normal ordering from the right. For example, $:\mathcal{O}_1\mathcal{O}_2\mathcal{O}_3:\, \equiv\, :\mathcal{O}_1 : \mathcal{O}_2 \mathcal{O}_3 ::$\,.  Generally  the ordering of operators in the normal-ordered product matters. However, in our case since the coefficients in the OPE \eqref{Nf4OPE} are all proportional to the identity operator, we can commute the operators in the normal-ordered product freely. For example, $:H^{ia}H^{jb} : = :H^{jb} H^{ia}:$ and $:H^{ia}H^{jb}H^{kc}: = :H^{jb}H^{ia}H^{kc}: =:H^{ia}H^{kc}H^{jb}:$. It is because of these symmetries that the $SU(2)$ doublet of the normal-ordered product $:H^{ia}H^{jb}H^{kc}:$ transform as $\mathbf{8}_v\oplus\mathbf{160}_v$ under $SO(8)$.} $:H^{ia}H^{jb}H^{kc}:$ that transforms in the representation $\mathbf{R}$ of $SO(8)$ and  in the doublet of $SU(2)$. Let us organize these dimension $3/2$ operators into primaries and descendants of the current algebra. The level 1 current algebra descendant of the dimension $1/2$ primary $H^{ia}(z)$ is
\begin{align}
\begin{split}
J^{[ij]}_{-1} \cdot H^{ka}(0) &= \oint {dz\over 2\pi i}  {1\over z}\, J^{[ij]}(z) H^{ka}(0) 
\\
&=\mathcal{N}\left[ \, - \delta^{ik} \, \partial H^{ja}(0)  + \delta^{jk} \, \partial H^{ia}(0)
+\epsilon_{bc}   :H^{ib} H^{jc} H^{ka}:(0)\,\right]\,.
\end{split}
\end{align}
This descendant, when decomposed into irreducible representations, contains $(H^3)_{\mathbf{160}_v}$ and a particular linear combination of  $(H^3)_{\mathbf{8}_v}$ and $ \partial H^{ia}$:
\begin{align}
-  \epsilon_{bc}   :H^{ja} H^{jb} H^{ic} + 7 \partial H^{ia}\,.
\end{align}

After having determined the descendants, we need to identify the primaries at dimension $3/2$.  To begin with, it is easy to see that $(\rho^{\pm A} H^{ib})(T^A)^a_{~b}$ is annihilated  by all the $J^{[ij]}_{+n}$ with $n>0$, and is hence a primary of the current algebra.  Next, one can straightforwardly check that the following combination
\begin{align}
\epsilon_{bc} :H^{j a}H^{j b} H^{ic}: (0) -3\partial H^{ia} (0) 
\end{align}
is a current algebra primary.


 In summary, we have organized the dimension $1/2$ and $3/2$ Wilson line defect Schur operators into current algebra descendants and primaries,
\begin{align}
\left.\begin{array}{|c|c|c|c|}
\hline \text{Dimension} & \text{Chiral Algebra Operator} & \widehat{so(8)}_{-2}&$SO(8)$ \text{ Rep} \\\hline 1/2 & H^{ia} & \text{Primary}& \mathbf{8}_v \\\hline 3/2 &\epsilon_{bc} H^{j a}H^{j b} H^{ic}  -3\partial H^{ia} 
& \text{Primary} & \mathbf{8}_v\\\hline 3/2  & (\rho^{\pm A} H^{ib})(T^A)^a_{~b} & \text{Primary} & \mathbf{8}_v\\\hline 3/2  &-  \epsilon_{bc}   H^{ja} H^{jb} H^{ic} + 7 \partial H^{ia}& \text{Descendant} & \mathbf{8}_v\\\hline 3/2  & (H^3)_{\mathbf{160}_v} & \text{Descendant}& \mathbf{160}_v \\\hline \end{array}\right.
\end{align}

Returning to the doublet half Wilson line index \eqref{doubletwilson}, we have, up to $q^{3/2}$,
\begin{align}
\mathcal{I}_{L_{0,1}}  =-  q^{1\over2} \chi_{\mathbf{8}_v}+q^{3\over2} \chi_{\mathbf{8}_v}  +\mathcal{O}(q^{5\over2})\,.
\end{align}
Recall that $\chi_{\mathbf{8}_v}  = 8 +168 q+ \cdots$.  It is now clear that the first term in $\mathcal{I}_{L_{0,1}}$ is the contribution from the current algebra primary $H^{ia}$ and its descendants, while the second term is the contribution from the bosonic   primary $\epsilon_{bc} H^{j a}H^{j b} H^{ic}  -3\partial H^{ia}$ and the two fermionic primaries $ (\rho^{\pm A} H^{ib})(T^A)^a_{~b}$ as well as their descendants.  This analysis gives a direct understanding of why the non-vacuum characters of the chiral algebra appears in the line defect indices in the case of Lagrangian theories.

Similarly, the other half Wilson line defect indices are also related to the $\widehat{so(8)}_{-2}$ affine characters, 
\begin{align}
\begin{split}
\mathcal{I}_{L_{0,2}}(q,\eta_i=1)&=- q  \, \chi_{[-2,0,0,0,0]} 
+(1-q)\sum_{k=1}^\infty (-1)^{k+1} q^{k(k+1)\over2}(1-q^{2k})\, \chi_{[-2k-2,2k,0,0,0]}\,,\\
\mathcal{I}_{L_{0,3}}(q,\eta_i=1)& =   q^{3\over 2} (1-q^2 )   \chi_{[-3,1,0,0,0]} -q^{3\over 2} (1-q+q^3-q^6 ) \chi_{[-5,3,0,0,0]}\\
& +q^{7\over 2} (1-q^2+q^5) \chi_{[-7,5,0,0,0]}- q^{13\over2} (1-q^3)  \chi_{[-9,7,0,0,0]} +q^{21\over2} \chi_{[-11,9,0,0,0]}+ \mathcal{O}(q^{25\over2})\,,\\
\mathcal{I}_{L_{0,4}}(q,\eta_i=1) &=  q^3 \chi_{[-2,0,0,0,0]}  -q^2(1-q^2-q^3 +q^5) \chi_{[-4,2,0,0,0]}\\
&
+q^2(1-q+q^5 -q^6 -q^8 +q^{10} )\chi_{[-6,4,0,0,0]}  -q^4(1-2q^2 +q^3 +q^8) \chi_{[-6,4,0,0,0]} \\
&
+q^7 (1-q^2-q^3+q^4)  \chi_{[-10,8,0,0,0]} -q^{11} \chi_{[-12,10,0,0,0]} +\mathcal{O}(q^{13}) \,,
\end{split}
\end{align}
where $\mathcal{I}_{L_{0,n}}$ is the line defect index for a half Wilson line in the $(n+1)$-dimensional representation of $SU(2)$
 For the half Wilson line defect indices $\mathcal{I}_{L_{0,2}}$ in the $\mathbf{3}$, we conjectured the above exact relation and  checked it to $\mathcal{O}(q^{14})$. For the other two cases we recorded our observation above without closed form formulas.  Here we have normalized the $\widehat{so(8)}_{-2}$ characters to start from order $q^0$.

\subsection{Verlinde Algebra from Line Defects}\label{sec:generalverlinde}

In this subsection we give a precise proposal on how the fusion rules in the two-dimensional chiral algebra can be realized from the four-dimensional line defect indices in the case of Argyres-Douglas theories, generalizing the results of \cite{Cecotti:2010fi}.

In all  known chiral algebras for the Argyres-Douglas theories,\footnote{Throughout this paper, we will only consider Argyres-Douglas theories whose BPS quivers are $ADE$ Dynkin diagrams.  The chiral algebras of  more general Argyres-Douglas theories have been explored in \cite{Xie:2016evu}.} there is always a distinguished finite set of primaries whose characters form a modular vector. For example, the chiral algebra of the $A_{2n}$ Argyres-Douglas theory is the $(2,2n+3)$ Virasoro minimal model.  As another example that we will encounter in this section, the chiral algebra for the $A_3$ Argyres-Douglas theory is the affine Lie algebra $\widehat{su(2)}_{-\frac43}$. There are three distinguished representations of $\widehat{su(2)}_{-\frac43}$, called the admissible representation, whose characters are linearly mapped to each other under the modular transformation.  We will focus on these primaries $\Phi_\alpha$  with this nice modular property.  $\Phi_0$ is chosen to be the identity. 

For the Argyres-Douglas theories, we claim that the  generators for the UV electromagnetic line defects can be labeled by the non-identity primaries $\Phi_{\alpha\neq0 }$ with a (finite) degeneracy labeled by $i$, 
\begin{align}\label{Lalphai}
L_{\alpha i}\,.
\end{align}
In particular, this implies that the number of the generators for the electromagnetic line defects is an \textit{integer multiple} of the number of non-identity  primaries in the chiral algebra.  We will test this claim explicitly in the $A_2,A_3,A_4$ Argyres-Douglas theories. 

In both the Argyres-Douglas theories and in the $SU(2)$ superconformal QCD, we observe  that  the  line defect indices $\mathcal{I}_{L_{\alpha i}}$ are related to the  characters $\chi_\alpha(q,z)$ of the chiral algebra by 
\begin{align}
\mathcal{I}_{L_{\alpha i}} (q,z)= \sum_{\beta\in \text{modules}} v^{\beta}_{\alpha i} ( q,z) \, \chi_\beta(q,z)\,,
\end{align} 
where $z$  denotes collectively all the flavor fugacities and $v^{\beta}_{\alpha i} (q,z)$ are some  polynomials in $q$ and $z$.  The sum in the modules above is finite for the Argyres-Douglas theory,  but infinite for $SU(2)$ superconformal QCD.  Similarly we observe that  the  line defect index with the insertion of two half lines $L_{\alpha i } ,L_{\beta i}$ (in the phase order, say, $\vartheta_{\alpha i}<\vartheta_{\beta j}$) can be decomposed into  characters of the chiral algebra,
\begin{align}
\mathcal{I}_{ L_{\alpha i } L_{\beta j }} (q,z)= \sum_{\gamma\in\text{modules}} v^{\gamma}_{\alpha i,\beta j} (q,z) \, \chi_\gamma(q,z)\,,
\end{align}
with some   polynomials $v^{\gamma}_{\alpha i,\beta j} (q,z)$. Note that $v^{\gamma}_{\alpha i,0} (q,z)=v^{\gamma}_{0,\alpha i} (q,z)=v^\gamma_{\alpha i}(q,z)$ by definition.

In all the Argyres-Douglas theories we have investigated, we find that when setting $q=z=1$, the coefficients $v^{\gamma}_{\alpha i }( 1,1)$ and $v^{\gamma} _{\alpha i,\beta j} (1,1)$ are independent of the degeneracy index $i$, while they still  depend on the index $\alpha$, which labels the primaries in the 2$d$ chiral algebra.\footnote{In fact, the polynomials $v^{\gamma}_{\alpha i}(q,z)$ with general $q,z$ are also independent of $i$ in all the examples we have tested. However, $v^{\gamma}_{\alpha i,\beta j} (q,z)$ \textit{does} depend on $i,j$ in a nontrivial way if $q\neq1$ or $z\neq1$.} We can therefore define
\begin{align}
\begin{split}
&V^\gamma_\alpha \equiv v^{\gamma}_{\alpha i}( q=1,z=1)\,,\\
&V^\gamma_{\alpha\beta } \equiv v^{\gamma}_{\alpha i,\beta j}( q=1,z=1)\,.
\end{split}
\end{align}
 Note that $V^{\gamma}_{\alpha 0}=V^{\delta}_{0\alpha }=V^\gamma_\alpha$ by definition. We also observe that $V^\gamma_{\alpha\beta}$ is symmetric in $\alpha$ and $\beta$, whereas $v^{\gamma}_{\alpha i,\beta j} ( q,z)$ is not in general.  In the following subsections we will determine the coefficients $V^\beta_\alpha(q,z)$ and $V^\gamma_{\alpha\beta}(q,z)$ for the $A_2,A_3,A_4$ Argyres-Douglas theories. 
 
Our main observation is that in the Argyres-Douglas theories, where the index $\alpha$ runs over finitely many modules, the coefficients $V^\beta_\alpha$ and $V^\gamma_{\alpha \beta}$ obey the Verlinde algebra of the associated chiral algebra,
\begin{align}\label{verlinde}
\boxed{
\,V^\gamma_{\alpha\beta}=\sum_{\alpha',\beta'\in \mathrm{modules}} \, \mathcal{N}_{\alpha' \beta'}^{\,\gamma} \,V^{\alpha'}_\alpha V^{\beta'}_\beta\,,}
\end{align}
where $\mathcal{N}_{\alpha' \beta'}^{\,\gamma} $ are the fusion coefficients of the 2$d$ chiral algebra. Importantly, we will \textit{define} the fusion coefficients $\mathcal{N}^{\,\gamma}_{\alpha \beta}$ by the Verlinde formula  \cite{Verlinde:1988sn}
\begin{align}\label{VF}
\mathcal{N}^{\,\gamma}_{\alpha\beta}  = \sum_{\delta\in \mathrm{modules}}{ \mathcal{S}^\delta_{~\alpha} \mathcal{S}^\delta_{~\beta} \mathcal{\bar S}^{\delta \gamma} \over \mathcal{S}^\delta_{~0}}\,,
\end{align}
where $\mathcal{S}$ is the modular transformation matrix and $\mathcal{\bar S}^{\delta \gamma}  = \mathcal{S}^\delta_{~\beta}\mathcal{C}^{\beta\gamma } $ with $\mathcal{C} = \mathcal{S}^2$. 
This definition circumvents certain subtleties in the fusion rules in non-rational CFTs.  For chiral algebra whose $\mathcal{S}$ matrix is not well-defined (for example, the $\widehat{so(8)}_{-2}$ in the $SU(2)$ with $N_f=4$ flavors theory), we will not attempt to define the fusion coefficients and we do not have a proposal for the relation between the 4$d$ line defect product and the 2$d$ fusion rule.

We can phrase the above relation in a more illuminating way. For each class of 4$d$ line defects labeled by $\alpha$, $\{L_{\alpha i}\}_i$, we associate it to an element $[L_\alpha]$ of the Verlinde algebra using the coefficients $V^\beta_\alpha$,
\begin{align}
[ L_\alpha] \equiv \sum_{\beta \in\text{modules}}  V^\beta_\alpha \, [\Phi_\beta]\,.
\end{align}
Similarly we define $[L_\alpha L_\beta]$ as
\begin{align}
[L_\alpha L_\beta] \equiv  \sum_{\gamma \in\text{modules}} V^\gamma_{\alpha\beta} \, [\Phi_\gamma]\,.
\end{align}
Now our main observation \eqref{verlinde} can be written as
\begin{align}\label{verlinde2}
\boxed{
\,[L_\alpha L_\beta ] =  [L_\alpha ]\times [L_\beta]\,,
}
\end{align}
where the product $\times$ is the fusion product in the 2$d$ Verlinde algebra. A similar proposal was made and verified in various examples in \cite{Cecotti:2010fi}.

\subsubsection{$A_2$ Argyres-Douglas Theory}

The  chiral algebra associated to the 4$d$ $A_2$ Argyres-Douglas theory is the $(2,5)$ Virasoro minimal model \cite{Beem:2013sza, Beem:2014zpa,Cordova:2015nma}.  The primaries of the $(2,5)$ minimal model are the identity $\Phi_0=1$ and a non-identity primary $\Phi_1\equiv \Phi_{1,2}$ with weight $-1/5$, hence the index $\alpha$ in the previous section runs over $0,1$. On the other hand there are five non-unity generators $L_i$ for the UV line defects with the same line defect indices, hence the degeneracy labeled by $i$ in the previous section is five here, $i=1,\cdots,5$. 

Recall that the Schur index  without any insertion of line defects equals to the vacuum character of the (2,5) Virasoro minimal model \cite{Cordova:2015nma},
\begin{align}
\mathcal{I}(q) = \chi_{0}(q)\,.
\end{align}
We find that the line defect index\footnote{As shown in Section \ref{sec:LineAD}, the five line defect indices $\mathcal{I}_{L_i}(q)$ are all equal and will be denoted simply as $\mathcal{I}_L(q)$.} $\mathcal{I}_{L}(q)$ is related to the character for the  non-identity primary $\Phi_1$ with weight $h_{1,2}=-1/5$ in the following way
\begin{align}
\label{adsumeq}
\mathcal{I}_L (q)=  q^{-{1\over2}} \chi_{0} (q) -  q^{-{1\over2}} \chi_{1} (q)\,.
\end{align}
We have normalized the characters such that they start from 1. 
For completeness, we record the two characters in the (2,5) minimal model character (see \eqref{minimalcharacter})
\begin{align}
\begin{split}
&\chi_0(q) =1 + q^2+q^3+q^4 +q^5 + 2q^6+ 2q^7 + 3q^8 +3q^9 + 4q^{10} +4q^{11} +\cdots\,,\\ 
&\chi_{1}(q)
=1+q+q^2 +q^3+2q^4 +2q^5 +3q^6 +3q^7 +4q^8+5q^9+6q^{10}+\cdots\,.
\end{split}
\end{align}
 A similar observation was made in \cite{Cecotti:2010fi} (see, in particular, (9.51)) in the case of  the inverse of the quantum KS operator, but the precise linear combination of line defects are different. Incidentally, the characters for the vacuum $\Phi_0$ and the non-identity primary $\Phi_{1}$ in the $(2,5)$ minimal model are known to be the two Rogers-Ramanujan functions $H(q)$ and $G(q)$, respectively.

The coefficients $V^{\alpha}_\beta$ can be read off to be
\begin{align}
V^\alpha_0 = ( 1,0)\,,~~~~~~~~~~~V^\alpha_1 = (1,-1)\,.
\end{align}
The trace of two line defects are given in \eqref{LL}.  For example, 
\begin{align}
\mathcal{I}_{L_i L_{i+2}} = \, \mathcal{I} +q^{1\over2} \mathcal{I}_L 
=2 \chi_{0}(q) - \chi_{1}(q)\,.
\end{align}
One can easily verify that the coefficients $V^\gamma_{\alpha\beta}$ are indeed independent of the degeneracy index $i$ and are given by
\begin{align}
\begin{split}
&V^\alpha_ {11} = (2,-1)\,,
\end{split}
\end{align}
together with $V^\beta_{ \alpha 0}$ given by $V^\beta _{\alpha}$.

On the other hand, the Verlinde algebra for the $(2,5)$ minimal model is
\begin{align}
[\Phi_1 ] \times[ \Phi_1] =[ 1]+ [\Phi _1]\,.
\end{align}
One can easily check that \eqref{verlinde} is satisfied. Indeed, from the coefficients $V^\beta_\alpha$ and $V^\gamma_{\alpha\beta}$ we have
\begin{align}
[L] = [1] - [\Phi_1]\,,
\end{align}
and
\begin{align}
[LL] = 2 [1] -[\Phi_1]\,,
\end{align}
and the equivalent statement \eqref{verlinde2} is satisfied
\begin{align}
[LL] = [L]\times  [L]\,.
\end{align}


\subsubsection{$A_3$ Argyres-Douglas Theory}

The chiral algebra associated to  the $A_3$ Argyres-Douglas theory is the affine Lie algebra $\widehat{su(2)}_{- {4\over3}}$  \cite{Beem:2013sza, Beem:2014zpa,Buican:2015ina,Cordova:2015nma}.  The weights of $\widehat{su(2)}_k$ are labeled by its Dynkin labels $[\lambda_0,\lambda_1]$ with $\lambda_0+\lambda_1=k$. There are three representations in $\widehat{su(2)}_{-{4\over3}}$, whose highest weights are
\begin{align}\label{admissible}
\Phi_0 = [-{4\over3},0]\,,~~~~\Phi_1 = [-{2\over3} ,-{2\over3} ]\,,~~~~\Phi_2 = [0,-{4\over3}]\,.
\end{align}
 that are known to be  \textit{admissible} \cite{kac1988modular} (see also \cite{philippe1997conformal}). Admissible representations have the nice property that their characters transform linearly into each other under modular transformation, so the $\mathcal{S}$ matrix of modular transformation is well-defined. Note that the first highest weight  above $[-{4\over3} ,0]$ is that for the vacuum module, whose Dynkin label of the finite $SU(2)$ Lie algebra is zero, $\lambda_1=0$. The latter two representations are conjugate to each other.  The index $\alpha$ in \eqref{Lalphai} runs over $0,1,2$, which labels the primaries. As we saw in Section \ref{sec:A3}, there are six non-flavor line defects $L_{1i}\equiv A_i$ and $L_{2i}\equiv B_i$, with $i=1,2,3$.   
 
The characters for the three admissible representations can be computed using the Kazhdan-Lusztig formula as reviewed in Appendix \ref{sec:AffineCharacter} (they can also be found in Chapter 18 of \cite{philippe1997conformal}),
\begin{align}
\begin{split}
&\chi_{0}(q,z)={\sum_{m=0}^\infty (-1)^m \, 
{  z^{  2m+1}  -  z^{-(2m+1)} \over z-z^{-1} }  \,q^{  {3m(m+1)\over 2} }\over  \prod_{n=1}^\infty (1-q^n)(1-z^2 q^n) (1-z^{-2}q^n)}\,,\\
&\chi_{1 } (q,z)
={ 1+ \sum_{n=1}^\infty (-1)^n \left(z^{-2n} q^{ {n\over 2} (3n-1)}+z^{2n}q^{ {n\over 2} (3n+1)}\right) \over (1-z^{-2}) \prod_{n=1}^\infty (1-q^n )(1-z^2 q^n )(1-z^{-2}q^n)}\,,\\
&\chi_{2} (q,z)={ 1+ \sum_{n=1}^\infty (-1)^n \left(z^{2n} q^{ {n\over 2} (3n-1)}+z^{-2n}q^{ {n\over 2} (3n+1)}\right) \over (1-z^{-2}) \prod_{n=1}^\infty (1-q^n )(1-z^2 q^n )(1-z^{-2}q^n)}\,.
\end{split}
\end{align}
Note that for the latter two modules, there are infinitely many states at each grade created by the zero modes  $J^A_0$ of the Kac-Moody algebra, due to the fact that finite $SU(2)$ Dynkin labels $\lambda_1$ are negative fractional. Hence the two characters diverge as $1/(1-z^{-2})$ as $z\rightarrow 1$.  The vacuum character $\chi_{0 }(q,z)$ has been computed previously in \cite{Buican:2015ina,Cordova:2015nma}.

 We find that the line defect indices are related to the  characters of $\widehat{su(2)}_{-{4\over3}}$ as follows,
\begin{align}
\begin{split}
&\mathcal{I} (q,z)= \chi_{0  } (q,z)\,,\\
&\mathcal{I}_A (q,z)= q^{-{1\over2}} z^{-1} \, \left[\, - \chi_{1} (q,z)+ \chi_{2}(q,z) \, \right]\,,\\
&\mathcal{I}_B(q,z) = q^{-{1\over2}}  \, \left[\,\chi_{0}(q,z) - \chi_{1} (q,z)+ z^{-2}\chi_2(q,z) \, \right]\,,
\end{split}
\end{align}
where $\chi_\alpha(q,z)$ is the character for the primary $\Phi_\alpha$.  The first line is the  Schur index without any insertion of line defects $\mathcal{I}(q,z)$, which equals to the vacuum character $\chi_0(q,z)$ of $\widehat{su(2)}_{-{4\over3}}$ \cite{Beem:2013sza, Beem:2014zpa,Buican:2015ina,Cordova:2015nma}. 
Hence the coefficients $V^\beta_\alpha$ are 
\begin{align}
\begin{split}
&V^\alpha_0 = (1,0,0)\,,~~~~~V^\alpha_1 = (0,-1,1)\,,~~~~~V^\alpha_2 = (1,-1,1)\,.
\end{split}
\end{align}

The Schur indices for two (half) line defects are given in \eqref{LLA3}. We have (no sum in the indices)
\begin{align}
\begin{split}
&\mathcal{I}_{A_iA_i}(q,z)= (1+q^{-1}) \chi_{ 0} (q,z)-q^{-1}\, \chi_{1}(q,z) +q^{-1} z^{-2} \, \chi_{2}(q,z)\,,\\
&\mathcal{I}_{B_iB_i}(q,z)= (1+q^{-1}+q^{-2})\chi_{0}(q,z)-\left[ \, q^{-1}(1+z^{-2}) +q^{-2}\,\right]\chi_{1}(q,z)\\
&~~~~~~~~~ +\left[ \, q^{-1}(1+z^{-2}) +q^{-2}z^{-2}\,\right]\chi_{2}(q,z)\,,\\
&\mathcal{I}_{A_iB_i}(q,z)= (z+z^{-1}) \chi_{ 0}(q,z) -(1+q^{-1})z^{-1}\,
\chi_{1}(q,z) +(1+q^{-1})z^{-1}\,\chi_{ 2}(q,z)\,.
\end{split}
\end{align}
Hence the coefficients $V^\gamma_{\alpha\beta}$ are 
\begin{align}
\begin{split}
&V^\alpha_{11} = (2,-1,1)\,,\\
&V^\alpha_{22} = (3,-3,3)\,,\\
&V^\alpha_{12}=V^\alpha_{21}=(2,-2,2)\,,
\end{split}
\end{align}
together with $V^\beta_{\alpha 0} = V^\beta_{ \alpha}$. 

On the other hand, the $\mathcal{S}$ matrix for these three admissible representations is \cite{philippe1997conformal}
\begin{align}
\mathcal{S}^\alpha_{~\beta} 
=-{1\over \sqrt{3}} \left(\begin{array}{ccc}1 & -1 & 1 \\-1 & e^{4\pi i\over3} & -e^{2\pi i\over3} \\1 & -e^{2\pi i\over3} & e^{4\pi i\over3}\end{array}\right)\,.
\end{align} 
The conjugation matrix $\mathcal{C}=\mathcal{S}^2$ is given by
\begin{align}
\mathcal{C}^\alpha_{~\beta} = \left(\begin{array}{ccc}1 & 0 & 0 \\0 & 0 & -1 \\0 & -1 & 0\end{array}\right)\,.
\end{align}
Note that $\Phi_1=[-{2\over3} ,-{2\over3}]$ and $\Phi_2=[0,-{4\over3}]$ are conjugate to each other. 
The fusion rules obtained from the Verlinde formula \eqref{VF} are
\begin{align}\label{A3fusion}
\begin{split}
&[\Phi_1 ]\times [\Phi_1 ]=[\Phi_2]\,\\
&[\Phi_2] \times [\Phi_2]  =- [\Phi_1 ]\,\\
&[\Phi_1 ]\times [\Phi_2]= - [\Phi_0 ]\,.
\end{split}
\end{align}
Note that the minus sign in the fusion rule signals the negative central charge of the affine Lie algebra $\widehat{su(2)}_{-{4\over3}}$. 

 From the coefficients $V^\beta_\alpha$ and $V^\gamma_{\alpha\beta}$, we have
\begin{align}
\begin{split}
&[A] = - [\Phi_1]+[\Phi_2]\,\\
&[B] = [\Phi_0 ] - [\Phi_1 ] +[\Phi_2]\,,
\end{split}
\end{align}
and
\begin{align}
\begin{split}
&[AA] = 2 [\Phi_0 ] -[\Phi_1]  +[\Phi_2]\,,\\
&[BB]  = 3[\Phi_0 ] - 3[\Phi_1] +3 [\Phi_2]\,,\\
&[AB]  = 2[\Phi_0 ] - 2[\Phi_1 ] +2 [\Phi_2]\,.
\end{split}
\end{align}
One can check straightforwardly that our proposal \eqref{verlinde2} is satisfied, i.e. $[AA]=[A]\times [A]$, $[BB]=[B]\times [B]$, and $[AB] = [A]\times [B]$ using the fusion rules \eqref{A3fusion}. 


\subsubsection{$A_4$ Argyres-Douglas Theory}

The  chiral algebra associated to the 4$d$ $A_4$ Argyres-Douglas theory is the $(2,7)$ Virasoro minimal model \cite{Beem:2013sza, Beem:2014zpa,Cordova:2015nma}.  The primaries of the $(2,7)$ minimal model are the identity $\Phi_{1,1}=1$ and two non-identity primaries $\Phi_{1,2}$ and $\Phi_{1,3}$ with weight $-2/7$ and $-3/7$, respectively. On the other hand there are 14 non-unity generators for the UV line defects grouped into $A_i$ and $B_i$ with $i=1,\cdots,7$.  The line defect indices  are related to the characters of the $(2,7)$ Virasoro minimal model by
\begin{align}
\begin{split}
&\mathcal{I}(q) = \chi_{(1,1)}(q)\,,\\
&\mathcal{I}_A(q)=  - q^{-1} \chi_{(1,2)} (q) +q^{-1} \chi_{(1,3)}(q) \,,\\
&\mathcal{I}_B(q)=  q^{-{1\over2}} \chi_{(1,1)} (q) - q^{-{1\over2}} \chi_{(1,2)}(q)\,.
\end{split}
\end{align}
The characters of the $\Phi_{s,r}$  primary with $1\le s\le p-1$ and $1\le r \le p'-1$ in the $(p,p')$ Virasoro minimal model is given by (see, for example, \cite{philippe1997conformal})
\begin{align}\label{minimalcharacter}
\begin{split}
\chi_{(s,r)}(q)  = q^{- {(r p-s p')^2 - (p-p')^2\over 4pp'} + {1\over 24}(1-{6(p-p')^2\over pp'}) } \left(\, K^{(p,p')}_{s,r} (q)   - K^{(p,p')}_{-s,r}(q)\, \right)
\end{split}
\end{align}
where
\begin{align}
\begin{split}
K^{(p,p')}_{s,r} ( q)  ={q^{-{1\over24}} \over (q)_\infty} \sum_{n\in \mathbb{Z} }  q^{  (2pp' n+ pr -p' s)^2 \over 4pp' }\,.
\end{split}
\end{align}
Again we have normalized the character to start from 1.  Note that we have the following identification between primaries, $\Phi_{s,r} = \Phi_{ p -s, p'-r}$. 

For the purpose of demonstrating the Verlinde algebra, we only need the following Schur indices of two line defects  (no sum in the indices)
\begin{align}
\begin{split}
&\mathcal{I}_{A_i A_i}(q)=q^{-3} (1+q) \chi_{(1,1)} (q) - q^{-3} \chi_{(1,2)}(q)\,,\\
&\mathcal{I}_{B_i B_i} (q)= q^{-2} (1+q) \chi_{(1,1)} (q) -q^{-2} (1+q)  \chi_{(1,2)}(q) +q^{-1} \chi_{(1,3)}(q)\,,\\
&\mathcal{I}_{ A_5 B_6} (q)= q^{-1/2} \chi_{(1,1)} (q) -2q^{-1/2} \chi_{(1,2)} (q)+q^{-1/2} \chi_{(1,3)}(q)\,.
\end{split}
\end{align}

From the above relations between the line defect indices and the characters, we define $\Phi_{1,3}$
\begin{align}
\begin{split}
&[A]  = - [\Phi_{1,2} ] + [\Phi_{1,3}]\,,\\
&[B] =  [\Phi_{1,1} ]  - [\Phi_{1,2}]\,.
\end{split}
\end{align}
and
\begin{align}
\begin{split}
&[AA] = 2 [\Phi_{1,1}] - [\Phi_{1,2}]\,,\\
&[BB] = 2[\Phi_{1,1} ] -2 [\Phi_{1,2}] + [\Phi_{1,3}]\,,\\
&[AB] =  [\Phi_{1,1} ] -2 [\Phi_{1,2}]  + [\Phi_{1,3}]\,.
\end{split}
\end{align}
The other Schur indices of two line defects (e.g. $\mathcal{I}_{ A_1 B_2}$) can be similarly shown to give the same definitions for $[AA]$, $[BB]$, $[AB]$.

The fusion rule in the $(2,7)$ Virasoro minimal model,
\begin{align}\label{27fusion}
\begin{split}
&[\Phi_{1,2} ] \times [\Phi_{1,2}] = [\Phi_{1,1}] + [\Phi_{1,3}]\,,\\
&[\Phi_{1,3}]\times [\Phi_{1,3}] = [\Phi_{1,1}] + [\Phi_{1,2}] + [\Phi_{1,3}]\,,\\
&[\Phi_{1,2}] \times [\Phi_{1,3}] =  [\Phi_{1,2}]  + [\Phi_{1,3}]\,.
\end{split}
\end{align}
We have omitted the trivial fusion rules between the identity $\Phi_{1,1}=1$ with others. It is straightforward to check that 
\begin{align}
[AA]=[A]\times [A] \, ,~~~~ [BB] = [B]\times [B]\,,~~~~ [AB] = [A]\times [B]\,,
\end{align}
 where $\times $ is the fusion product in the $(2,7)$ Virasoro minimal model given in \eqref{27fusion}. To conclude, we find that the indices of products of 4$d$ line defects  are reproduced by the  Verlinde algebra of the $(2,7)$ Virasoro minimal model.

  \section*{Acknowledgements} 
We thank Tomoyuki Arakawa, Chris Beem, Chih-Kai Chang, Heng-Yu Chen, Thomas Dumitrescu, Sarah Harrison, Leonardo Rastelli, Cumrun Vafa, Herman Verlinde, Masahito Yamazaki for interesting discussions. CC is supported by a Schmidt fellowship at the Institute for Advanced Study and DOE grant DE-SC0009988.  The research of DG was supported by the Perimeter Institute for Theoretical Physics. Research at Perimeter Institute is supported by the Government of Canada through Industry Canada and by the Province of Ontario through the Ministry of Economic Development \& Innovation. SHS would like to thank National Taiwan University, University of Amsterdam, Tata Institute of Fundamental Research, Perimeter Institute for Theoretical Physics for their hospitality during various stages of this work.  We thank the 2015 Simons workshop in Mathematics and Physics and the Simons Center for Geometry and Physics for hospitality.

 \appendix

\section{Supercharges of Line Defects and Chiral Algebras}\label{sec:supercharge}

In this appendix we will work out the supercharges shared by the line defects and the chiral algebra. We will see that both the full line defects and the half line defects share the same set of supercharges \eqref{fancyQ} with the chiral algebra plane.

We will follow the convention in \cite{Beem:2013sza} for the 4$d$ $\mathcal{N}=2$ superconformal algebra. $A,B,\cdots =1,2$ will denote the doublet index of  $SU(2)_R$. $\alpha,\beta,\cdots= +,-$ and $\dot\alpha,\dot\beta,\cdots= \dot +,\dot -$   will denote the doublet indices of $SU(2)_{1}\times SU(2)_2 = SO(4)_{\text{rotation}}$.  All the doublet indices will be raised and lowered by $\epsilon^{12}=\epsilon_{21}=+1$.
 The nonzero anticommutators between the sixteen fermionic generators $\{Q^A_{~\alpha},~ \tilde Q_{A\dot \alpha},~S_A^{~\alpha}, ~\tilde S^{A\dot \alpha}\}$ in the 4$d$ $\mathcal{N}=2$ superconformal algebra are
\begin{align}
\begin{split}
&\{Q^A_{~\alpha}  ,  \tilde Q_{B\dot \beta} \} =2 \delta^A_B \sigma^\mu_{\alpha \dot \beta} P_\mu = \delta^A_B P_{\alpha \dot \beta}\,,\\
&\{\tilde S^{A\dot \alpha} , S_B^{~\beta} \} =2 \delta^A_B \bar\sigma^{\mu\dot\alpha\beta} K_\mu  =  \delta^A_B  K^{\dot \alpha \beta}\,,\\
&\{Q^A_{~\alpha} , S_B^{~\beta} \} = {1\over 2} \delta^A_B \delta^\beta_\alpha D +\delta^A_B M_\alpha^{~\beta} - \delta_\alpha^\beta R^A_{~B}\, ,\\
&\{\tilde S^{A\dot \alpha} ,\tilde Q_{B\dot \beta} \} ={1\over 2}\delta^A_B \delta^{\dot \alpha}_{\dot \beta}D 
+\delta^A_B M^{\dot \alpha}_{~\dot \beta} +\delta^{\dot \alpha}_{\dot\beta} R^A_{~B}\,.
\end{split}
\end{align}
The $SU(2)_R$ generators $R^\pm, R$ and the $U(1)_r$ generator $r$ sit inside $R^A_{~B}$ as
\begin{align}
R^1_{~2}  = R^+\, ,~~~R^2_{~1}  = R^-\, ,~~~ R^1_{~1} = {1\over 2}r +R \, ,~~~R^2_{~2} = {1\over 2}r -R\,,
\end{align}
where $[R^+,R^-] = 2R$ and $[R,R^\pm ] = \pm R^\pm$.

\subsection{Supercharges Preserved by Full Lines}

The eight supercharges  preserved by an infinitely extended line defect (a full line) pointing in the direction $n^\mu$ in $\mathbb{R}^4$ are \cite{Gaiotto:2010be}
\begin{align}
\begin{split}\label{LineQ}
&G^A_{~\alpha} \equiv \xi^{-1} Q^A_{~\alpha} + \xi  n_\mu \sigma^\mu_{\alpha\dot\alpha} \tilde Q^{A\dot \alpha}\,,\\
&H_A^{~\alpha} \equiv \xi S_A^{~\alpha} - \xi^{-1}  n_\mu \bar \sigma^{\mu\dot \alpha \alpha} \tilde S_{A\dot \alpha}\,,
\end{split}
\end{align}
where $\xi$ is a phase related to the $u$ and $\vartheta$ in the previous sections by $u= \xi^{-2}=e^{i\vartheta}$. Here $\bar \sigma^{\mu \dot \alpha  \alpha} =  \epsilon^{\dot\alpha\dot\beta} \epsilon^{\alpha\beta} \sigma^\mu_{\dot \beta  \beta}$.

To make sure the above linear combinations are the correct supercharges preserved by the line defect, one has to check, for example,  their anticommutators do not contain the $U(1)_r$ generator $r$, the translations, nor the special conformal transformations along other directions than $n_\mu$. Let us check this for a  few anticommutators.\footnote{The  parentheses  and the square brackets denote the symmetrization and antisymmetrization of indices, respectively. That is, $T^{(\alpha\beta) } = \frac12 (T^{\alpha\beta} + T^{\beta\alpha})$ and $T^{[\alpha\beta] } = \frac12 (T^{\alpha\beta} -T^{\beta\alpha})$.} 
\begin{align}
\begin{split}
&\{ G^A_{~\alpha} , G^B_{~\beta} \} =  4 \epsilon^{AB} n_\mu P_\nu \epsilon^{\dot \alpha \dot \beta} \sigma^\mu_{[\alpha|\dot \alpha} \sigma^\nu _{\beta]\dot \beta} = 4\epsilon^{AB} \epsilon_{\alpha\beta}  n^\mu P_\mu \,,\\
&\{H_A^{~\alpha},  H_B^{~\beta} \} 
   = 4  \epsilon_{AB} \epsilon_{\dot \alpha\dot \gamma} \bar\sigma^{\mu\dot\alpha[ \beta}\bar \sigma^{\nu \dot \gamma |\alpha]} n_\mu K_\nu = -4 \epsilon_{AB} \epsilon^{\alpha\beta}  n^\mu K_\mu\, ,
   \end{split}
\end{align}
where we have used $ \epsilon^{\dot \alpha \dot \beta} \sigma^\mu_{[\alpha|\dot \alpha} \sigma^\nu _{\beta]\dot \beta} = \delta^{\mu\nu} \epsilon_{\alpha\beta}$ and $ \epsilon_{\dot \alpha\dot \gamma} \bar\sigma^{\mu\dot\alpha[ \beta}\bar \sigma^{\nu \dot \gamma |\alpha]}  = \delta^{\mu\nu} \epsilon_{\beta\alpha}$. Also, let us check that there is no $r$ in the anticommutator between $G^A_{~\alpha}$ and $H_A^{~\alpha}$.
\begin{align}
\begin{split}
\{G^A_{~\alpha} , H_B^{~\beta} \} &= 
\{Q^A_{~\alpha} , S_B^{~\beta} \} -( n_\mu\sigma^\mu_{\alpha\dot \alpha})(n_\nu \bar \sigma^{\nu\dot\beta\beta} ) \epsilon^{AC}\epsilon^{\dot \alpha \dot \gamma}\epsilon_{BD} \epsilon_{\dot\beta\dot\delta}
\{\tilde Q_{C \dot\gamma} ,  \tilde S^{D \dot \delta} \}\\
&= -\delta^\beta_\alpha (R^A_{~B} + \epsilon^{AC} \epsilon_{BD} R^D_{~C} ) + \text{rotations and dilation}
\end{split}
\end{align}
where we have used $\sigma^{(\mu}_{\alpha\dot \alpha} \, \bar \sigma^{\nu)\dot\alpha\beta}= - \delta^{\mu\nu} \delta_\alpha^\beta$. Indeed, the righthand side does not contain the $U(1)_r$ generator.

Finally, we would like to determine the supercharges preserved by the Schur operators in the presence of a full line defect. To do so, it is convenient to fix our conventions for the Pauli matrices to be
\begin{align}
\sigma^1 _{\alpha\dot\beta} = \left(\begin{array}{cc}0& 1 \\1 & 0\end{array}\right),~~~
\sigma^2 _{\alpha\dot\beta} = \left(\begin{array}{cc}0 & -i \\i & 0\end{array}\right),~~~
\sigma^3 _{\alpha\dot\beta} = \left(\begin{array}{cc}1 & 0 \\0 & -1\end{array}\right),~~~
\sigma^4 _{\alpha\dot\beta} = \left(\begin{array}{cc}i & 0 \\0 & i\end{array}\right).
\end{align}
We have $\bar\sigma^\mu= (   -\sigma^1, -\sigma^2 ,-\sigma^3,\sigma^4)$ numerically.
Further, we will choose our line defect to be along the 1-direction, i.e., $n_{\mu}= (1,0,0,0)$, and $\xi=0$. 

The ordinary Schur index without insertions of line defects receives contributions only from operators that are annihilated by four supercharges, which can be chosen to be $Q^1_{~-}$, $\tilde Q_{2\dot -}$, $S_1^{~-}$, $\tilde S^{2\dot-}$ \cite{Gadde:2011ik,Gadde:2011uv}. In the presence of a full line defect, two out of the four supercharges above are shared by the eight supercharges \eqref{LineQ} that are preserved by a full line defect,\footnote{As we will see in Section \ref{sec:ShareQ}, these are the two supercharges shared by the chiral algebra plane and the line defects. }
\begin{align}\label{GH1-}
\begin{split}
&G^1_{~-} = Q^1_{~-} + \tilde Q_{2\dot -}\,,\\
&H_1^{~- } = S_1^{~-} + \tilde S^{2\dot -}\,.
\end{split}
\end{align}
Their anticommutators are
\begin{align}
\begin{split}
\{G^1_{~- } , G^1_{~-} \} &= 0 \, ,~~~~~\{ H_1^{~-} , H_1^{~-} \} =0\,,\\
\{ G^1_{~-} , H_1^{~-} \} 
& = E- ( j_1+j_2 ) -2R  \,.
\end{split}
\end{align}
Here $E$, $j_1$, $j_2$ are the eigenvalue of the dilation charge $D$, ${M}_+^{~+}$, $M^{\dot + }_{~\dot +}$, respectively. Note that $\mathcal{M} \equiv j_1+j_2$ is the rotation of the plane orthogonal to the line defect.

In summary, the Schur operators in the presence of a line defect are annihilated by the two supercharges \eqref{GH1-}, and they obey the following condition,
\begin{align}\label{hatL0}
\text{Line defect Schur operator}:~\hat L_0\equiv{1\over 2} \left(  E- ( j_1+j_2 )\right) -R=0\, .
\end{align}
 Recall that the ordinary Schur operators without the line defect obey an additional condition $\mathcal{Z}\equiv r+ (j_1-j_2)=0$.

More generally, we consider multiple full line defects $L_i$  with  phases 
\begin{align}
\xi_i=e^{-i\vartheta_i/2}
\end{align}
 and pointing in the directions $(n_i)^\mu$. To preserve supersymmetry, we have to put them on the 12-plane (the plane orthogonal to the chiral algebra plane) with orientations
\begin{align}\label{orientation}
(n_i)^\mu  =(  \cos \vartheta_i ,\,\sin \vartheta_i ,\,0,\,0 ).
\end{align}
Note that $(n_i)_\mu \sigma^\mu_{\alpha \dot \alpha} = \left(\begin{array}{cc}0 & e^{-i \vartheta_i} \\e^{i \vartheta_i} & 0\end{array}\right)$ and $(n_i)_\mu \bar\sigma^{\mu\dot\alpha \alpha} = -\left(\begin{array}{cc}0 & e^{-i \vartheta_i} \\e^{i \vartheta_i} & 0\end{array}\right)$. 

The 4 supercharges shared by all the (full) line defects are
\begin{align}
\begin{split}
&G^A_{~ -} = Q^A_{~-} + \tilde Q^{A \dot +}\,,\\
&H_A^{~-} = S_A^{~-} + \tilde S_{A\dot +}\,.
\end{split}
\end{align}
which in particular include two supercharges $G^1_{~-}$ and $H_1^{~-}$ \eqref{GH1-} that are used to define the line defect Schur index.

\subsection{Supercharges Preserved by  Half Lines}

In this subsection we will check that half line defects preserve the same two supercharges \eqref{GH1-} that are used to define the line defect Schur index. 
To simplify the notations, we define
\begin{align}
G_\alpha\equiv G^1_\alpha\,,~~~~~H^{\alpha}\equiv H_1^{~\alpha}\,,~~~~~\Delta \equiv  E-2R\,.
\end{align}
They satisfy the following hermicity conditions  $(G_\alpha)^\dagger = H^\alpha ,\,\Delta^\dagger = \Delta\,.$

The superalgebra preserved by a half line pointing along the $1$-direction is
\begin{align}
\begin{split}
&\{G_\alpha, G_\beta\} = \{H^\alpha , H^\beta\}=0\,,~~~~~~~~~~~\{G_\alpha , H^\beta\} =  M_\alpha^{~\beta}  + \delta_\alpha^\beta \Delta\,,\\
&[\Delta,G_\alpha] = -{1\over 2}G_\alpha\,,~~~~~~~~~~~~~~~~~~~~~~~~[\Delta,H^\alpha ] = {1\over 2} H^\alpha\,,\\
&[M_\alpha^{~\beta} , G_\gamma] = \delta_\gamma^\beta G_\alpha -{1\over2 } \delta_\alpha^\beta G_\gamma\,,~~~~~~~~~~[M_\alpha^{~\beta} , H^\gamma] = - \delta^\gamma_\alpha H^\beta +{1\over2 } \delta_\alpha^\beta H^\gamma\,,\\
&[M_\alpha^{~\beta} , M_\gamma^{~\delta}] = \delta_\gamma ^\beta M_\alpha^{~\delta} - \delta_\alpha^{\delta} M_\gamma^{~\beta}\,,~~~~~~~[M_\alpha^{~\beta} ,  \Delta] = 0\,.
\end{split}
\end{align}
Here $M_\alpha^{~\beta}$ are the generators of the $SO(3)$ rotating the $\mathbb{R}^3$ transverse to the half line.  Note that in contrast to the case of a full line defect, the translation $P^1$ and special conformal transformation $K^1$ are no longer symmetries of the configuration, hence do not show up in the above algebra.  The preserved four supercharges are $G^1_\alpha$ and $H_1^{~\alpha}$, which include  \eqref{GH1-}.

 More generally, if we include multiple half lines with phases $\xi_i =e^{- i \vartheta_i /2}$ and orientations given as in \eqref{orientation}, the only preserved rotation symmetry is the one rotating the 34-plane, $M_-^{~-}$ (whose eigenvalue is $j_1+j_2$ in the notation before). The preserved superalgebra is
 \begin{align}
 \begin{split}\label{multiplehalf}
 &\{ G_- , G_- \} =  \{ H^- , H^- \} =0\,,~~~~~\{G_- , H^- \} =  \Delta+ M_-^{~-}  \,,\\
 &[\Delta , G_-  ]  = -{1\over2}G_- \,,~~~~~~~[\Delta , H^- ] = {1\over 2} H^-\,,\\
 &[M_-^{~- } , G_- ] = {1\over 2}G_- \,,~~~~~~[M_-^{~-} ,H^-] = - {1\over 2} H^-\,,~~~~~~[M_-^{~-},\Delta]=0\,,
 \end{split}
 \end{align}
The preserved supercharges $G^1_-$ and $H_1^{~-}$ \eqref{GH1-} are precisely those used to define the line defect Schur index.  Incidentally, $2\hat L_0 = \Delta+ M_-^{~-}$ where $\hat L_0={1\over2}(E-2R - (j_1+j_2))=0$ is satisfied by all the line defect Schur operators \eqref{hatL0}.

\subsection{Supercharges Shared by the Chiral Algebra and Line Defects}\label{sec:ShareQ}

In this subsection we will determine the supercharges shared by the chiral algebra plane and the (full or half) line defects. 
The chiral algebra operators live in the cohomology of the following four supercharges \cite{Beem:2013sza}
\begin{align}
\begin{split}\label{ChiralAlgebraQ}
&\mathtt{Q}_1 \equiv Q^1_{~-} + \tilde S^{2\dot -}\,,~~~\mathtt{Q}_2 \equiv  S_1^{~-}- \tilde Q_{2\dot -}\,,\\
&\mathtt{Q}_1^\dagger  \equiv  S_1^{~-}+ \tilde Q_{2\dot -}\,,~~~\mathtt{Q}_2^\dagger \equiv Q^1_{~-} - \tilde S^{2\dot -}\,.
\end{split}
\end{align}
We  find  two  supercharges $G^1_{~-}$ and $H_1^{~-}$ (in the notation of \eqref{GH1-}) that are preserved by both (full or half) line defects and the chiral algebra,
\begin{align}
\begin{split}\label{fancyQ}
&G^1_{~-} = {1\over2} (\mathtt{Q}_1 +\mathtt{Q}_1^\dagger  - \mathtt{Q}_2 +\mathtt{Q}_2^\dagger)
= Q^1_{~-}  + \tilde Q_{2\dot -} \,,\\
&H_1^{~-} = {1\over2} (\mathtt{Q}_1 +\mathtt{Q}_1^\dagger  + \mathtt{Q}_2 -\mathtt{Q}_2^\dagger)
= S_1^{~-}  + \tilde S^{2\dot -}\,,
\end{split}
\end{align}  
They satisfy the hermicity conditions $(G^1_{~-})^\dagger = H_1^{~-}$.  These are precisely the two supercharges preserved by multiple half line defects lying on the 12-plane \eqref{multiplehalf}.  Let us record the anticommutators of the supercharges $G^1_{~-}$ and $H_1^{~-}$ again,
\begin{align}
\begin{split}
&\{ G^1_{~-} , G^1_{~-} \} = \{ H_1^{~-} , H_1^{~-} \}=0\,,~~~~\{ G^1_{~-}  , H_1^{~-} \}  = 2 \hat L_0\,,
\end{split}
\end{align}
where $\hat L_0={1\over2}(E-2R - (j_1+j_2))=  0$ is the defining condition for the line defect Schur operators \eqref{hatL0}. Note that $\hat L_0$ involves the rotation $\mathcal{M}= j_1+ j_2$ on the chiral algebra plane. Since $\hat L_0$ shows up on the righthand side of the anticommutator of supercharges preserved by the line defect, the  defect must lie on the plane \textit{transverse} to the chiral algebra plane, with which it  intersects  at a point. In the convention in \eqref{ChiralAlgebraQ}, we have chosen the chiral algebra plane to be the 34-plane where $x^1=x^2=0$ \cite{Beem:2013sza}, and the line defects lie on the 12-plane where $x^3=x^4=0$. See Figure \ref{fig:IncidenceGeometry2} for the incidence geometry of  line defects and the chiral algebra plane.

Notice that since $G^1_{~-}$ and $H_1^{~-}$  only involve of $Q$'s and $S$'s, respectively, the $\hat L_\pm$ generators cannot be both exact under either of them.  Thus it seems difficult to construct  the chiral algebra for the defect operators by generalizing \cite{Beem:2013sza}.  

On the other hand, we can consider the following linear combinations of $G^1_{~-}$ and $H_1^{~-}$,
\begin{align}
\begin{split}
&\mathfrak{Q}_1 \equiv G^1_{~-} +H_1^{~-}=  \mathtt{Q}_1  +\mathtt{Q}_1^\dagger\,,~~~~~\mathfrak{Q}_2\equiv-G^1_{~-} + H_1^{~-} =  \mathtt{Q}_2  -\mathtt{Q}_2^\dagger\,,
\end{split}
\end{align}
such that $\hat L_{\pm1}$ are both $\mathfrak{Q}_1$-exact and $\mathfrak{Q}_2$-exact,
\begin{align}
\begin{split}
&\{ \mathfrak{Q}_1 , \tilde Q_{1\dot -}\} 
=\{ \mathfrak{Q}_2 , -  Q^2_{~ -}\} 
 = \hat L_{-1}  =  P_{-\dot -} + R^2_{~1}\,,\\
&\{ \mathfrak{Q}_1 ,  S_2^{~-} \} 
=\{ \mathfrak{Q}_2 ,\tilde  S^{1\dot -}\} 
= \hat L_{+1} =  K^{\dot --} -R^1_{~2}\,.
\end{split}
\end{align}
However, the disadvantage of these combinations  $\mathfrak{Q}_i$ is that they are not nilpotent, but square to  $\hat L_0$
\begin{align}
&\{ \mathfrak{Q}_1, \mathfrak{Q}_1 \} = 4\hat L_0\,,~~~~~\{ \mathfrak{Q}_2, \mathfrak{Q}_2 \} =-4\hat L_0\,.
\end{align}
Note that $\mathfrak{Q}_1$ and $\mathfrak{Q}_2$ anticommute $\{ \mathfrak{Q}_1 , \mathfrak{Q}_2 \} = 0$.

 \section{Framed Quivers for the Argyres-Douglas Theories}
 
 In this Appendix we derive the core charges and the framed BPS quivers for the generators of line defects in the $A_3$ and $A_4$ Argyres-Douglas theories, following closely the example of the $A_2$ Argyres-Douglas theory in \cite{Cordova:2013bza}.

Given a fixed point on the moduli space and a choice of the half plane, the BPS quiver, if exists, is unique.  The charges $\{\gamma_i\}$ of the nodes of the  quiver are called a \textit{seed}.  Associated to this seed is a cone $\mathcal{C}$ on the charge lattice $\Gamma$,
\begin{align}
\mathcal{C} = \left\{  \sum_i a_i \gamma_i \in \Gamma \, \Big| \, a_i \in \mathbb{R}_{\ge 0}\right\}\,,
\end{align}
generated by the seed with non-negative coefficients.  We also define the dual cone $\check{\mathcal{C}}$ in the charge lattice $\Gamma$ as
\begin{align}
\check{\mathcal{C}} = \left\{  \check\gamma\in \Gamma \, \Big|\,  \langle \check \gamma, \gamma \rangle\ge0\right\}\,.
\end{align}
One important feature of the dual cone is that the UV line defects in $\check{\mathcal{C}}$ satisfy a universal OPE,
\begin{align}
L_{\gamma_1} L_{\gamma_2}  =  q^{\frac12 \langle \gamma_1,\gamma_2\rangle } L_{\gamma_1+\gamma_2}\,,~~~~~~~~\gamma_1,\,\gamma_2 \in \check{\mathcal{C}}\,.
\end{align}
where $\gamma_i$'s are the core charges of the defect $L_i$.  

Starting from an initial seed, we can generate other seeds by mutation and obtain their associated dual cones.  It follows that on the charge lattice for the defects, there are many distinct dual cones $\check{\mathcal{C}}$, inside which the defect OPE takes the simple form above. For the Argyres-Douglas theories considered in this paper, the dual cones cover the full charge lattice.  This simplifies the study of defect OPE significantly.  In particular,  the defect OPEs are  completely encoded in the OPEs between   those defects whose core charges lie at the boundaries of the dual cones.  These distinguished defects will be called the generators of defects.  In the rest of this Appendix, we will compute the core charges of these generators and their associated framed BPS quiver in the $A_3$ and $A_4$ Argyres-Douglas theories.

 \subsection{$A_3$ Argyres-Douglas Theory}\label{sec:A3Cones}

As shown in  Figure \ref{fig:A3Seed}, from the initial seed $\{\gamma_1,\gamma_2,\gamma_3\}$, we generate fourteen seeds by mutation.  Each of the fourteen seeds  is associated to  a dual cone $\check{\mathcal{C}}$ in the space of line defects.  Out of the fourteen seeds, eight of them give rise to degenerate dual cones that are of higher codimensions.  The remaining six non-degenerate dual cones are
\begin{align}
\begin{split}
&\check{\mathcal{C} }_{\{ \gamma_1, \gamma_2 , \gamma_3 \} }  = \{a_1 \gamma_1 + a_2 \gamma_2  + a_3 \gamma_3   \,  \Big|  \,    a_1 +a_3 \ge 0, \,  a_2\le 0  \} \,,\\
&\check{\mathcal{C} }_{\{ \gamma_1, -\gamma_2 , \gamma_3 \} }  = \{a_1 \gamma_1 + a_2 \gamma_2  + a_3 \gamma_3   \,  \Big|  \, 0\ge   a_1 +a_3 \ge a_2  \} \,,\\
&\check{\mathcal{C} }_{\{ - \gamma_1, \gamma_1 + \gamma_2 +\gamma_3 ,  -  \gamma_3 \} }  = \{a_1 \gamma_1 + a_2 \gamma_2  + a_3 \gamma_3   \,  \Big|  \,    a_1 +a_3 \ge 0, \,  a_2\ge 0  \} \,,\\
&\check{\mathcal{C} }_{\{ \gamma_2+\gamma_3, -\gamma_1 - \gamma_2  -\gamma_3, \gamma_1+\gamma_2 \} }  = \{a_1 \gamma_1 + a_2 \gamma_2  + a_3 \gamma_3   \,  \Big|  \,    a_1 +a_3 \le 0, \,  a_2\ge 0  \} \,,\\
&\check{\mathcal{C} }_{\{- \gamma_1, -\gamma_2 , -\gamma_3 \} }  = \{a_1 \gamma_1 + a_2 \gamma_2  + a_3 \gamma_3   \,  \Big|  \,   a_2\ge  a_1 +a_3\ge2 a_2  \} \,,\\
&\check{\mathcal{C} }_{\{ -\gamma_1-\gamma_2 , \gamma_2 , -\gamma_2- \gamma_3 \} }  = \{a_1 \gamma_1 + a_2 \gamma_2  + a_3 \gamma_3   \,  \Big|  \,   0 \ge 2  a_2\ge a_1 +a_3  \} \,.
\end{split}
\end{align}
For example, to obtain the dual cone of the seed $\{ \gamma_1 ,  - \gamma_2  ,  \gamma_3\}$, we first find those charges which have positive Dirac pairing with the seed, and then (right) mutate them back to the original seed to see how this dual cone is embedded in the charge lattice.  Explicitly, we have
\begin{align}
\begin{split}
\check{\mathcal{C} }_{\{ \gamma_1, -\gamma_2 , \gamma_3 \} } &  =  
\mu_{R_2} \left(
\left\{
a_1\gamma_1 + a_2 \gamma_2 +a_3 \gamma_3 \,\Big| \, a_1+a_3\le 0\,, \,a_2\le0
\right\}
\right)\\
& = \left\{
a_1\gamma_1 + a_2 \gamma_2 +a_3 \gamma_3 \,\Big| \,  0 \ge a_1+a_3\ge a_2
\right\}\,.
\end{split}
\end{align}
Note that since $\gamma_1-\gamma_3$ is a flavor node, only the combination $a_1  +a_3$ shows up but not  $a_1$ and $a_3$ individually.  We show the two-dimensional projection $(a_2,a_1+a_3)$ of the geometry of the dual cones in Figure \ref{fig:A3dualcone}.  The six boundaries of the dual cones are two-dimensional half-planes,  each generated by a flavor charge $\gamma_1-\gamma_3$ and an electromagnetic core charge $A_i$, $B_i$ with $i=1,2,3$. The core charges, defined as the images of the \textbf{RG} map \cite{Cordova:2013bza},  are
\begin{align}
\begin{split}
&\textbf{RG}(A_1)  = \gamma' \,,~\,~~~~~\textbf{RG} (A_2)  = -\gamma' -\gamma_2\,,~\,\,\,~~~\textbf{RG}(A_3)  =-\gamma'\,,\\
 &\textbf{RG}(B_1)  = -\gamma_2 \,,~~~~\textbf{RG} (B_2)  = -2\gamma' -\gamma_2\,,~~~~\textbf{RG}(B_3)  =\gamma_2\,,
 \end{split}
\end{align}
where $\gamma'$ is any charge vector of the form $x \gamma_1 + (1-x) \gamma_3$.  The associated framed BPS quivers for these six defects are given in Figure \ref{fig:A3framedquiver}.  By applying the mutation method to these framed quivers, we obtain the generating functions for these defects \eqref{generatingA3}.

\newpage

\begin{figure}[h!]
\begin{center}
\includegraphics[width=.96\textwidth]{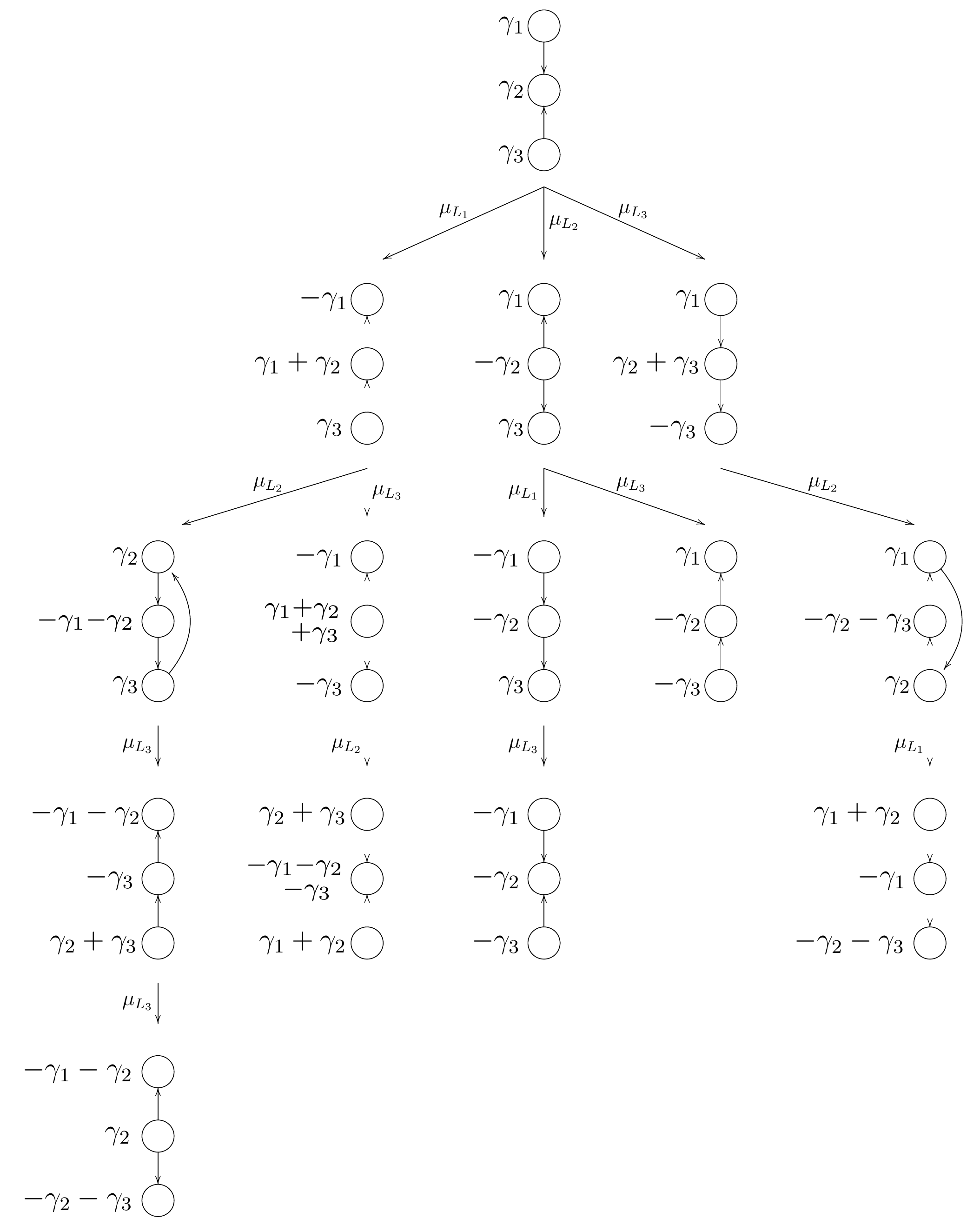}
\end{center}
\caption{ The fourteen seeds of the $A_3$ Argyres-Douglas theory. $\mu_{L_i}$ denotes the left mutation with respect to the $i$-th node counting from the top.}\label{fig:A3Seed}
\end{figure}

\newpage

\begin{figure}[h]
\begin{center}
\includegraphics[width=.45\textwidth]{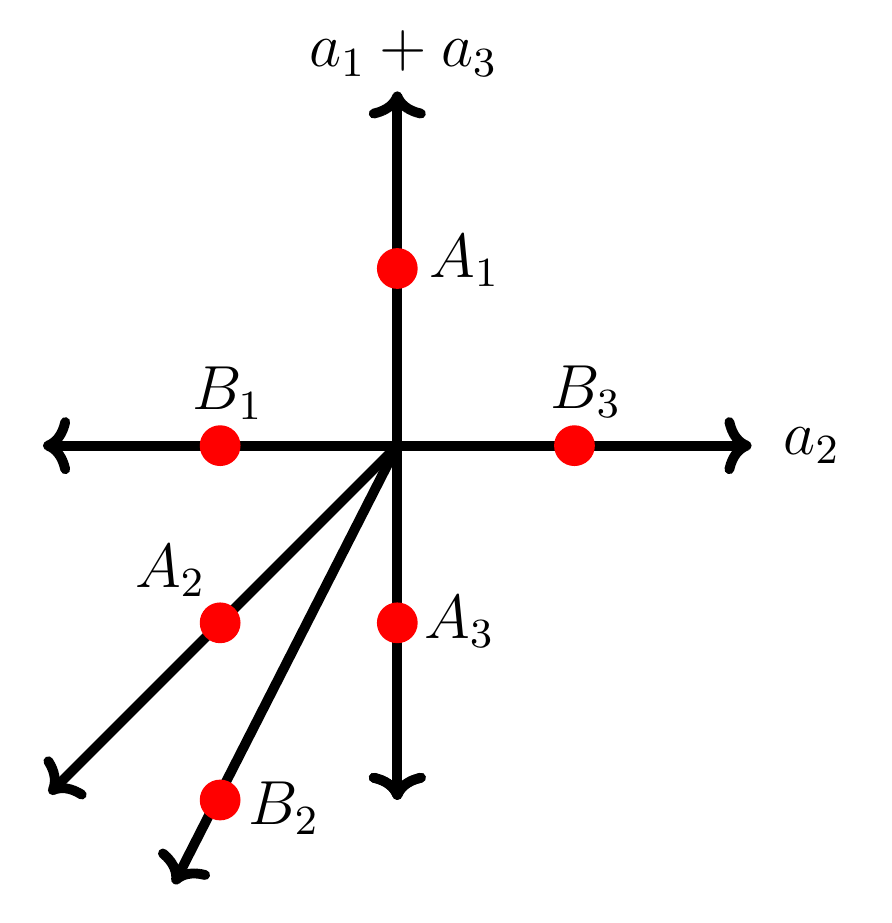}
\end{center}
\caption{The six dual cones of the $A_3$ Argyres-Douglas theory. Here we show the two-dimensional projection $(a_2, a_1+a_3)$ of the three-dimensional space $\Gamma \otimes _\mathbb{Z} \mathbb{R}$, which is identified as $\mathbb{R}\oplus \mathbb{R} \oplus \mathbb{R}$ by expressing the seed as $\gamma = a_1\gamma_1 + a_2 \gamma_2 +a_3\gamma_3$. The black arrows are the projection of the boundary half-planes  of the dual cones. The red dots are the generators $A_i$, $B_i$ for the  line defects.}\label{fig:A3dualcone}

 \subfloat[$A_1$]{\xy  0;<1pt,0pt>:<0pt,-1pt>::
(-40,0) *+{}*\cir<8pt>{} ="1",
(0,0) *+{}*\cir<8pt>{} ="2",
(40,0) *+{}*\cir<8pt>{} ="3",
(0,-40) *+{\color{white}{1}}*\frm{-} ="4",
(-40, 15) *+{\gamma_{1}} ="a",
(0, 15) *+{\gamma_{2}} ="b",
(40, 15) *+{\gamma_{3}} ="c",
(0, -55) *+{\gamma'} ="d",
\ar @{->} "1"; "2"
\ar @{<-} "2"; "3"
\ar @{->} "4" ;"2"
\endxy}
~~~~~~~~~~~~
\subfloat[$A_2$]{\xy  0;<1pt,0pt>:<0pt,-1pt>::
(-40,0) *+{}*\cir<8pt>{} ="1",
(0,0) *+{}*\cir<8pt>{} ="2",
(40,0) *+{}*\cir<8pt>{} ="3",
(0,-40) *+{\color{white}{1}}*\frm{-} ="4",
(-40, 15) *+{\gamma_{1}} ="a",
(0, 15) *+{\gamma_{2}} ="b",
(40, 15) *+{\gamma_{3}} ="c",
(0, -55) *+{- \gamma'-\gamma_2} ="d",
\ar @{->} "1"; "2"
\ar @{<-} "2"; "3"
\ar @{<-} "4" ;"2"
\ar @{->} "4" ;"1"
\ar @{->} "4" ;"3"
\endxy}
~~~~~~~~~~~~
\subfloat[$A_3$]{\xy  0;<1pt,0pt>:<0pt,-1pt>::
(-40,0) *+{}*\cir<8pt>{} ="1",
(0,0) *+{}*\cir<8pt>{} ="2",
(40,0) *+{}*\cir<8pt>{} ="3",
(0,-40) *+{\color{white}{1}}*\frm{-} ="4",
(-40, 15) *+{\gamma_{1}} ="a",
(0, 15) *+{\gamma_{2}} ="b",
(40, 15) *+{\gamma_{3}} ="c",
(0, -55) *+{-\gamma'} ="d",
\ar @{->} "1"; "2"
\ar @{<-} "2"; "3"
\ar @{<-} "4" ;"2"
\endxy}\\
 \subfloat[$B_1$]{\xy  0;<1pt,0pt>:<0pt,-1pt>::
(-40,0) *+{}*\cir<8pt>{} ="1",
(0,0) *+{}*\cir<8pt>{} ="2",
(40,0) *+{}*\cir<8pt>{} ="3",
(0,-40) *+{\color{white}{1}}*\frm{-} ="4",
(-40, 15) *+{\gamma_{1}} ="a",
(0, 15) *+{\gamma_{2}} ="b",
(40, 15) *+{\gamma_{3}} ="c",
(0, -55) *+{-\gamma_2} ="d",
\ar @{->} "1"; "2"
\ar @{<-} "2"; "3"
\ar @{->} "4" ;"1"
\ar @{->}"4";"3"
\endxy}
~~~~~~~~~~~~
\subfloat[$B_2$]{\xy  0;<1pt,0pt>:<0pt,-1pt>::
(-40,0) *+{}*\cir<8pt>{} ="1",
(0,0) *+{}*\cir<8pt>{} ="2",
(2,-5) *+{}="2a",
(-2,-5) *+{}="2b",
(40,0) *+{}*\cir<8pt>{} ="3",
(0,-40) *+{\color{white}{1}}*\frm{-} ="4",
(2,-36) *+{} ="4a",
(-2,-36) *+{} ="4b",
(-40, 15) *+{\gamma_{1}} ="a",
(0, 15) *+{\gamma_{2}} ="b",
(40, 15) *+{\gamma_{3}} ="c",
(0, -55) *+{- 2\gamma'-\gamma_2} ="d",
\ar @{->} "1"; "2"
\ar @{<-} "2"; "3"
\ar @{->} "4" ;"1"
\ar @{->} "4" ;"3"
\ar @{<-} "4a" ;"2a"
\ar @{<-} "4b" ;"2b"
\endxy}
~~~~~~~~~~~~
\subfloat[$B_3$]{\xy  0;<1pt,0pt>:<0pt,-1pt>::
(-40,0) *+{}*\cir<8pt>{} ="1",
(0,0) *+{}*\cir<8pt>{} ="2",
(40,0) *+{}*\cir<8pt>{} ="3",
(0,-40) *+{\color{white}{1}}*\frm{-} ="4",
(-40, 15) *+{\gamma_{1}} ="a",
(0, 15) *+{\gamma_{2}} ="b",
(40, 15) *+{\gamma_{3}} ="c",
(0, -55) *+{\gamma_2} ="d",
\ar @{->} "1"; "2"
\ar @{<-} "2"; "3"
\ar @{->}"1";"4"
\ar @{->}"3";"4"
\endxy}
\caption{The framed quivers for the six generators $A_i$ and $B_i$ ($i=1,2,3$) of the $A_3$ Argyres-Douglas theory.  The core charges are labeled above the framed nodes (the square nodes).  $\gamma'$ is any charge vector of the form $\gamma'  = x\gamma_1  + (1-x) \gamma_3$ with real $x$.}\label{fig:A3framedquiver}
 \end{figure}

\clearpage

 \subsection{$A_4$ Argyres-Douglas Theory}\label{sec:A4Cones}

 For the $A_4$ Argyres-Douglas theory, there are 42 seeds and their associated dual cones are listed below.
\begin{align*}
&
\check{\mathcal{C}}_{\{\gamma_1 , \gamma_2 , \gamma_3 , \gamma_4 \} }=
\left \{\,
\sum_{i=1}^4 a_i \gamma_i \, \Big| \, a_2\le 0\,,\, a_1+a_3 \ge0 \,,\, a_2+a_4\le 0\,,\, a_3\ge0\,
\right\}\,,\\
&
\check{\mathcal{C}}_{\{\gamma_1 , \gamma_2 , \gamma_3 , - \gamma_4 \} }=
\left \{\,
\sum_{i=1}^4 a_i \gamma_i \, \Big| \, a_2\le 0\,,\, a_1+a_3 \ge0 \,,\,a_3\ge a_2+a_4\,,\, a_3\le0\,
\right\}\,,\\
&
\check{\mathcal{C}}_{\{\gamma_1 , \gamma_2+\gamma_3  , - \gamma_3 , \gamma_3+  \gamma_4 \} }=
\left \{\,
\sum_{i=1}^4 a_i \gamma_i \, \Big| \, a_2\le 0\,,\, a_1+a_3 \ge0 \,,  a_2+a_4\ge0\,,\, a_3\ge0\,
\right\}\,,\\
&
\check{\mathcal{C}}_{\{\gamma_1 ,- \gamma_2 , \gamma_3 ,  \gamma_4 \} }=
\left \{\,
\sum_{i=1}^4 a_i \gamma_i \, \Big| \,a_1+a_3\ge a_2\,,\, a_1+a_3 \le0 \,,\,a_1+a_3\ge a_2+a_4\,,\, a_3\ge0\,
\right\}\,,\\
&
\check{\mathcal{C}}_{\{- \gamma_1 ,\gamma_1+  \gamma_2 , \gamma_3 ,  \gamma_4 \} }=
\left \{\,
\sum_{i=1}^4 a_i \gamma_i \, \Big| \, a_2\ge 0\,,\, a_1+a_3 \ge0 \,,\, a_2+a_4\le0\,,\, a_3\ge0\,
\right\}\,,\\
&
\check{\mathcal{C}}_{\{\gamma_1 ,- \gamma_2 , -\gamma_3 , \gamma_3+ \gamma_4 \} }=
\left \{\,
\sum_{i=1}^4 a_i \gamma_i \, \Big| \, a_1+a_3\ge a_2\,,\, a_2+a_4 \le0 \,,\,a_2+a_4\ge a_1+a_3\,,\, a_3\ge0\,
\right\}\,,\\
&
\check{\mathcal{C}}_{\{ -  \gamma_1 , - \gamma_2 , \gamma_3 , \gamma_4 \} }=
\left \{\,
\sum_{i=1}^4 a_i \gamma_i \, \Big| \, a_2\le 0\,,\, a_1+a_3 \le a_2 \,,\,a_1+ a_3\ge a_2+a_4\,,\, a_3\ge0\,
\right\}\,,\\
&
\check{\mathcal{C}}_{\{\gamma_1 ,-  \gamma_2 , \gamma_3 , - \gamma_4 \} }=
\left \{\,
\sum_{i=1}^4 a_i \gamma_i \, \Big| \, a_1+a_3\ge a_2\,,\, a_1+a_3 \le0 \,,\,a_1+2a_3\ge a_2+a_4\,,\, a_3\le0\,
\right\}\,,\\
&
\check{\mathcal{C}}_{\{- \gamma_1 , \gamma_1+ \gamma_2+\gamma_3 , -\gamma_3 , \gamma_3+\gamma_4 \} }=
\left \{\,
\sum_{i=1}^4 a_i \gamma_i \, \Big| \, a_2\ge 0\,,\, a_1+a_3 \ge0 \,,\, a_2+a_4\ge0\,,\, a_3\ge0\,
\right\}\,,\\
&
\check{\mathcal{C}}_{\{\gamma_2 , -\gamma_1-  \gamma_2 , \gamma_3 ,  \gamma_4 \} }=
\left \{\,
\sum_{i=1}^4 a_i \gamma_i \, \Big| \, a_2\ge 0\,,\, a_1+a_3 \le0 \,,\,a_1+a_3\ge a_2+a_4\,,\, a_3\ge0\,
\right\}\,,\\
&
\check{\mathcal{C}}_{\{- \gamma_1 , \gamma_1+ \gamma_2 , \gamma_3 ,  -\gamma_4 \} }=
\left \{\,
\sum_{i=1}^4 a_i \gamma_i \, \Big| \, a_2\ge 0\,,\, a_1+a_3 \ge0 \,,\,  a_3\ge a_2+a_4\,,\, a_3\le0\,
\right\}\,,\\
&
\check{\mathcal{C}}_{\{\gamma_1 , \gamma_2+  \gamma_3 , - \gamma_3 ,  -\gamma_4 \} }=
\left \{\,
\sum_{i=1}^4 a_i \gamma_i \, \Big| \, a_2\le 0\,,\, a_1+a_3 \ge0 \,,\, a_3\le a_2+a_4\,,\, a_2+a_4\le0\,
\right\}\,,\\
&
\check{\mathcal{C}}_{\{\gamma_1 , \gamma_2+  \gamma_3 , \gamma_4 ,  -\gamma_3-\gamma_4 \} }=
\left \{\,
\sum_{i=1}^4 a_i \gamma_i \, \Big| \, a_2\le 0\,,\, a_1+a_3 \ge0 \,,\,  a_2+a_4 \ge 0\,,\, a_3\le0\,
\right\}\,,
\end{align*}

\begin{align*}
&
\check{\mathcal{C}}_{\{\gamma_1 , -\gamma_2-  \gamma_3 , \gamma_2 ,  \gamma_3+ \gamma_4 \} }=
\left \{\,
\sum_{i=1}^4 a_i \gamma_i \, \Big| \, a_1+a_3\ge a_2\,,\, a_1+a_3 \le0 \,,\,a_2+a_4\ge 0\,,\, a_3\ge0\,
\right\}\,,\\
&
\check{\mathcal{C}}_{\{\gamma_1 , -  \gamma_2 , -\gamma_3 , - \gamma_4 \} }=
\left \{\,
\sum_{i=1}^4 a_i \gamma_i \, \Big| \, a_1+a_3\ge a_2\,,\, a_3 \ge a_2+a_4\ge a_1+2a_3\,,\, a_1+a_3\ge a_2+a_4\,
\right\}\,,\\
&
\check{\mathcal{C}}_{\{-\gamma_1 , -  \gamma_2 , \gamma_3 ,  -\gamma_4 \} }=
\left \{\,
\sum_{i=1}^4 a_i \gamma_i \, \Big| \, a_2\le 0\,,\, a_1+a_3 \le a_2 \,,\,a_1+2a_3\ge a_2+a_4\,,\, a_3\le0\,
\right\}\,,\\
&
\check{\mathcal{C}}_{\{ - \gamma_1 ,  \gamma_1+\gamma_2+\gamma_3 , \gamma_4 ,  -\gamma_3- \gamma_4 \} }=
\left \{\,
\sum_{i=1}^4 a_i \gamma_i \, \Big| \, a_2\ge 0\,,\, a_1+a_3 \ge0 \,,\,   a_2+a_4\ge0  \,,\, a_3\le0\,
\right\}\,,\\
&
\check{\mathcal{C}}_{\{\gamma_2 +\gamma_3 , -\gamma_1-  \gamma_2-\gamma_3  , \gamma_1+\gamma_2  ,  \gamma_3+ \gamma_4 \} }=
\left \{\,
\sum_{i=1}^4 a_i \gamma_i \, \Big| \, a_2\ge 0\,,\, a_1+a_3 \le0 \,,\,  a_2+a_4\ge0 \,,\, a_3\ge0\,
\right\}\,,\\
&
\check{\mathcal{C}}_{\{\gamma_2+\gamma_3 , -\gamma_1-  \gamma_2 ,-\gamma_3 ,  \gamma_3+ \gamma_4 \} }=
\left \{\,
\sum_{i=1}^4 a_i \gamma_i \, \Big| \, a_2\ge 0\,,\, a_1+a_3 \le a_2+a_4 \,,\, a_2+a_4\le 0\,,\, a_3\ge0\,
\right\}\,,\\
&
\check{\mathcal{C}}_{\{- \gamma_1 , \gamma_1+  \gamma_2+\gamma_3 , -\gamma_3 ,  -\gamma_4 \} }=
\left \{\,
\sum_{i=1}^4 a_i \gamma_i \, \Big| \, a_2\ge 0\,,\, a_1+a_3 \ge0 \,,\,   a_2+a_4\le 0 \,,\, a_3\le a_2+a_4\,
\right\}\,,\\
&
\check{\mathcal{C}}_{\{\gamma_2 , -\gamma_1-  \gamma_2 , \gamma_3 ,  -\gamma_4 \} }=
\left \{\,
\sum_{i=1}^4 a_i \gamma_i \, \Big| \, a_2\ge 0\,,\, a_1+a_3 \le0 \,,\,a_1+2 a_3\ge a_2+a_4\,,\, a_3\le0\,
\right\}\,,\\
&
\check{\mathcal{C}}_{\{\gamma_1 , -  \gamma_2 , -\gamma_3 -\gamma_4 ,  \gamma_4 \} }=
\left \{\,
\sum_{i=1}^4 a_i \gamma_i \, \Big| \, a_1+a_3\ge a_2\,,\, a_3 \ge a_2+a_4 \,,\,a_1+a_3\le a_2+a_4\,,\, a_3\le0\,
\right\}\,,\\
&
\check{\mathcal{C}}_{\{\gamma_1 , -\gamma_2-  \gamma_3 , \gamma_2 ,  -\gamma_4 \} }=
\left \{\,
\sum_{i=1}^4 a_i \gamma_i \, \Big| \, 0\ge a_1+a_3 \ge a_2\,,\,a_3\le a_2+a_4\,,\, a_1+a_3\ge a_2+a_4\,
\right\}\,,\\
&
\check{\mathcal{C}}_{\{\gamma_1 , -\gamma_2-  \gamma_3 , \gamma_2+\gamma_3+\gamma_4 ,-\gamma_3-  \gamma_4 \} }=
\left \{\,
\sum_{i=1}^4 a_i \gamma_i \, \Big| \, 0\ge a_1+a_3\ge a_2\,,\,  a_2+a_4\ge 0\,,\, a_3\le0\,
\right\}\,,\\
&
\check{\mathcal{C}}_{\{ - \gamma_2 - \gamma_3 , -\gamma_1 , \gamma_1+\gamma_2 , \gamma_3+  \gamma_4 \} }=
\left \{\,
\sum_{i=1}^4 a_i \gamma_i \, \Big| \, a_2\le 0\,,\, a_1+a_3 \le a_2 \,,\, a_2+a_4\ge 0 \,,\, a_3\ge0\,
\right\}\,,\\
&
\check{\mathcal{C}}_{\{- \gamma_1 , -  \gamma_2 , - \gamma_3 ,\gamma_3+   \gamma_4 \} }=
\left \{\,
\sum_{i=1}^4 a_i \gamma_i \, \Big| \,  a_2 \ge a_1+a_3 \ge 2a_2 +a_4 \,,\,a_1+a_3\le a_2+a_4\,,\, a_3\ge0\,
\right\}\,,\\
&
\check{\mathcal{C}}_{\{ -  \gamma_1 , -\gamma_2-  \gamma_3 , \gamma_1+\gamma_2 ,-  \gamma_4 \} }=
\left \{\,
\sum_{i=1}^4 a_i \gamma_i \, \Big| \,  0\ge a_2 \ge a_1+a_3 \ge a_2+a_4\ge a_3\,
\right\}\,,\\
&
\check{\mathcal{C}}_{\{\gamma_1 , \gamma_4 , -\gamma_2-\gamma_3 -\gamma_4 ,  \gamma_2 \} }=
\left \{\,
\sum_{i=1}^4 a_i \gamma_i \, \Big| \, 0 \ge a_2+a_4\ge a_1+a_3 \ge a_2 \,,\, a_3\le a_2+a_4\,
\right\}\,,\\
\end{align*}

\begin{align*}
&
\check{\mathcal{C}}_{\{ - \gamma_2 -\gamma_3 , -\gamma_1 ,\sum_{i=1}^4 \gamma_i,-\gamma_3-   \gamma_4 \} }=
\left \{\,
\sum_{i=1}^4 a_i \gamma_i \, \Big| \,  0\ge a_2\ge a_1+a_3  \,,\, a_2+a_4\ge0\,,\, a_3\le0\,
\right\}\,,\\
&
\check{\mathcal{C}}_{\{ - \gamma_1, -  \gamma_2 ,-  \gamma_3 -\gamma_4,  \gamma_4 \} }=
\left \{
\sum_{i=1}^4 a_i \gamma_i \, \Big| \, a_1+a_3\le  a_2\,,\, a_1+2a_3 \ge 2a_2+a_4 \,,\,a_1+a_3\le a_2+a_4\,,\, a_3\le0
\right\},\\
&
\check{\mathcal{C}}_{\{-\gamma_1 , -  \gamma_2 , -\gamma_3 ,-  \gamma_4 \} }=
\left \{\,
\sum_{i=1}^4 a_i \gamma_i \, \Big| \, a_2\ge a_1+a_3 \ge a_2+a_4\,,\, a_2+a_4 \ge0 a_1+2a_3\ge 2a_2+a_4\,
\right\}\,,\\
&
\check{\mathcal{C}}_{\{\gamma_2 +\gamma_3, -\gamma_1-  \gamma_2-\gamma_3 , \sum_{i=1}^4\gamma_i ,  -\gamma_3-\gamma_4 \} }=
\left \{\,
\sum_{i=1}^4 a_i \gamma_i \, \Big| \, a_2\ge 0\,,\, a_1+a_3 \le0 \,,\,  a_2+a_4\ge0 \,,\, a_3\le0\,
\right\}\,,\\
&
\check{\mathcal{C}}_{\{\gamma_2 +\gamma_3, \gamma_4 ,- \gamma_3-\gamma_4 , - \gamma_1-\gamma_2 \} }=
\left \{\,
\sum_{i=1}^4 a_i \gamma_i \, \Big| \, a_2\ge 0\,,\, a_3 \ge a_2+a_4 \ge a_1+a_3\,,\, a_3\le0\,
\right\}\,,\\
&
\check{\mathcal{C}}_{\{ - \gamma_2  -\gamma_3, -\gamma_1-  \gamma_2 , \gamma_2 ,\gamma_3+   \gamma_4 \} }=
\left \{\,
\sum_{i=1}^4 a_i \gamma_i \, \Big| \, a_2\le 0\,,\, a_1+a_3 \le 2a_2+a_4 \,,\, a_2+a_4\le 0\,,\, a_3\ge0\,
\right\}\,,\\
&
\check{\mathcal{C}}_{\{\gamma_2+\gamma_3  , -\gamma_1-  \gamma_2 -\gamma_3, \gamma_1+\gamma_2 ,-  \gamma_4 \} }=
\left \{\,
\sum_{i=1}^4 a_i \gamma_i \, \Big| \, a_2\ge 0\,,\, 0\ge a_1+a_3 \ge a_2+a_4 \ge a_3\,
\right\}\,,\\
&
\check{\mathcal{C}}_{\{\gamma_2 +\gamma_3, -\gamma_1-  \gamma_2 , -\gamma_3 ,  -\gamma_4 \} }=
\left \{\,
\sum_{i=1}^4 a_i \gamma_i \, \Big| \, a_2\ge 0\,,\, a_1+a_3 \ge a_2+a_4\ge a_1+2a_3 \,,\, a_3\ge a_2+a_4\,
\right\}\,,\\
&
\check{\mathcal{C}}_{\{- \gamma_1 , \gamma_4 ,-\gamma_2- \gamma_3-\gamma_4 ,  \gamma_1+\gamma_2 \} }=
\left \{
\sum_{i=1}^4 a_i \gamma_i \Big|  a_2\ge  a_1+a_3 \ge 2a_2 +a_4 \,,\,a_1+a_3\le a_2+a_4\,,\, a_3\le a_2+a_4
\right\},\\
&
\check{\mathcal{C}}_{\{- \gamma_2-\gamma_3 , \gamma_2+  \gamma_3+\gamma_4 , -\sum_{i=1}^4\gamma_i,  \gamma_1+\gamma_2 \} }=
\left \{\,
\sum_{i=1}^4 a_i \gamma_i \, \Big| \, a_2\le 0\,,\, a_1+a_3 \le 2a_2+a_4 \,,\,0\ge a_2+a_4\ge a_3\,
\right\}\,,\\
&
\check{\mathcal{C}}_{\{\gamma_2+\gamma_3 , \gamma_4 , -\sum_{i=1}^4\gamma_i ,  \gamma_1+\gamma_2\} }=
\left \{\,
\sum_{i=1}^4 a_i \gamma_i \, \Big| \, a_2\ge 0\,,\, 0 \ge a_2+a_4 \ge a_1+a_3 \,,\,a_3\le a_2+a_4\,,\, a_3\ge0\,
\right\}\,,\\
&
\check{\mathcal{C}}_{\{- \gamma_2-\gamma_3 , -\gamma_1-  \gamma_2 , \gamma_2+\gamma_3+\gamma_4 ,  -\gamma_3-\gamma_4 \} }=
\left \{\,
\sum_{i=1}^4 a_i \gamma_i \, \Big| \, a_2\le 0\,,\, a_1+a_3 \le 2 a_2+ a_4 \,,\, 0 \ge a_3\ge a_2+a_4\,
\right\}\,,\\
&
\check{\mathcal{C}}_{\{\gamma_4 , -\gamma_2-  \gamma_3 -\gamma_4, \gamma_2, - \gamma_1-\gamma_2 \} }=
\left \{
\sum_{i=1}^4 a_i \gamma_i  \Big|  a_3\ge a_2+a_4 \ge a_1+a_3 \ge 2a_2+a_4\,,\, a_1+2a_3 \le 2a_2+a_4
\right\},\\
&
\check{\mathcal{C}}_{\{- \gamma_2 -\gamma_3 , -\gamma_1-  \gamma_2 , \gamma_2 ,  -\gamma_4 \} }=
\Big \{\,
\sum_{i=1}^4 a_i \gamma_i \, \Big| \, a_2\le 0\,,\, a_3 \ge a_2+a_4 \,,\, a_1+a_3\ge a_2+a_4\,,\, a_1+2a_3\le 2a_2+a_4\,
\Big\}\,.
\end{align*}

\clearpage

Each of the 42 dual cones is generated by four boundary rays, and there are in total fourteen distinct boundary rays which are the generators of line defects in the $A_4$ Argyres-Douglas theory. We will denote them by $A_i$ and $B_i$ with $i=1,\cdots,7$. Their core charges are
\begin{align}
&\textbf{RG}(A_1)  =  -\gamma_2 +\gamma_4 \,,~~~~~~~~ 
\textbf{RG}(A_2)  =  \gamma_2 -\gamma_4 \,,~~~~\,\,\,~
\textbf{RG}(A_3)  =  - \gamma_1- \gamma_2 +\gamma_4 \,,~~~~\notag\\
&\textbf{RG}(A_4)  =  \gamma_1-\gamma_3 -\gamma_4 \,,~~~~
\textbf{RG}(A_5)  = - \gamma_1+ \gamma_3\,,~~~~
\textbf{RG}(A_6)  =  \gamma_1 -\gamma_3 \,,~~~~\notag\\
&\textbf{RG}(A_7)  =  -\gamma_1-\gamma_4 \,,~~~~\\
&\textbf{RG}(B_1)  =   -\gamma_1 \,,~~~~~~~~~~~~~\,~
\textbf{RG}(B_2)  =  -\gamma_3 -\gamma_4 \,,~~~~
\textbf{RG}(B_3)  = - \gamma_2 -\gamma_3 \,,~~~~\notag\\
&
\textbf{RG}(B_4)  =-  \gamma_1-\gamma_2 \,,~~~~~~~~
\textbf{RG}(B_5)  =   -\gamma_4 \,,~~~~\,\,\,\,\,~~~~
\textbf{RG}(B_6)  =  \gamma_1 \,,~~~~~~~\notag\\
&\textbf{RG}(B_7)  =  \gamma_4 \,.\notag
\end{align}
Their associated framed quivers are shown in Figure \ref{fig:A4framedquiver}. By applying the mutation method to these framed quivers, we obtain the generating functions for these defects \eqref{A4generating}.

\begin{figure}
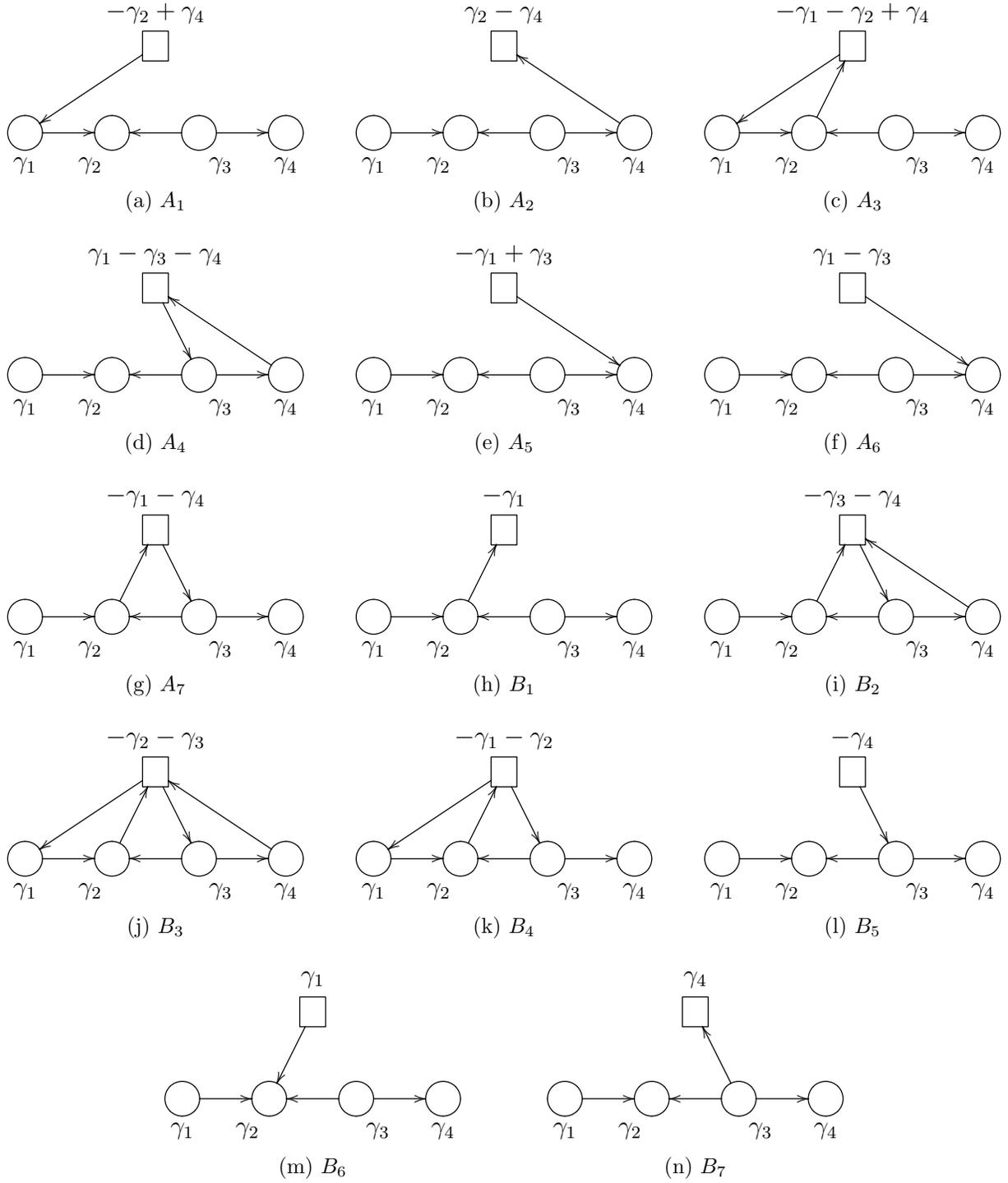

\centering
 \subfloat[$A_1$]{\xy  0;<1pt,0pt>:<0pt,-1pt>::
(-60,0) *+{}*\cir<8pt>{} ="1",
(-20,0) *+{}*\cir<8pt>{} ="2",
(20,0) *+{}*\cir<8pt>{} ="3",
(60,0) *+{}*\cir<8pt>{} ="4",
(0,-40) *+{\color{white}{1}}*\frm{-} ="5",
(-60, 15) *+{\gamma_{1}} ="a",
(-30, 15) *+{\gamma_{2}} ="b",
(30, 15) *+{\gamma_{3}} ="c",
(60, 15) *+{\gamma_{4}} ="d",
(0, -55) *+{-\gamma_2+\gamma_4} ="e",
\ar @{->} "1"; "2"
\ar @{<-} "2"; "3"
\ar @{->} "3"; "4"
\ar @{->} "5" ;"1"
\endxy}
~~~~~
\subfloat[$A_2$]{\xy  0;<1pt,0pt>:<0pt,-1pt>::
(-60,0) *+{}*\cir<8pt>{} ="1",
(-20,0) *+{}*\cir<8pt>{} ="2",
(20,0) *+{}*\cir<8pt>{} ="3",
(60,0) *+{}*\cir<8pt>{} ="4",
(0,-40) *+{\color{white}{1}}*\frm{-} ="5",
(-60, 15) *+{\gamma_{1}} ="a",
(-30, 15) *+{\gamma_{2}} ="b",
(30, 15) *+{\gamma_{3}} ="c",
(60, 15) *+{\gamma_{4}} ="d",
(0, -55) *+{\gamma_2-\gamma_4} ="e",
\ar @{->} "1"; "2"
\ar @{<-} "2"; "3"
\ar @{->} "3"; "4"
\ar @{->} "4" ;"5"
\endxy}
~~~~~
\subfloat[$A_3$]{\xy  0;<1pt,0pt>:<0pt,-1pt>::
(-60,0) *+{}*\cir<8pt>{} ="1",
(-20,0) *+{}*\cir<8pt>{} ="2",
(20,0) *+{}*\cir<8pt>{} ="3",
(60,0) *+{}*\cir<8pt>{} ="4",
(0,-40) *+{\color{white}{1}}*\frm{-} ="5",
(-60, 15) *+{\gamma_{1}} ="a",
(-30, 15) *+{\gamma_{2}} ="b",
(30, 15) *+{\gamma_{3}} ="c",
(60, 15) *+{\gamma_{4}} ="d",
(0, -55) *+{-\gamma_1-\gamma_2+\gamma_4} ="e",
\ar @{->} "1"; "2"
\ar @{<-} "2"; "3"
\ar @{->} "3"; "4"
\ar @{->} "5" ;"1"
\ar @{->} "2";"5"
\endxy}\\
\subfloat[$A_4$]{\xy  0;<1pt,0pt>:<0pt,-1pt>::
(-60,0) *+{}*\cir<8pt>{} ="1",
(-20,0) *+{}*\cir<8pt>{} ="2",
(20,0) *+{}*\cir<8pt>{} ="3",
(60,0) *+{}*\cir<8pt>{} ="4",
(0,-40) *+{\color{white}{1}}*\frm{-} ="5",
(-60, 15) *+{\gamma_{1}} ="a",
(-30, 15) *+{\gamma_{2}} ="b",
(30, 15) *+{\gamma_{3}} ="c",
(60, 15) *+{\gamma_{4}} ="d",
(0, -55) *+{\gamma_1-\gamma_3-\gamma_4} ="e",
\ar @{->} "1"; "2"
\ar @{<-} "2"; "3"
\ar @{->} "3" ;"4"
\ar @{->} "5"; "3"
\ar @{->} "4" ;"5"
\endxy}
~~~~~
\subfloat[$A_5$]{\xy  0;<1pt,0pt>:<0pt,-1pt>::
(-60,0) *+{}*\cir<8pt>{} ="1",
(-20,0) *+{}*\cir<8pt>{} ="2",
(20,0) *+{}*\cir<8pt>{} ="3",
(60,0) *+{}*\cir<8pt>{} ="4",
(0,-40) *+{\color{white}{1}}*\frm{-} ="5",
(-60, 15) *+{\gamma_{1}} ="a",
(-30, 15) *+{\gamma_{2}} ="b",
(30, 15) *+{\gamma_{3}} ="c",
(60, 15) *+{\gamma_{4}} ="d",
(0, -55) *+{-\gamma_1+\gamma_3} ="e",
\ar @{->} "1"; "2"
\ar @{<-} "2"; "3"
\ar @{->} "3"; "4"
\ar @{->} "5" ;"4"
\endxy}
~~~~~
\subfloat[$A_6$]{\xy  0;<1pt,0pt>:<0pt,-1pt>::
(-60,0) *+{}*\cir<8pt>{} ="1",
(-20,0) *+{}*\cir<8pt>{} ="2",
(20,0) *+{}*\cir<8pt>{} ="3",
(60,0) *+{}*\cir<8pt>{} ="4",
(0,-40) *+{\color{white}{1}}*\frm{-} ="5",
(-60, 15) *+{\gamma_{1}} ="a",
(-30, 15) *+{\gamma_{2}} ="b",
(30, 15) *+{\gamma_{3}} ="c",
(60, 15) *+{\gamma_{4}} ="d",
(0, -55) *+{\gamma_1-\gamma_3} ="e",
\ar @{->} "1"; "2"
\ar @{<-} "2"; "3"
\ar @{->} "3"; "4"
\ar @{->} "5" ;"4"
\endxy}\\
\subfloat[$A_7$]{\xy  0;<1pt,0pt>:<0pt,-1pt>::
(-60,0) *+{}*\cir<8pt>{} ="1",
(-20,0) *+{}*\cir<8pt>{} ="2",
(20,0) *+{}*\cir<8pt>{} ="3",
(60,0) *+{}*\cir<8pt>{} ="4",
(0,-40) *+{\color{white}{1}}*\frm{-} ="5",
(-60, 15) *+{\gamma_{1}} ="a",
(-30, 15) *+{\gamma_{2}} ="b",
(30, 15) *+{\gamma_{3}} ="c",
(60, 15) *+{\gamma_{4}} ="d",
(0, -55) *+{-\gamma_1-\gamma_4} ="e",
\ar @{->} "1"; "2"
\ar @{<-} "2"; "3"
\ar @{->} "3"; "4"
\ar @{->} "2" ;"5"
\ar @{->} "5";"3"
\endxy}~~~~~
\subfloat[$B_1$]{\xy  0;<1pt,0pt>:<0pt,-1pt>::
(-60,0) *+{}*\cir<8pt>{} ="1",
(-20,0) *+{}*\cir<8pt>{} ="2",
(20,0) *+{}*\cir<8pt>{} ="3",
(60,0) *+{}*\cir<8pt>{} ="4",
(0,-40) *+{\color{white}{1}}*\frm{-} ="5",
(-60, 15) *+{\gamma_{1}} ="a",
(-30, 15) *+{\gamma_{2}} ="b",
(30, 15) *+{\gamma_{3}} ="c",
(60, 15) *+{\gamma_{4}} ="d",
(0, -55) *+{-\gamma_1} ="e",
\ar @{->} "1"; "2"
\ar @{<-} "2"; "3"
\ar @{->} "3"; "4"
\ar @{->} "2" ;"5"
\endxy}
~~~~~
\subfloat[$B_2$]{\xy  0;<1pt,0pt>:<0pt,-1pt>::
(-60,0) *+{}*\cir<8pt>{} ="1",
(-20,0) *+{}*\cir<8pt>{} ="2",
(20,0) *+{}*\cir<8pt>{} ="3",
(60,0) *+{}*\cir<8pt>{} ="4",
(0,-40) *+{\color{white}{1}}*\frm{-} ="5",
(-60, 15) *+{\gamma_{1}} ="a",
(-30, 15) *+{\gamma_{2}} ="b",
(30, 15) *+{\gamma_{3}} ="c",
(60, 15) *+{\gamma_{4}} ="d",
(0, -55) *+{-\gamma_3-\gamma_4} ="e",
\ar @{->} "1"; "2"
\ar @{<-} "2"; "3"
\ar @{->} "3"; "4"
\ar @{->} "2" ;"5"
\ar @{->} "5"; "3"
\ar @{->}"4";"5"
\endxy}
\\\subfloat[$B_3$]{\xy  0;<1pt,0pt>:<0pt,-1pt>::
(-60,0) *+{}*\cir<8pt>{} ="1",
(-20,0) *+{}*\cir<8pt>{} ="2",
(20,0) *+{}*\cir<8pt>{} ="3",
(60,0) *+{}*\cir<8pt>{} ="4",
(0,-40) *+{\color{white}{1}}*\frm{-} ="5",
(-60, 15) *+{\gamma_{1}} ="a",
(-30, 15) *+{\gamma_{2}} ="b",
(30, 15) *+{\gamma_{3}} ="c",
(60, 15) *+{\gamma_{4}} ="d",
(0, -55) *+{-\gamma_2-\gamma_3} ="e",
\ar @{->} "1"; "2"
\ar @{<-} "2"; "3"
\ar @{->} "3"; "4"
\ar @{->} "4" ;"5"
\ar @{->} "5";"1"
\ar @{->} "2";"5"
\ar @{->} "5";"3"
\endxy}
~~~~~
\subfloat[$B_4$]{\xy  0;<1pt,0pt>:<0pt,-1pt>::
(-60,0) *+{}*\cir<8pt>{} ="1",
(-20,0) *+{}*\cir<8pt>{} ="2",
(20,0) *+{}*\cir<8pt>{} ="3",
(60,0) *+{}*\cir<8pt>{} ="4",
(0,-40) *+{\color{white}{1}}*\frm{-} ="5",
(-60, 15) *+{\gamma_{1}} ="a",
(-30, 15) *+{\gamma_{2}} ="b",
(30, 15) *+{\gamma_{3}} ="c",
(60, 15) *+{\gamma_{4}} ="d",
(0, -55) *+{- \gamma_1-\gamma_2} ="e",
\ar @{->} "1"; "2"
\ar @{<-} "2"; "3"
\ar @{->} "3"; "4"
\ar @{->} "5" ;"1"
\ar @{->} "2";"5"
\ar @{->} "5";"3"
\endxy}
~~~~~
\subfloat[$B_5$]{\xy  0;<1pt,0pt>:<0pt,-1pt>::
(-60,0) *+{}*\cir<8pt>{} ="1",
(-20,0) *+{}*\cir<8pt>{} ="2",
(20,0) *+{}*\cir<8pt>{} ="3",
(60,0) *+{}*\cir<8pt>{} ="4",
(0,-40) *+{\color{white}{1}}*\frm{-} ="5",
(-60, 15) *+{\gamma_{1}} ="a",
(-30, 15) *+{\gamma_{2}} ="b",
(30, 15) *+{\gamma_{3}} ="c",
(60, 15) *+{\gamma_{4}} ="d",
(0, -55) *+{-\gamma_4} ="e",
\ar @{->} "1"; "2"
\ar @{<-} "2"; "3"
\ar @{->} "3"; "4"
\ar @{->} "5" ;"3"
\endxy}\\
\subfloat[$B_6$]{\xy  0;<1pt,0pt>:<0pt,-1pt>::
(-60,0) *+{}*\cir<8pt>{} ="1",
(-20,0) *+{}*\cir<8pt>{} ="2",
(20,0) *+{}*\cir<8pt>{} ="3",
(60,0) *+{}*\cir<8pt>{} ="4",
(0,-40) *+{\color{white}{1}}*\frm{-} ="5",
(-60, 15) *+{\gamma_{1}} ="a",
(-30, 15) *+{\gamma_{2}} ="b",
(30, 15) *+{\gamma_{3}} ="c",
(60, 15) *+{\gamma_{4}} ="d",
(0, -55) *+{\gamma_1} ="e",
\ar @{->} "1"; "2"
\ar @{<-} "2"; "3"
\ar @{->} "3"; "4"
\ar @{->} "5" ;"2"
\endxy}
~~~~~~~~~
\subfloat[$B_7$]{\xy  0;<1pt,0pt>:<0pt,-1pt>::
(-60,0) *+{}*\cir<8pt>{} ="1",
(-20,0) *+{}*\cir<8pt>{} ="2",
(20,0) *+{}*\cir<8pt>{} ="3",
(60,0) *+{}*\cir<8pt>{} ="4",
(0,-40) *+{\color{white}{1}}*\frm{-} ="5",
(-60, 15) *+{\gamma_{1}} ="a",
(-30, 15) *+{\gamma_{2}} ="b",
(30, 15) *+{\gamma_{3}} ="c",
(60, 15) *+{\gamma_{4}} ="d",
(0, -55) *+{\gamma_4} ="e",
\ar @{->} "1"; "2"
\ar @{<-} "2"; "3"
\ar @{->} "3"; "4"
\ar @{->} "3" ;"5"
\endxy}

\caption{The framed quivers associated to the fourteen generators of line defects $A_i$, $B_i$ in the $A_4$ Argyres-Douglas theory.  The core charges are labeled above the framed nodes (the square nodes).}\label{fig:A4framedquiver}
\end{figure}

 \section{Affine Characters of Kac-Moody Algebra at Negative Level}\label{sec:AffineCharacter}

In this appendix we review a generalization  of the Weyl-Kac formula, known as the Kazhdan-Lusztig conjecture \cite{kazhdan1979representations}, for  affine characters of Kac-Moody algebra at negative levels, following \cite{DeVos:1995an}.  We will compute the affine characters for several modules in $\widehat{su(2)}_{-{4\over3}}$ and $\widehat{so(8)}_{-2}$, which are the chiral algebras of the $A_3$ Argyres-Douglas theory and the $SU(2)$ with $N_f=4$ flavors theory, respectively.

\subsection{Generalities on Affine Lie Algebra}

We begin by reviewing some basic facts about affine Lie algebra (see, for example, \cite{philippe1997conformal}). Let $\mathfrak{g}$ be an affine Lie algebra associated with a finite dimensional simple Lie algebra $\underline{\mathfrak{g}}$ of rank $r$.  An affine weight $\lambda$ of $\mathfrak{g}$ will be denoted by
\begin{align}\label{eigenvalue}
\lambda = ( \underline{\lambda} ;  k ; n)\,,
\end{align}
where $\underline{\lambda}$ is a weight of the finite dimensional Lie algebra $\underline{\mathfrak{g}}$. $k$ is the level of the weight and $n$ is the eigenvalue with respect to $-L_0$.  The inner product between weights is given by
\begin{align}
( \lambda_1, \lambda_2 ) =  (\underline{ \lambda}_1,\underline{ \lambda }_2)  + k_1 n _2 + k_2 n_1\,. 
\end{align}

The simple roots of the affine Lie algebra $\mathfrak{g}$ consist of
\begin{align}
\begin{split}
&\alpha_0 = ( - \theta ; 0 ;1)  =   -\theta + \delta\,,\\
&\alpha_i  =  (  \underline{\alpha}_i ;  0  ;  0 ),~~~~i=1,\cdots,r\,,
\end{split}
\end{align}
where $\underline\alpha_i$'s are the simple roots of $\underline{\mathfrak{g}}$.  $\theta$ is the highest root of $\underline{\mathfrak{g}}$ normalized such that $|\theta|^2=2$.  $\delta$ is   defined as $\delta = ( 0  ; 0 ;1)$. 
Since $( \delta, \delta)=0$, $n\delta$ is called an imaginary root for all $n$. The other roots are said to be real.

The set of positive roots $\Delta_+$ of the affine Lie algebra is
\begin{align}
\Delta_+ =  \{ \,  \alpha+ n\delta \, |  \, n>0,~ \alpha \in \underline{\Delta} \,\}\cup \{ \alpha| \alpha\in \underline{\Delta}_+\}\,,
\end{align}
where $\underline{\Delta}$ and $\underline{\Delta}_+$ are the sets of roots and positive roots of the finite dimensional Lie algebra $\underline{\mathfrak{g}}$, respectively.  The set of real positive roots $\Delta_+^{re}$ is defined as 
\begin{align}
\Delta_+^{re}  = \Delta_+/ \{ n\delta\}  =  \{\,  \alpha+ n\delta \, |\, n>0,~\alpha\in \underline{\Delta},~\alpha\neq0\,\} \cup \{ \alpha| \alpha\in \underline{\Delta}_+\}\,.
\end{align}

The Cartan matrix $A_{ij}$ of the affine Lie algebra $\mathfrak{g}$ is defined as $A_{ij}  =  (\alpha_i , \alpha_j^\vee)$ with $0\le i,j\le r$. Note in particular, $A_{00}=2$ and $A_{0i}= - (\theta, \underline{\alpha}_i^\vee )  $.

The marks $a_i$ and comarks $a_i^\vee$ ($i=1,\cdots ,r$) for the finite Lie algebra $\underline{\mathfrak{g}}$  are defined as
\begin{align}
\theta = \sum_{i=1}^r a_i \alpha_i =  \sum_{i=1}^r a_i^\vee \alpha_i^\vee\,.
\end{align}
For the affine Lie algebra $\mathfrak{g}$, the mark and comark of the extra simple root $\alpha_0$ is defined to be 1, $a_0 = a_0^\vee=1$.   The dual Coxeter number is defined as $h^\vee = 1+ \sum_{i=1}^r a_i^\vee =  \sum_{i=0}^r a_i^\vee$.

%

We will sometimes label an affine weight $\lambda$ by its Dynkin labels,
\begin{align}\label{dynkin}
\lambda=  [ \lambda_0 , \lambda_1, \cdots, \lambda_r] \,,
\end{align}
where $\lambda_ i  = ( \lambda,  \alpha_i^\vee)$.  Note that the Dynkin labels have $r+1$ components, one less than the $(\underline{\lambda} , k , n)$ notation.  In other words, the Dynkin labels do not completely specify an affine weight, but up to an imaginary root $n \delta$. Note that the level $k$ of a weight is related to the Dynkin labels by
\begin{align}\label{level}
k  = \sum_{i=0}^r a_i^\vee \lambda_i\,.
\end{align}
Finally, the affine Weyl vector is defined as $\rho  = [1,1,\cdots,1]$. 

\subsubsection*{Weyl Reflections}

Let $W$ be the Weyl group of the affine Lie algebra $\mathfrak{g}$. The Weyl reflection $s_\alpha\in W$ generated by real affine root $\alpha$ is given by
\begin{align}
s_\alpha  =  \lambda - (\lambda, \alpha^\vee)  \alpha\,.
\end{align}
The shifted Weyl reflection $\circ$ generated by a real affine root $\alpha$ is defined as
\begin{align}
s_\alpha \circ \lambda =  s_\alpha( \lambda+\rho ) -\rho = \lambda - (\lambda+\rho ,\alpha^\vee)\alpha\,.
\end{align}
The Weyl group $W$ is generated by $s_{\alpha_i}$, where $\alpha_i$'s are the simple roots, with the following relations
\begin{align}
\begin{split}
&s_i^2=1\,,\\
&s_i s_j  = s_j s_i ,~~~~~\,~~~~\text{if}~~A_{ij}=0\,,\\
&(s_i s_j )^{m_{ij}}=1,~~~~~~~\text{if}~~i\neq j\,,
\end{split}
\end{align}
where $m_{ij}=2,3,4,6,\infty$ if the number of lines joining the $i$-th and $j$-th node is $0,1,2,3,4$.

\subsubsection*{The case of $\widehat{su(2)}$}

Let us collect some properties of the affine Lie algebra $\widehat{su(2)}$.  The affine Dynkin diagram has two nodes connected by 4 lines. The Cartan matrix is
\begin{align}
A_{ij}=\left(\begin{array}{cc}2 &- 2 \\ - 2 & 2\end{array}\right)\,.
\end{align}
The highest root of $su(2)$ is $\theta = [2]$. Hence the marks and comarks of $\widehat{su(2)}$ are $a_i = a_i^\vee= (1,1).$ The level of an affine weight $\lambda$ is then given by 
\begin{align}
k = \lambda_0  +\lambda_1\,.
\end{align}
The dual Coxeter number is $h^\vee=2$.

The affine Weyl group of $\widehat{su(2)}$ is generated by $s_0\,,s_1$ satisfying $(s_i)^2=1$. 
The affine Weyl group elements are
\begin{align}
W = \{ s_0, s_1,s_1s_0 ,s_0s_1,\cdots \}\,.
\end{align}

\subsubsection*{The case of $\widehat{so(8)}$}

Let us collect some properties of the affine Lie algebra $\widehat{so(8)}$.  The Cartan matrix is
\begin{align}
A_{ij}=\left(\begin{array}{cccc}2 & -1 & 0 & 0 \\-1 & 2 & -1 & -1 \\0 & -1 & 2 & 0 \\0 & -1 & 0 & 2\end{array}\right)\,,
\end{align}
where the central node in the affine Dynkin diagram is $\alpha_2$.  The highest root of $so(8)$ is $\theta = [0,1,0,0].$ Hence the marks and comarks of $\widehat{so(8)}$ are $a_i = a_i^\vee= (1,1,2,1,1)$.  The level of an affine weight $\lambda$ is then given by 
\begin{align}
k = \lambda_0  +\lambda_1 + 2\lambda_2 + \lambda_3+\lambda_4\,.
\end{align}
The dual Coxeter number is $h^\vee=6$.

The affine Weyl group of $\widehat{so(8)}$ is generated by $s_0\,,s_1\,,s_2\,,s_3\,,s_4$ satisfying
\begin{align}
\begin{split}
&(s_i)^2=1,~~~~~~~~~i=0,1,\cdots,4\,,\\
&s_i s_j = s_j s_i ,~~~~~~i,j  = 0,1,3,4\,,\\
&(s_2 s_i )^3 =1,~~~~~~i=0,1,3,4\,.
\end{split}
\end{align}

\subsection{Affine Characters and the Kazhdan-Lusztig Polynomials}

In this subsection we present the formula for affine characters following \cite{DeVos:1995an} (see also \cite{Beem:2013sza}). We will assume 
\begin{align}
k+h^\vee>0\,,
\end{align}
which is indeed the case for $\widehat{su(2)}_{-{4\over3}}$ and $\widehat{so(8)}_{-2}$.

To every weight $\lambda$, we define a subset $\Delta^{re}_{+,\lambda}$ of the real positive roots of $\mathfrak{g}$ to be
\begin{align}
\Delta^{re}_{+,\lambda} = \{ \,  \alpha\in \Delta^{re}_+ \, | \, ( \lambda, \alpha^\vee ) \in \mathbb{Z}\,\}\,,
\end{align}
and let $W_\lambda$ be the subgroup of the affine Weyl group $W$ generated by $s_{\alpha}$ with $\alpha \in \Delta^{re}_{+,\lambda}$.  In the case when $\lambda$ is integral (i.e. all the Dynkin labels are integers), $W_\lambda=W$.  

Let $\lambda$ be the highest affine weight of a module, then in the orbit
\begin{align}
W_\lambda \circ \lambda
\end{align}
there is exactly one element $\Lambda$ such that the Dynkin labels of $\Lambda+\rho$ are all non-negative,
\begin{align}
( \Lambda+\rho , \alpha_i^\vee) \ge0\,,
\end{align}
where $\alpha_i$'s are the simple roots of the affine Lie algebra $\mathfrak{g}$.  

Let $\text{ch}\,M(\mu)$ be the character of the Verma module  with highest weight $\mu$,
\begin{align}
\text{ch} \,M(\mu)  = {e^\mu  \over \prod_{\alpha \in \Delta_+}(1-e^{-\alpha} )^{\text{mult}(\alpha)}}\,,
\end{align}
where $\Delta_+$ is the set of positive roots for the affine Lie algebra $\mathfrak{g}$ and mult$(\alpha)$ is the multiplicity of the root $\alpha$.  Suppose $\mu=(\underline{\mu}; k ; n)$, then $e^\mu$  is understood as
\begin{align}
e^\mu  = q^{-n}\, \eta^{\underline{\mu}}\,,~~~~
\end{align}
where we have defined a compact notation $\eta^{\underline{\mu}} =   \prod_{i=1}^r\eta_i^{\underline{\mu}_i} $. Here $\eta_i$ are the fugacities and $\underline{\mu}_i$'s are the Dynkin labels of the weight $\underline{\mu}$ of the finite Lie algebra $\mathfrak{\underline{g}}$.  
The character $\text{ch}\, L( \lambda)$ of the irreducible module with the highest weight $\lambda  =  w \circ \Lambda$ is then given by
\begin{align}\label{main}
\text{ch} \, L( w\circ \Lambda)  =  \sum_{\substack{ w'\in W_\Lambda / W_\Lambda^0\\ w'\ge w}} m_{w,w'} \, \text{ch}\, M (w'\circ \Lambda)\,.
\end{align}
Here $W_\Lambda^0$ is the subgroup of $W_\Lambda$ that leaves $\Lambda$ invariant.

To define the order $>$ on the coset $W_\Lambda/W_\Lambda^0$, we first define the Bruhat order on the Weyl group $W_\Lambda$. An arbitrary element $w$ in $W_\Lambda $ can be written as $w=s_{i_1}\cdots s_{i_k}$. An expression of minimal length is called reduced.\footnote{Note that given an element in the Weyl group, the reduced expression may not be unique.} Let $w,w'\in W_\Lambda$, then we write
\begin{align}
w<w'
\end{align}
if the reduced expression for $w$ can be obtained by dropping simple reflections from a reduced expression for $w'$.  The resulting relation $w\le w'$ is called the Bruhat order.

 The order on the coset space $W_\Lambda/W_\Lambda^0$ is then defined as
\begin{align}
w \le w' \, ,~~~~~\text{with}~~w,w'\in W_\Lambda /W_\Lambda^0~~~~~\text{iff}~~~~~\underline w \le \underline w' \,, ~~~~~\text{with}~~\underline w,\underline w' \in W_\Lambda\,,
\end{align}
where $\underline w$ is the minimal  representative of $w$ in the coset, defined by $\ell(\underline w s)>\ell(\underline w)$ for all $s\in W_\Lambda^0$.  Here $\ell$ is the length of a reduced expression of a Weyl group element $s$.  The determination of the multiplicities $m_{w,w'}$ is the content of the Kazhdan-Lusztig conjecture.

\subsubsection{The Kazhdan-Lusztig Conjecture}

The Kazhdan-Lusztig conjecture states that the multiplicities $m_{w,w'}$ are given by the inverse Kazhdan-Lusztig polynomials $\tilde Q^I_{w,w'}(q)$ for the coset $W_\Lambda/W_\Lambda^0$ evaluated at $q=1$,\footnote{The  $I$ superscript means that it is for the coset $W_\Lambda/W_\Lambda^0$ but not for the Weyl group $W_\Lambda$.}
 \begin{align}
 m_{w,w'}  =  \tilde Q^I_{w,w'}(1)\,.
 \end{align}
The inverse  Kazhdan-Lusztig polynomials $\tilde Q^I_{w,w'}(q)$ for the coset $W_\Lambda/W_\Lambda^0$ are in turn related to those $Q_{w,w'}(q)$ of $W_\Lambda$ by
\begin{align}\label{cosetKL}
\tilde Q^I _{x,y}  (q)=\sum_{z\in [y]} Q_{\bar x,z }(q) (-1)^{\ell(\bar x)}(-1)^{\ell(z)}\,,
\end{align}
where $\bar z$ and $\underline z$ are the maximal and minimal representative of the coset $[z]$ of $z$.  For rational $k$ with $k+h^\vee>0$ , which is indeed the case for $\widehat{su(2)}_{-{4\over3}}$ and $\widehat{so(8)}_{-2}$, the Kazhdan-Lusztig conjecture has been proven in \cite{kashiwara1996} (see also \cite{Kashiwara1990,casian1990kazhdan,kashiwara1990kazhdan,kashiwara1995} for earlier works).

  The inverse Kazhdan-Lusztig polynomials for the Weyl group $W_\Lambda$ are determined using a recurrence relation (see, for example, \cite{DeVos:1995an}).  In the case of $\widehat{so(8)}_{-2}$, we will use  the C code \texttt{Coxeter} \cite{du2002computing} to compute the inverse Kazhdan-Lusztig polynomials $Q_{w,w'}(q)$.

\subsubsection*{Weyl-Kac Character Formula}

Consider the special case $\lambda$ itself is a dominant weight, i.e. $(\lambda, \alpha^\vee_i)\ge 0$ for all the simple roots $\alpha_i$.  In this case $\Lambda=\lambda$ and $w=e$ is the identity element in the affine Weyl group $W$. Since $\lambda$ is integral, $W_\Lambda=W_\lambda = W$.  Further, the subgroup $W_\Lambda^0$ is trivial and the coset $W/W_\lambda^0$ is the full Weyl group $W$. In this case we have $\tilde Q_{ e,y } = (-1)^{\ell(y)}$.\footnote{We drop the superscript $I$ because in this case the coset $W/W_\lambda^0$ is the full Weyl group.}  Then the character \eqref{main} for the module with highest weight $\lambda$ reduces to the familiar Weyl-Kac formula.

\subsection{Affine Characters of $\widehat{su(2)}_{-{4\over3}}$}

In this subsection we apply the above formalism to compute the characters of admissible representations \eqref{admissible} of $\widehat{su(2)}_{-{4\over3}}$.  The vacuum character of $\widehat{su(2)}_{-{4\over3}}$ has been previously computed in \cite{Buican:2015ina} (see also \cite{philippe1997conformal}) so we will not repeat it here.  We will explicitly compute the character $\chi_1$ for the admissible representation with highest weight $\Phi_1 = [-{2\over3} ,-{2\over3} ]$.  The character $\chi_2$ for the other admissible representation $\Phi_2 = [0,-{4\over3}]$ is completely analogous.

The real positive roots associated to the highest weight $\lambda= [-{2\over3} ,-{2\over3}]$ is
\begin{align}
\Delta^{re}_{+,\lambda}  =  \{  \, (\alpha_1 ; 0 ; 3m+1) \,|\, m\ge0 \,\}\cup
\{ \, (-\alpha_1 ; 0 ; 3m+2)\, | \, m\ge 0 \,\}\,.
\end{align}
One can easily check that $\langle \lambda+\rho , \alpha^\vee\rangle >0$ for all $\alpha \in \Delta^{re}_{+,\lambda}$, hence there is no need to perform a further Weyl reflection $w$.  In the notations of the previous section, we have $\Lambda= \lambda$ and $w=e$.   Furthermore, the inverse Kazhdan-Lusztig polynomials are just signs in this case,  $\tilde Q_{e, y}=(-1)^{\ell(y)}$.   The character formula \eqref{main} reduces to
\begin{align}\label{KacWakimoto}
\text{ch} \, L(\lambda )  =  \sum_{\substack{ w'\in W_\lambda }} (-1)^{\ell(w')} \, \text{ch}\, M (w'\circ \lambda)\,.
\end{align}
This special case of the character formula is known as the Kac-Wakimoto formula \cite{kac1988modular}.

Let us take a closer look into $W_\lambda$, which is the subgroup of the  affine Weyl group that is generated by roots in $\Delta^{re}_{+,\lambda}$.  The simple roots $\tilde \alpha_0,\, \tilde \alpha_1$ of $\Delta^{re}_{+,\lambda}$ are
\begin{align}
\tilde \alpha_0  = 2\alpha_0 +\alpha_1\,,~~~~~~
\tilde \alpha_1  =  \alpha_0 +2\alpha_1\,.
\end{align}
$W_\lambda$ is then generated by the corresponding Weyl reflections $\tilde s_0,\, \tilde s_1$,
\begin{align}
\tilde s_0  =s_0 s_1s_0 \,,~~~~~\tilde s_1=  s_1s_0s_1\,,
\end{align}
where $s_0$ and $s_1$ are the Weyl reflections generated by the simple roots $\alpha_0$ and $\alpha_1$ of $\widehat{su(2)}$, respectively.  The elements in $W_\lambda$ includes $e$, $(\tilde s_0 \tilde s_1)^{n-1} \tilde s_0$, $(\tilde s_1 \tilde s_0 )^{n-1} \tilde s_1$, $(\tilde s_1 \tilde s_0)^n$, and $(\tilde s_0 \tilde s_1)^n$  with $n\ge1$.  After working out the shifted Weyl reflection on the highest weight $w\circ \lambda$,\footnote{For example, the shifted Weyl reflection of $\tilde s_0$ on $\lambda= [-\frac23 ,-\frac23]$ is
$$
\tilde s_0 \circ \lambda=  \lambda- \langle \lambda+\rho, \tilde\alpha_0\rangle\tilde \alpha_0
= \lambda - \tilde \alpha_0  = ({2\over 3}\alpha_1 ; - {4\over3} ; -2)\,,
$$
and it contributes to the numerator of \eqref{main} by $e^{\mu (\tilde s_0 \circ \lambda)}  = z^{4\over3} q^2$.
} we obtain the character $\chi_1(q,z)$ for  $\lambda=[-{2\over3} ,-{2\over3}]$, 
\begin{align}
\chi_1(q,z)  =  { 1+ \sum_{n=1}^\infty  (-1)^n  \left( z^{- 2n } q^{ {n\over 2} (3n-1)}  + z^{2n}  q^{{n\over2}(3n+1)}\right)\over  (1-z^{-2} ) \prod_{n=1}^\infty   (1-q^n ) (1-z^2 q^n ) (1-z^{-2} q^n)}\,.
\end{align}
Similarly the character $\chi_2(q,z)$ for $[0,-{4\over3}]$ is
\begin{align}
&\chi_{2} (q,z)={ 1+ \sum_{n=1}^\infty (-1)^n \left(z^{2n} q^{ {n\over 2} (3n-1)}+z^{-2n}q^{ {n\over 2} (3n+1)}\right) \over (1-z^{-2}) \prod_{n=1}^\infty (1-q^n )(1-z^2 q^n )(1-z^{-2}q^n)}\,.
\end{align}

\subsection{Affine Characters of $\widehat{so(8)}_{-2}$}

In this subsection we will record the answers of the affine characters for several highest weight modules in $\widehat{so(8)}_{-2}$. The line defect indices of the $SU(2)$ with $N_f=4$ flavors theory turn out to be linear combinations of these affine characters.  The computation of these characters are done with the help of \texttt{Mathematica} and the C code \texttt{Coxeter} \cite{du2002computing} which computes the Kazhdan-Lusztig polynomials efficiently.  The vacuum character of $\widehat{so(8)}_{-2}$ has been previously computed in \cite{Beem:2013sza}.

Recall that to apply the Kazhdan-Lusztig formula for a given  module with highest affine weight $\lambda$, we need to find an element $w$ of the affine Weyl group $W$ such that $\Lambda = w^{-1} \circ \lambda$ has all affine Dynkin labels no smaller than -1.  We list the affine Dynkin labels, dimensions\footnote{A nontrivial module of an affine Lie algebra is of course infinite dimensional. Here by dimension we mean the dimension of the finite Lie algebra representation for the level zero states.}, $\Lambda$, $w$ of the highest weight modules that we will compute their characters below:
\begin{align*}
\left.\begin{array}{|c|c|c|c|}\hline \text{Affine Dynkin Label} & \text{Dimension} & \Lambda & w \\\hline [-2,0,0,0,0] & \textbf{1} & [0,0,-1,0,0] & s_0 \\\hline [-3,1,0,0,0] & \textbf{8}_v & [0,0,0,-1,-1] & s_0s_2 \\\hline [-4,2,0,0,0] & \textbf{35}_v & [0,0,-1,0,0] & s_0s_2s_3s_4 \\\hline [-5,3,0,0,0] & \textbf{112}_v & [-1,-1,0,0,0] & s_0s_2s_3s_4s_2 \\\hline [-6,4,0,0,0] & \textbf{294}_v & [0,0,-1,0,0] & s_0s_2s_3s_4s_2s_1s_0 \\\hline [-7,5,0,0,0] & \textbf{672}_v & [0,0,0,-1,-1] & s_0s_2s_3s_4s_2s_1s_0s_2 \\\hline [-8,6,0,0,0] & \textbf{1386}_v & [0,0,-1,0,0] & s_0s_2s_3s_4s_2s_1s_0s_2s_3s_4 \\\hline [-9,7,0,0,0] & \textbf{2640}_v & [-1,-1,0,0,0] & s_0s_2s_3s_4s_2s_1s_0s_2s_3s_4s_2 \\\hline [-10,8,0,0,0] & \textbf{4719}_v & [0,0,-1,0,0] & s_0s_2s_3s_4s_2s_1s_0s_2s_3s_4s_2s_1s_0 \\\hline [-11,9,0,0,0] & \textbf{8008}_v & [0,0,0,-1,-1] & s_0s_2s_3s_4s_2s_1s_0s_2s_3s_4s_2s_1s_0s_2 \\\hline [-12,10,0,0,0] & \textbf{13013}_v & [0,0,-1,0,0] & s_0s_2s_3s_4s_2s_1s_0s_2s_3s_4s_2s_1s_0s_2s_3s_4 \\\hline \end{array}\right.
\end{align*}

We will denote the affine character of a highest weight module with affine Dynkin labels $[a_0,a_1,a_2,a_3,a_4]$ by $\chi_{[a_0,a_1,a_2,a_3,a_4]}$. We record several affine characters of $\widehat{so(8)}_{-2}$  with flavor fugacities set to be 1 below:
\begin{align*}
\chi_{[-2,0,0,0,0]} &= 1+28q + 329q^2+ 2632 q^3 + 16380 q^4 + 85764q^5 + 393 589 q^6\\
&+1 628 548 q^7 + 6 190 527 q^8 + 21 921 900 q^9 + 73070 291 q^{10}  + 231 118 384 q^{11}\\
& + 698 128 389 q^{12}  +2024433460 q^{13}+5659730075 q^{14}+\mathcal{O}(q^{15})\,,\\
{\chi}_{[-3,1,0,0,0]}& = 8+168q + 1904q^2+15512q^3+101696q^4+569072q^5+2817640q^6  \\
&
+ 12642016 q^7+52275216q^8+201716032q^9
 +  733326440  q^{10}\\
 &   +   2530609536 q^{11}+
\mathcal{O}(q^{12})\,,\\
{\chi}_{[-4,2,0,0,0]}&= 35+630q + 6524q^2+ 49490 q^3  +305795q^4 +1625060q^5+ 7683550  q^6 \\
&
+ 33058956  q^7  + 131529944   q^8
+   489700512q^9   + 1721754391 q^{10}  +     5757937528 q^{11}\\
&  + 18421706924  q^{12}+ 56652322636 q^{13}+168128863196 q^{14}
+\mathcal{O}(q^{15})\,,
\end{align*}
\begin{align*}
{\chi}_{[-5,3,0,0,0]}& = 112+1840q + 17920q^2+130480q^3+783440q^4 + 4080272q^5\\
&+19021296 q^6+81047568q^7+320390944 q^8 + 1188177312q^9\\
& +4169249728q^{10} +13936198304 q^{11}+\mathcal{O}(q^{12})\,,\\
{\chi}_{[-6,4,0,0,0]}& = 294+ 4557q + 42516q^2+  299103  q^3+  1744106q^4 + 8852963q^5  \\
&   +40 326 132 q^6  + 168 224 525 q^7 +652 089 872 q^8 + 2 374 316 228 q^9 + 8188532296 q^{10}\\
&+26926301206 q^{11}+84872860408q^{12}+
\mathcal{O}(q^{13})\,,\\
{\chi}_{[-7,5,0,0,0]} &= 672+ 10016q + 90608q^2+621264 q^3+3547040q^4+
17690960 q^5    \\
&+ 79410464 q^6  +327212704 q^7 + 1255299568  q^8 + 4530910720 q^9 +  \mathcal{O}(q^{10})\,,\\
{\chi}_{[-8,6,0,0,0]} &= 1386+ 20097q + 177716q^2
+1194963 q^3+6707204 q^4+32946053 q^5 \\
& + 145853498 q^6   +593383028 q^7+
2249609656q^8 +  8030084594 q^9\\
& +27204209116 q^{10}+
\mathcal{O}(q^{11})\,,\\
{\chi}_{[-9,7,0,0,0]} &= 2640+37520q+  326144  q^2+2160144 q^3+11962832  q^4 +58063376  q^5 \\
&+  254318288  q^6+ 1024821136 q^7  +\mathcal{O}(q^8)\,,\\
{\chi}_{[-10,8,0,0,0]}&= 4719+66066 q+566748 q^2+3709524 q^3+20324192q^4 +97685672  q^5\\
&+424021332  q^6  +1694405948q^7+  \mathcal{O}(q^8)\,,\\
{\chi}_{[-11,9,0,0,0]} &= 8008+110824 q+  940912   q^2+\mathcal{O}(q^3)\,,\\
{\chi}_{[-12,10,0,0,0]} &= 13013+178464 q+  \mathcal{O}(q^2)\,.
\end{align*}

\bibliography{DefectIndicesFinal}{}
\bibliographystyle{utphys}
\end{document}